\documentclass{article}
\usepackage{amssymb}

\usepackage{graphicx}
\usepackage{amsmath}

\newtheorem{theorem}{Theorem}

\newtheorem{definition}[theorem]{Definition}

\newtheorem{lemma}[theorem]{Lemma}

\newtheorem{proposition}[theorem]{Proposition}
\newtheorem{remark}[theorem]{Remark}

\newenvironment{proof}[1][Proof]{\textbf{#1.} }{\ \rule{0.5em}{0.5em}}
\input{tcilatex}

\begin{document}

\author{V. S. Borisov \thanks{%
The Pearlstone Center for Aeronautical Engineering Studies, Department of
Mechanical Engineering, Ben-Gurion University of the Negev, Beer-Sheva,
Israel. E-mail: viatslav@bgu.ac.il} \and M. Mond \thanks{%
The Pearlstone Center for Aeronautical Engineering Studies, Department of
Mechanical Engineering, Ben-Gurion University of the Negev, Beer-Sheva,
Israel. E-mail: mond@bgu.ac.il}}
\title{On monotonicity, stability, and construction of central schemes for
hyperbolic conservation laws with source terms (Revised Version)}
\maketitle

\begin{abstract}
The monotonicity and stability of difference schemes for, in general,
hyperbolic systems of conservation laws with source terms are studied. The
basic approach is to investigate the stability and monotonicity of a
non-linear scheme in terms of its corresponding scheme in variations. Such
an approach leads to application of the stability theory for linear equation
systems to establish stability of the corresponding non-linear scheme. The
main methodological innovation is the theorems establishing the notion that
a non-linear scheme is stable (and monotone) if the corresponding scheme in
variations is stable (and, respectively, monotone). Criteria are developed
for monotonicity and stability of difference schemes associated with the
numerical analysis of systems of partial differential equations. The theorem
of Friedrichs (1954) is generalized to be applicable to variational schemes
with non-symmetric matrices. A new modification of the central
Lax-Friedrichs (LxF) scheme is developed to be of the second order accuracy.
A monotone piecewise cubic interpolation is used in the central schemes to
give an accurate approximation for the model in question. The stability and
monotonicity of the modified scheme are investigated. Some versions of the
modified scheme are tested on several conservation laws, and the scheme is
found to be accurate and robust. As applied to hyperbolic conservation laws
with, in general, stiff source terms, it is constructed a second order
scheme based on operator-splitting techniques.
\end{abstract}


\section{Introduction\label{Introduction}}

We are mainly concerned with the stability and monotonicity \cite{Borisov
and Sorek 2004} of difference schemes for hyperbolic systems of conservation
laws with source terms. Such systems are used to describe many physical
problems of great practical importance in magneto-hydrodynamics, kinetic
theory of rarefied gases, linear and nonlinear waves, viscoelasticity,
multi-phase flows and phase transitions, shallow waters, etc. (see, e.g., 
\cite{Bereux and Sainsaulieu 1997}, \cite{Caflisch at el. 1997}, \cite
{Godlewski and Raviart 1996}, \cite{Jin Shi 1995}, \cite{Kulikovskii et al.
2001}, \cite{Kurganov and Tadmor 2000}, \cite{LeVeque 2002}, \cite{Monthe
2003}, \cite{Naldi and Pareschi 2000}, \cite{Pareschi Lorenzo 2001}, \cite
{Pareschi and Russo 2005}). We will consider a 1-D system of the
conservation laws written in the following form 
\begin{equation}
\frac{\partial \mathbf{u}}{\partial t}+\frac{\partial }{\partial x}\mathbf{f}%
\left( \mathbf{u}\right) =\frac{1}{\tau }\mathbf{q}\left( \mathbf{u}\right)
,\quad x\in \mathbb{R},\ 0<t\leq T_{\max };\quad \left. \mathbf{u}\left(
x,t\right) \right| _{t=0}=\mathbf{u}^{0}\left( x\right) ,  \label{INA10}
\end{equation}
where $\mathbf{u}=\left\{ u_{1},u_{2},\ldots ,u_{M}\right\} ^{T}$ is a
vector-valued function from $\mathbb{R}$ $\times $ $[0,+\infty )$ into an
open subset $\Omega _{\mathbf{u}}\subset \mathbb{R}^{M}$, $\mathbf{f}\left( 
\mathbf{u}\right) =\left\{ f_{1}\left( \mathbf{u}\right) ,f_{2}\left( 
\mathbf{u}\right) ,\ldots ,f_{M}\left( \mathbf{u}\right) \right\} ^{T}$ is a
smooth function (flux-function) from $\Omega _{\mathbf{u}}$ into $\mathbb{R}%
^{M}$, $\mathbf{q}\left( \mathbf{u}\right) =\left\{ q_{1}\left( \mathbf{u}%
\right) ,q_{2}\left( \mathbf{u}\right) ,\ldots ,q_{M}\left( \mathbf{u}%
\right) \right\} ^{T}$ denotes the source term, $\tau >0$ denotes the
stiffness parameter, $\mathbf{u}^{0}\left( x\right) $\ is either of compact
support or periodic. We will assume that $\tau =const$ without loss of
generality. We will assume that the system (\ref{INA10}) is hyperbolic and,
hence, the Jacobian matrix of $\mathbf{f}\left( \mathbf{u}\right) $
possesses $M$ linearly independent eigenvectors (see, e.g., \cite{Godlewski
and Raviart 1996}). In addition, it is assumed in this paper that 
\begin{equation}
\underset{\mathbf{u\in }\Omega _{\mathbf{u}}}{\sup }\left\| \mathbf{A}%
\right\| _{2}\leq \lambda _{\max }<\infty ,\quad \mathbf{A=}\frac{\partial 
\mathbf{f}\left( \mathbf{u}\right) }{\partial \mathbf{u}},  \label{IN80}
\end{equation}
and all eigenvalues, $\xi _{k}=\xi _{k}\left( \mathbf{u}\right) ,$ of the
Jacobian matrix $\mathbf{G}$ ($\mathbf{=}\partial \mathbf{q}\left( \mathbf{u}%
\right) \diagup \partial \mathbf{u}$) have non-positive real parts, i.e. 
\begin{equation}
R_{e}\xi _{k}\left( \mathbf{u}\right) \leq 0,\quad \forall k,\quad \forall 
\mathbf{u\in }\Omega _{\mathbf{u}}.  \label{INA20}
\end{equation}

Here and in what follows $\left\| \mathbf{M}\right\| _{p}$ denotes the
matrix norm of a matrix $\mathbf{M}$ induced by the vector norm $\left\| 
\mathbf{v}\right\| _{p}$ $=\left( \sum_{i}\left| v_{i}\right| ^{p}\right)
^{1/p}$, and $\left\| \mathbf{M}\right\| $ denotes the matrix norm induced
by a prescribed vector norm. $\mathbb{R}$ and $\mathbb{C}$ denote the fields
of real and complex numbers, respectively, and $\mathbb{K}$ denotes either
of these fields.

In the numerical solution of the, in general, stiff ($\tau \ll 1$) system (%
\ref{INA10}), one is seeking to establish a numerical scheme that would be
robust enough to eradicate spurious oscillations, i.e. a monotone scheme 
\cite{Borisov and Sorek 2004}. At the present time there are several notions
of scheme monotonicity. The notion of `monotonicity preserving scheme'
originally appeared in Godunov \cite{Godunov 1959}. Such a scheme transforms
a monotone increasing (or decreasing) function $v\left( x\right) $ of space
coordinate $x$ at a time level $t$ into a monotone increasing (or
decreasing, respectively) function $\widehat{v}\left( x\right) $ at the next
time level $t+\Delta t$. Thus, with the use of a monotonicity preserving
scheme, any discontinuity in the initial monotone data can be smeared in
succeeding time steps but cannot become oscillatory. Nowadays monotonicity
preserving schemes are also known as, e.g., monotonicity conserving
iterations (or methods) \cite{Horvath 2002}, monotone schemes (e.g., \cite
{Abgrall and Roe 2003}, \cite{Kalitkin 1978}, \cite{Morton 1996}),
monotonicity preserving methods \cite{LeVeque 2002}, and Godunov-monotone
schemes \cite{Borisov and Sorek 2004}. Harten et al. \cite{Harten et al.
1976} put forward their own definition of scheme monotonicity as follows: a
difference scheme, $\widehat{v}_{i}=H\left( v_{i-k},v_{i-k+1},\ldots
,v_{i+k}\right) $, is said to be monotone if $H$ is a monotone increasing
function of each of its arguments. The following definition is due to
Samarskiy: a scheme is regarded as monotone if the boundary maximum
principle is maintained \cite{Samarskiy 1965} (see also, e.g., \cite[p. 183]
{Samarskii 2001}, \cite{Borisov and Sorek 2004}, \cite{Boglaev 2006}). A
difference scheme may also be referred to as monotone if a maximum
principle, e.g., the boundary maximum principle, the region maximum
principle, the maximum principle for inverse column entries, the maximum
principle for the absolute values, etc', holds for this scheme \cite{Borisov
V.S. 2003}. A further important notion of difference scheme monotonicity
was, in fact, done in \cite{Ostapenko 1998} (see also \cite{Borisov and
Sorek 2004}). A scheme will be referred to as monotone if it is monotonicity
preserving \cite{Godunov 1959} and transforms a ``$\wedge $-function'' (or ``%
$\vee $-function'') into a ``$\mu $ - function'' (or into an ``$\eta $ -
function'', respectively). Here and in what follows, a scalar grid function $%
v_{i}$ will be referred to as $\wedge $-function (or $\vee $-function) if
there exist grid nodes $m$ and $n$ such that $m\leq n$; $v_{m}>v_{m-1}$ and $%
v_{i}\geq v_{i-1}$ ($v_{m}<v_{m-1}$ and $v_{i}\leq v_{i-1}$ for the $\vee $%
-function) if $i<m$; $v_{i}=const$ if $m\leq i\leq n$; $v_{n}>v_{n+1}$ and $%
v_{i}\geq v_{i+1}$ ($v_{n}<v_{n+1}$ and $v_{i}\leq v_{i+1}$ for the $\vee $%
-function) if $i>n$. Simply stated, the $\wedge $-function (or $\vee $%
-function) is a scalar grid function $v_{i}$ that has only one generalized
local maximum \cite{Ostapenko 1998} (or generalized local minimum \cite
{Ostapenko 1998}, respectively). The set of $\mu $-functions (or $\eta $%
-functions) is the union of $\wedge $-functions ($\vee $-functions,
respectively) and the set of monotone functions.

Hereinafter, for the sake of convenience, a monotonicity preserving scheme
will be referred to as G-monotone for short, a scheme monotone in terms of
Harten et al. \cite{Harten et al. 1976} will be referred to as H-monotone, a
scheme will be referred to as S-monotone if a maximum principle holds for
this scheme (see, e.g., \cite{Samarskiy 1965}, \cite[p. 183]{Samarskii 2001}%
, \cite{Borisov and Sorek 2004}, \cite{Borisov V.S. 2003}), and a scheme
being monotone from the standpoint of Ostapenko \cite{Ostapenko 1998} will
be referred to as GO-monotone \cite{Borisov and Sorek 2004}.

For studying G-monotonicity of non-linear schemes the notion of total
variation (TV, see, e.g., \cite{Godlewski and Raviart 1996}, \cite
{Kulikovskii et al. 2001}, \cite{LeVeque 2002}) turns out to be an useful
tool, since any total variation diminishing (TVD) scheme is G-monotone 
\cite[p. 168]{Godlewski and Raviart 1996}, \cite{Harten 1983}, \cite[p. 110]
{LeVeque 2002}. H-monotone as well as TVD schemes appear to be attractive
for at least the following reasons: (i) H-monotone schemes are TVD \cite
{Harten 1983} and, hence, G-monotone \cite{Harten 1983}; (ii) any H-monotone
scheme, being TVD, converges to the physically relevant solution, i.e.
solutions of H-monotone schemes do satisfy the entropy condition \cite
{Harten et al. 1976}, \cite{Harten 1983}; (iii) The notion of a TVD method
is sufficient to prove convergence \cite[p. 148]{LeVeque 2002}; (iv) There
exist simple sufficient conditions for a scalar scheme to be TVD \cite[p.
169]{Godlewski and Raviart 1996}. Besides, it is widely believed that TVD
methods are free from spurious oscillations (e.g., \cite{Chacravarthy and
Osher 1983}, \cite{Harten 1983}, \cite{JIN AND LEVERMORE 1996}, \cite
{LeVeque 2002}, \cite{Liu and Tadmor 1998}, \cite{Nessyahu and Tadmor 1990}%
); and thus, for the above-stated reasons, TVD schemes are in common
practice (see, e.g., \cite{Godlewski and Raviart 1996}, \cite{Kulikovskii et
al. 2001}, \cite{Kurganov and Tadmor 2000}, \cite{LeVeque 2002}, \cite
{Morton 1996}, \cite{Morton 2001}, \cite{Naldi and Pareschi 2000}, \cite
{Pareschi Lorenzo 2001}, \cite{Pareschi and Russo 2005}, \cite{Pareschi et
al. 2005}, \cite{Serna and Marquina 2005}).

Let us note that Harten's theorem relative to G-monotonicity of H-monotone
schemes (i.e., H-monotonicity $\Rightarrow $ TVD $\Rightarrow $
G-monotonicity) was proven in \cite[p. 360]{Harten 1983} for a specific
class of schemes approximating a 1-D scalar partial differential equation
(PDE), and hence it does not always happen that an H-monotone scheme is
G-monotone. Actually, let us consider the following linear scheme: 
\begin{equation}
\widehat{v}_{i}=\alpha _{i}v_{i-1}+\beta _{i}v_{i}+\gamma _{i}v_{i+1},
\label{IN40}
\end{equation}
where $\alpha _{i}=1-2\varepsilon $, $\beta _{i}=\gamma _{i}=\varepsilon $
at $i=1$, $2$, $3$, $\ldots $, and $\alpha _{i}=\beta _{i}=\varepsilon $, $%
\gamma _{i}=1-2\varepsilon $ at $i=0$, $-1$, $-2$, $\ldots $, $\left(
0<\varepsilon <0.25\right) $. Scheme (\ref{IN40}) is H-monotone, since $%
\alpha _{i}$, $\beta _{i}$, $\gamma _{i}$ $>$ $0$ $\forall i$. Considering
the monotone increasing function $v_{i}$, namely $v_{i}=0$ for all $i\leq 0$
and $v_{i}=1$ for all $i>0$, we obtain that $\widehat{v}_{i}=0$ for all $i<0$%
, $\widehat{v}_{0}=1-2\varepsilon $, $\widehat{v}_{1}=2\varepsilon $, and $%
\widehat{v}_{i}=1$ for all $i>1$. We note that the grid function $\widehat{v}%
_{i}$ will not be monotone. Thus, H-monotonicity is not sufficient as well
as not necessary (see, e.g., \cite[Example 4.2]{Borisov and Sorek 2004})
condition for a scheme to be free of spurious oscillations.

Furthermore, as demonstrated in \cite{Borisov and Sorek 2004}, a
conservative scheme, H-monotone (hence, consistent with the entropy
condition \cite{Harten et al. 1976}), TVD (hence, G-monotone \cite{Harten
1983}), and S-monotone, be it non-linear \cite[pp. 1578-1580]{Borisov and
Sorek 2004} or even linear with constant coefficients \cite[p. 1561]{Borisov
and Sorek 2004}, can produce spurious oscillations comparable with the size
of the jump-discontinuity (cf. \cite{Harten 1987}). Thus, the notions of
TVD, S-monotonicity, and G-monotonicity are not sufficient for a numerical
scheme to be non-oscillatory. As is shown in \cite{Borisov and Sorek 2004},
the notion of GO-monotonicity is a very helpful tool for the construction of
non-oscillatory schemes. However, a GO-monotone scheme can be not TVD \cite
{Borisov and Sorek 2004} and, hence, not S-monotone, since the notion TVD
can be viewed as a concept within S-monotonicity \cite{Borisov and Sorek
2004}. Hence, if a numerical scheme will be GO-monotone as well as
S-monotone (GOS-monotone for short), then this scheme will be stable and, in
general, free of spurious oscillations \cite{Borisov and Sorek 2004}.

An approach to investigate non-linear difference schemes for S-monotonicity
in terms of corresponding variational schemes was suggested in \cite{Borisov
and Sorek 2004}, \cite{Borisov V.S. 2003}. The advantage of such an approach
is that the variational scheme will always be linear and, hence, enables the
investigation of the monotonicity for nonlinear operators using linear
patterns. It is proven for the case of explicit schemes that the
S-monotonicity of a variational scheme will guarantee that its original
scheme will also be S-monotone \cite{Borisov and Sorek 2004}. Analogous
theorem for the case of implicit schemes can be found in Section \ref
{non-linear schemes}, Theorem \ref{Monotonicity due to variation}. Since a
variational scheme carries such an important properties of its original
scheme as S-monotonicity and GO-monotonicity (\cite{Borisov and Sorek 2004},
see also Section \ref{non-linear schemes}, Theorem \ref{Monotonicity due to
variation}), the following definition will be useful in the investigation on
scheme monotonicity.

\begin{definition}
\label{Variational monotonicity}A numerical scheme is termed variationally
monotone if its variational scheme is monotone.
\end{definition}

The notion of TV turns out to be an effective tool, \cite[p. 148]{LeVeque
2002}, for studying the stability of non-linear schemes. Actually, the
following property 
\begin{equation}
\left\| \mathcal{N}\left( \mathbf{v}+\delta \mathbf{v}\right) -\mathcal{N}%
\left( \mathbf{v}\right) \right\| \leq \left( 1+\alpha \Delta t\right)
\left\| \delta \mathbf{v}\right\|   \label{IN50}
\end{equation}
is sufficient for stability of a two-step method \cite{LeVeque 2002},
however it is, in general, difficult to obtain. Here $\Delta t$ denotes the
time increment, $\alpha $ is a constant independent of $\Delta t$ as $\Delta
t\rightarrow 0$, $\mathbf{v}$ and $\delta \mathbf{v}$ are any two grid
functions ($\delta \mathbf{v}$ will often be referred to as the variation of
the grid function $\mathbf{v}$), $\mathcal{N}$ denotes the scheme operator.
At the same time, the stability of linearized version of the non-linear
scheme is generally not sufficient to prove convergence \cite{Harten 1984}, 
\cite{LeVeque 2002}. Instead, the TV-stability adopted in \cite{Harten 1984}
(see also \cite[s. 8.3.5]{LeVeque 2002}) makes it possible to prove
convergence (to say, TV-convergence) of non-linear scalar schemes with ease.
However, the TVD property is a purely scalar notion that cannot, in general,
be extended for non-linear systems of equations, as the true solution itself
is usually not TVD \cite{Godlewski and Raviart 1996}, \cite{LeVeque 2002}.
Moreover, one can see in \cite[pp. 1578-1581]{Borisov and Sorek 2004} that a
TVD scheme can be non-convergent in, at least, $L_{\infty }$, in spite that
the scheme is TV-stable (see, e.g., \cite[Theorem 12.2]{LeVeque 2002}). Such
a phenomenon is caused by the fact that TV is not a norm, but a semi-norm. 

Nowadays, there exists a few methods for stability analysis of some classes
of nonlinear difference schemes approximating systems of PDEs (see, e.g., 
\cite{Ganzha and Vorozhtsov 1996b}, \cite{Gil' 2007}, \cite{LeVeque 2002}, 
\cite{Morton 1996}, \cite{Naterer and Camberos 2008}, \cite{Samarskii 2001}
and references therein). It is noted in \cite{Gil' 2007} that the problem of
stability analysis is still one of the most burning problems, because of the
absence of its complete solution. In particular, as noted in \cite{Ganzha
and Vorozhtsov 1996b} in this connection, the vast majority of difference
schemes, currently in use, have still not been analyzed. LeVeque \cite
{LeVeque 2002} noted as well that, in general, no numerical method for
non-linear systems of equations has been proven to be stable. There is not
even a proof that the first-order Godunov method converges on general
systems of non-linear conservation laws \cite[p. 340]{LeVeque 2002}. Thus, a
different approach to testing scheme stability must be adopted to prove
convergence of non-linear schemes for systems of PDEs. The notion of scheme
in variations (or variational scheme \cite{Borisov and Sorek 2004}, \cite
{Borisov V.S. 2003}) has, in all likelihood, much potential to be an
effective tool for studying stability of nonlinear schemes. Such an approach
goes back to the one suggested by Lyapunov (1892), namely, to investigate
stability by the first approximation. This idea has long been exploited for
investigation of the stability of motion \cite{Gil' 1998}) as well as the
stability of difference equations \cite{Gil' 2007}. We establish the notion
that the stability of a scheme in variations implies the stability of its
original scheme (see Section \ref{non-linear schemes}, Theorem \ref
{Linear-nonlinear stability} and Remark \ref{Linear-nonlinear implicit
stability}).

Aiming to demonstrate potentialities such notions as `variational scheme'
and `GOS-monotonicity' in construction numerical schemes, we will develop a
novel version of the central Lax-Friedrichs (LxF) scheme (e.g., \cite[p. 170]
{Godlewski and Raviart 1996}) with second-order accuracy in space and time.
The stability of this scheme is proven in Section \ref{Second-order central
schemes}. We restrict our proof mainly to the case when the solutions to (%
\ref{INA10}) are smooth. Notice, such a property as smoothness is peculiar
not only to vanishing-viscosity (e.g., \cite{Godlewski and Raviart 1996}, 
\cite{LeVeque 2002}) solutions of a hyperbolic system. Since the stiffness
parameter, $\tau $, in the system (\ref{INA10}) is like a viscosity \cite
{LeVeque 2002}, a solution to the system of conservation laws (\ref{INA10})
is expected to be smooth provided that input data are sufficiently smooth.
This specific type of systems is interesting itself, as such systems are not
uncommon in practice. For a detailed discussion on this subject, including
the sufficient conditions for the global existence of smooth solutions, see 
\cite{Hanouzet and Natalini 2003}.

An extensive literature is devoted to central schemes, since these schemes
are attractive for various reasons: no Riemann solvers, characteristic
decompositions, complicated flux splittings, etc., must be involved in
construction of a central scheme (see, e.g., \cite{Balaguer and Conde 2005}, 
\cite{Kurganov and Tadmor 2000}, \cite{Kurganov and Levy 2000}, \cite
{LeVeque 2002}, \cite{Pareschi Lorenzo 2001}, \cite{Pareschi et al. 2005}
and references therein), and hence such schemes can be implemented as a
black-box solvers for general systems of conservation laws \cite{Kurganov
and Tadmor 2000}. Let us, however, note that the numerical domain of
dependence \cite[p. 69]{LeVeque 2002} for a central scheme approximating,
e.g., a scalar transport equation coincides with the numerical domain of
dependence for a standard explicit scheme approximating diffusion equations 
\cite[p. 67]{LeVeque 2002}. Such a property is inherent to central schemes
in contrast to, e.g., the first-order upwind schemes \cite[p. 73]{LeVeque
2002}. Hence, central schemes do not satisfy the long known principle (e.g., 
\cite[p. 304]{Anderson et al. 1984.}) that derivatives must be correctly
treated using type-dependent differences, and hence there is a risk for
every central scheme to exhibit spurious solutions. The results of
simulations in \cite{Nessyahu and Tadmor 1990} can be seen as an
illustration of the last assertion. Notice, all versions of the, so called,
Nessyahu-Tadmor (NT) central scheme, in spite of sufficiently small CFL
number ($Cr=0.475$), exhibit spurious oscillations in contrast to the
second-order upwind scheme ($Cr=0.95$). The first order LxF scheme exhibits
the excessive numerical viscosity. Thus, the central scheme should be chosen
with great care to reflect the true solution and to avoid significant but
spurious peculiarities in numerical solutions.

Let us note that LxF scheme -- the forerunner for central schemes \cite
{Balaguer and Conde 2005}, \cite{Kurganov and Tadmor 2000} -- does not
produce spurious oscillations. While, from the pioneering works of Nessyahu
and Tadmor \cite{Nessyahu and Tadmor 1990} and on, the higher order versions
of LxF scheme can produce spurious oscillations. The reason has to do with a
negative numerical viscosity introduced to obtain a higher order accurate
scheme. Let us illustrate it with the scheme (135) in \cite{Borisov and Mond
2007b}, which is $O(\Delta t+\left( \Delta x\right) ^{2})$ accurate. We
rewrite this scheme to read 
\begin{equation*}
\frac{1}{2}\frac{\mathbf{v}_{i+0.5}^{n+0.25}-\mathbf{v}_{i+0.5}^{n}}{%
0.25\Delta t}+\frac{\mathbf{f}\left( \mathbf{v}_{i+1}^{n}\right) -\mathbf{f}%
\left( \mathbf{v}_{i}^{n}\right) }{\Delta x}=
\end{equation*}
\begin{equation}
\frac{\Delta x^{2}}{\Delta t}\frac{\mathbf{v}_{i}^{n}-2\mathbf{v}%
_{i+0.5}^{n}+\mathbf{v}_{i+1}^{n}}{\Delta x^{2}}-\frac{\Delta x^{2}}{4\Delta
t}\frac{\mathbf{d}_{i+1}^{n}-\mathbf{d}_{i}^{n}}{\Delta x},  \label{INA30}
\end{equation}
where $\mathbf{d}_{i}^{n}$ denotes the derivative of the interpolant at $%
x=x_{i}$. Notice, the second term in the right-hand side of (\ref{INA30})
is, in fact, the negative numerical viscosity. Without this term, Scheme (%
\ref{INA30}) would be of the first order, namely LxF scheme. Let us note
that there is a possibility to increase the scheme's order of accuracy, up
to $O((\Delta t)^{2}+\left( \Delta x\right) ^{2})$, by introducing into
Scheme (\ref{INA30}) an additional non-negative numerical viscosity. Such an
approach is similar to the vanishing viscosity method \cite{Godlewski and
Raviart 1996}, \cite{LeVeque 2002}, and hence possesses its advantages, yet
it appears to be free of the disadvantages of this method, since the
additional viscosity term is not artificial. With this approach, the second
order scheme is developed in Section \ref{Second-order central schemes},
where sufficient conditions for stability as well as a necessary condition
for S-monotonicity of the scheme are found. The developed scheme is tested
on several conservation laws in Section \ref{Exemplification and discussion}.

A stable numerical scheme may yield spurious results when applied to a stiff
hyperbolic system with relaxation (see, e.g., \cite{Ahmad and Berzins 2001}, 
\cite{Aves Mark A. et al. 2000}, \cite{Bereux and Sainsaulieu 1997}, \cite
{Caflisch at el. 1997}, \cite{Du Tao et al. 2003}, \cite{Jin Shi 1995}, \cite
{Pember 1993}, \cite{Pember 1993a}). Specifically, spurious numerical
solution phenomena may occur when underresolved numerical schemes (i.e.,
insufficient spatial and temporal resolution) are used (e.g., \cite{Ahmad
and Berzins 2001}, \cite{Jin Shi 1995}, \cite{JIN AND LEVERMORE 1996}, \cite
{Naldi and Pareschi 2000}). However, during a computation, the stiffness
parameter may be very small, and, hence, to resolve the small stiffness
parameter, we need a huge number of time and spatial increments, making the
computation impractical. Hence, we are interested to solve the system, (\ref
{INA10}), with underresolved numerical schemes. It is significant that for
relaxation systems a numerical scheme must possess a discrete analogy to the
continuous asymptotic limit, because any scheme violating the correct
asymptotic limit leads to spurious or poor solutions (see, e.g., \cite
{Caflisch at el. 1997}, \cite{Jin Shi 1995}, \cite{Jin Shi et al. 2000}, 
\cite{Naldi and Pareschi 2000}, \cite{Pareschi Lorenzo 2001}). Most methods
for solving such systems can be described as operator splitting ones, \cite
{Du Tao et al. 2003}, or methods of fractional steps, \cite{Bereux and
Sainsaulieu 1997}. After operator splitting, one solves the advection
homogeneous system, and then the ordinary differential equations associated
with the source terms. As reported in \cite{Gosse L. 2000}, this approach is
well suited for the stiff systems. Let us however note that the schemes
based on the operator-splitting techniques are of the first order in time,
excluding rare cases such as, e.g., the Strang splitting \cite{LeVeque 2002}%
. Thus, following Pareschi \cite[p. 1396]{Pareschi Lorenzo 2001}, we
conclude that such schemes are, in general, not robust. Fortunately, as
applied to System (\ref{INA10}), the second order schemes can be constructed
on the basis of operator-splitting techniques with ease. We are mainly
concerned with such an approach in Section \ref{SOSOST}.

\section{Monotonicity and stability of difference schemes}

\subsection{Non-linear schemes\label{non-linear schemes}}

We consider a nonlinear implicit scheme 
\begin{equation}
\mathbf{H}_{i}\left( \mathbf{y}_{1},\ldots ,\mathbf{y}_{M}\right) =\mathbf{x}%
_{i},\quad i\in \omega \equiv \left\{ 1,2,\ldots ,M\right\} ,  \label{G10}
\end{equation}
where $\mathbf{y}_{i}$$\in $$L\equiv \mathbb{K}^{N}$ denotes the sought-for
vector-valued function of grid nodes, $\mathbf{x}_{i}$$\in $$L\equiv \mathbb{%
K}^{N}$ denotes the prescribed vector-valued function of grid nodes, $%
\mathbf{H}_{i}$$=$ $\left\{ H_{i,1},\ldots ,H_{i,N}\right\} ^{T}$ is a
vector-valued function with the range belonging to $\mathbb{K}^{N}$. If we
introduce the additional notation $\mathbf{y}$$\mathbf{=}$$\left\{ \mathbf{y}%
_{1}^{T},\ldots ,\mathbf{y}_{M}^{T}\right\} ^{T}$, $\mathbf{x}$$\mathbf{=}$$%
\left\{ \mathbf{x}_{1}^{T},\ldots ,\mathbf{x}_{M}^{T}\right\} ^{T}$, $%
\mathbf{H}$$\mathbf{=}$$\left\{ \mathbf{H}_{1}^{T},\ldots ,\mathbf{H}%
_{M}^{T}\right\} ^{T}$, then the scheme (\ref{G10}) can be represented in
the form 
\begin{equation}
\mathbf{H}\left( \mathbf{y}\right) =\mathbf{x,\quad x\in }L^{M},\ \mathbf{%
y\in }L^{M}.  \label{G20}
\end{equation}

\begin{theorem}
\label{Monotonicity due to variation}Let a nonlinear implicit scheme (\ref
{G10}) be written in the form (\ref{G20}), where $\mathbf{x\in }\Omega
_{x}\subset L^{M}$, $\Omega _{x}$ denotes a closed and bounded convex set.
Let the mapping $\mathbf{H}$ in (\ref{G20}) have a strong Fr\'{e}chet
derivative (strong F-derivative \cite[item 3.2.9]{Ortega and Rheinboldt 1970}%
), $\mathbf{H}^{\prime }\left( \mathbf{y}\right) $, at every $\mathbf{y}$ $%
\mathbf{\in }$ $int\left( \Omega _{y}\right) $ provided that $\mathbf{H}%
\left( \Omega _{y}\right) =\Omega _{x}$, and let $\mathbf{H}^{\prime }\left( 
\mathbf{y}\right) $ be nonsingular. Then, for any $\mathbf{x}$, $\mathbf{x+}%
\delta \mathbf{x}$ $\in $ $\Omega _{x}$ the scheme will be S-monotone if its
variational difference scheme is S-monotone.
\end{theorem}

\begin{proof}
The scheme (\ref{G20}) can be seen, \cite[p. 1126]{Borisov V.S. 2003}, as a
two-node implicit scheme, and its variational scheme becomes 
\begin{equation}
\delta \mathbf{x=H}^{\prime }\left( \mathbf{y}\right) \cdot \delta \mathbf{%
y,\quad H}^{\prime }\left( \mathbf{y}\right) \equiv \frac{\partial \mathbf{H}%
\left( \mathbf{y}\right) }{\partial \mathbf{y}}.  \label{G30}
\end{equation}
As $\mathbf{H}^{\prime }\left( \mathbf{y}\right) $ is nonsingular, we may
rewrite (\ref{G30}) in the form 
\begin{equation}
\delta \mathbf{y=}\left( \mathbf{H}^{\prime }\right) ^{-1}\left( \mathbf{y}%
\right) \cdot \delta \mathbf{x}.  \label{G35}
\end{equation}
Then, by (\ref{G35}) 
\begin{equation}
\left\| \delta \mathbf{y}\right\| \mathbf{=}\left\| \left( \mathbf{H}%
^{\prime }\right) ^{-1}\left( \mathbf{y}\right) \cdot \delta \mathbf{x}%
\right\| \leq \left\| \left( \mathbf{H}^{\prime }\right) ^{-1}\left( \mathbf{%
y}\right) \right\| \,\left\| \delta \mathbf{x}\right\| .  \label{G37}
\end{equation}
Let (\ref{G30}) be S-monotone, i.e. let 
\begin{equation}
\left\| \delta \mathbf{y}\right\| \leq \left\| \delta \mathbf{x}\right\| .
\label{G40}
\end{equation}
In view of (\ref{G37}) and (\ref{G40}) we obtain \cite[p. 1126]{Borisov V.S.
2003} that 
\begin{equation}
\left\| \left( \mathbf{H}^{\prime }\right) ^{-1}\left( \mathbf{y}\right)
\right\| \leq 1,\ \forall \mathbf{y\in }\Omega _{y}\subseteq L^{M}.
\label{G45}
\end{equation}
In view of the inverse function theorem \cite[item 5.2.1]{Ortega and
Rheinboldt 1970} we obtain from (\ref{G20}) that 
\begin{equation}
\mathbf{y}=\mathbf{H}^{-1}\left( \mathbf{x}\right) \mathbf{,\quad x\in }%
\Omega _{x}\subset L^{M},\ \mathbf{y\in }\Omega _{y}\subseteq L^{M}.
\label{G50}
\end{equation}
By virtue of the mean-value theorem \cite[item 3.2.3]{Ortega and Rheinboldt
1970} we obtain from (\ref{G50}) 
\begin{equation*}
\left\| d\mathbf{y}\right\| =\left\| \mathbf{H}^{-1}\left( \mathbf{x}+d%
\mathbf{x}\right) -\mathbf{H}^{-1}\left( \mathbf{x}\right) \right\| \leq 
\underset{0\leq t\leq 1}{\sup }\left\| \left( \mathbf{H}^{-1}\right)
^{\prime }\left( \mathbf{x}+td\mathbf{x}\right) \right\| \,\left\| d\mathbf{x%
}\right\|
\end{equation*}
\begin{equation}
\leq \underset{\mathbf{y}\in \Omega _{y}}{\sup }\left\| \left( \mathbf{H}%
^{-1}\right) ^{\prime }\left( \mathbf{H}\left( \mathbf{y}\right) \right)
\right\| \,\left\| d\mathbf{x}\right\| .  \label{G60}
\end{equation}
In view of the inverse function theorem \cite[item 5.2.1]{Ortega and
Rheinboldt 1970} we can write 
\begin{equation}
\left( \mathbf{H}^{-1}\right) ^{\prime }\left( \mathbf{H}\left( \mathbf{y}%
\right) \right) =\left( \mathbf{H}^{\prime }\right) ^{-1}\left( \mathbf{y}%
\right) .  \label{G70}
\end{equation}
By virtue of (\ref{G70}) and (\ref{G45}) we obtain from (\ref{G60}) that 
\begin{equation}
\left\| d\mathbf{y}\right\| \leq \left\| d\mathbf{x}\right\| .  \label{G80}
\end{equation}
The inequality (\ref{G80}) manifests the prove of the theorem.
\end{proof}

\begin{theorem}
\label{Linear-nonlinear stability}Let a non-linear explicit scheme be
written in the form\/ 
\begin{equation}
\widehat{\mathbf{v}}=\mathbf{H}(\mathbf{v}),\quad \widehat{\mathbf{v}}\in
L,\quad \mathbf{v}\in \Omega \subset L,  \label{STB10}
\end{equation}
where $\Omega $ denotes a closed and bounded convex set in a linear vector
space $L$. Then for any\/ $\mathbf{v},\mathbf{v}+\delta \mathbf{v}\in \Omega 
$ the scheme will be stable if its variational scheme is stable.
\end{theorem}

\begin{proof}
The variational scheme corresponding to the scheme (\ref{STB10}) reads 
\begin{equation}
\delta \widehat{\mathbf{v}}=\mathbf{H}^{\prime }(\mathbf{v})\cdot \delta 
\mathbf{v},\quad \mathbf{H}^{\prime }(\mathbf{v})\equiv \frac{\partial 
\mathbf{H}(\mathbf{v})}{\partial \mathbf{v}}.  \label{STB20}
\end{equation}
The linear scheme (\ref{STB20}) will be stable \cite[p. 145]{LeVeque 2002}
if $\Vert \mathbf{H}^{\prime }(\mathbf{v})\Vert \leq 1+\alpha \Delta t$ for
all $\mathbf{v}\in \Omega $, that is 
\begin{equation}
\sup_{\mathbf{v}\in \Omega }\Vert \mathbf{H}^{\prime }(\mathbf{v})\Vert \leq
1+\alpha \Delta t.  \label{STB25}
\end{equation}
By virtue of the mean-value theorem \cite[item 3.2.3]{Ortega and Rheinboldt
1970} we obtain from (\ref{STB10}) 
\begin{equation}
\Vert \mathbf{H}(\mathbf{v}+\delta \mathbf{v})-\mathbf{H}(\mathbf{v})\Vert
\leq \sup_{0\leq \zeta \leq 1}\Vert \mathbf{H}^{\prime }(\mathbf{v}+\zeta
\delta \mathbf{v})\Vert \,\Vert \delta \mathbf{v}\Vert \leq \sup_{\mathbf{v}%
\in \Omega }\Vert \mathbf{H}^{\prime }(\mathbf{v})\Vert \,\Vert \delta 
\mathbf{v}\Vert .  \label{STB30}
\end{equation}
In view of (\ref{STB25}) we conclude from (\ref{STB30}) that the inequality (%
\ref{IN50}) for (\ref{STB10}) will be fulfilled, and hence the original
non-linear scheme (\ref{STB10}) will be stable.
\end{proof}

\begin{remark}
\label{Linear-nonlinear implicit stability}Theorem \ref{Linear-nonlinear
stability} can be reformulated for implicit schemes with ease. The proof of
this theorem is identical to the proof of Theorem \ref{Monotonicity due to
variation}.
\end{remark}

\subsection{Linear schemes}

We will consider explicit linear schemes on a uniform grid with time step $%
\Delta t$ and spatial mesh size $\Delta x$. In view of the CFL condition 
\cite{LeVeque 2002}, we assume for the explicit schemes, that $\Delta
t=O\left( \Delta x\right) $. Moreover, we will also assume that $\Delta
x=O\left( \Delta t\right) $, since a central scheme generates a conditional
approximation to Eq. (\ref{INA10}) (see Section \ref{COS}). In such a case,
the following inequalities will be valid, for sufficiently small $\Delta t$
and $\Delta x$, 
\begin{equation}
\nu _{0}\Delta t\leq \Delta x\leq \mu _{0}\Delta t,\quad \nu _{0},\mu
_{0}=const,\ 0<\nu _{0}\leq \mu _{0}.  \label{CA05}
\end{equation}
Notice, for hyperbolic problems it is often assumed that $\Delta t$ and $%
\Delta x$ are related in a fixed manner (e.g., \cite[p. 140]{LeVeque 2002}, 
\cite[p. 120]{Richtmyer and Morton 1967}), i.e. it is assumed that $\Delta t$
and $\Delta x$ fulfill a more strong condition than (\ref{CA05}).

Let $\mathbf{v\equiv }\left\{ \ldots ,\mathbf{v}_{i}^{T},\ldots \right\}
^{T} $ (or $\mathbf{A}=\left\{ \mathbf{A}_{ij}\right\} $) be a partitioned 
\cite{Mirsky 1990} vector (or matrix, respectively), then we shall denote by 
$\left\langle \mathbf{v}\right\rangle $ the ordinary vector obtained from $%
\mathbf{v}$\ (or by $\left\langle \mathbf{A}\right\rangle $ the ordinary
matrix obtained from $\mathbf{A}$, respectively) by removing its partitions.
It is easy to see that 
\begin{equation}
\left\| \mathbf{v}\right\| _{\infty }\equiv \underset{i}{\max }\left\| 
\mathbf{v}_{i}\right\| _{\infty }=\left\| \left\langle \mathbf{v}%
\right\rangle \right\| _{\infty }.  \label{IN75}
\end{equation}

To start with, we obtain necessary conditions for some class of linear
schemes to be GOS-monotone. We consider the following explicit homogeneous
scheme 
\begin{equation}
\mathbf{z}_{i}=\sum\limits_{j}\mathbf{B}_{ij}\cdot \mathbf{y}_{j},\quad 
\mathbf{z}_{i},\mathbf{y}_{j}\in L,  \label{CA10}
\end{equation}
where $L$ denotes the linear vector space with the orthonormal basis $%
\left\{ \mathbf{b}_{l}\right\} _{1}^{M}$, $\mathbf{b}_{1}$ $=$ $\left\{
1,0,\ldots ,0\right\} ^{T}$, $\mathbf{b}_{2}$ $=$ $\left\{ 0,1,\ldots
,0\right\} ^{T}$, ..., $\mathbf{b}_{M}$ $=$ $\left\{ 0,0,\ldots ,1\right\}
^{T}$; $\mathbf{B}_{ij}\equiv \left\{ B_{ij}^{kl}\right\} $ is a square
matrix. It is assumed that any constant (i.e., $\mathbf{z}_{i}$ $=$ $\mathbf{%
y}_{j}$ $\equiv $ $\mathbf{c}$ $=$ $const$) is a solution to (\ref{CA10}).
Then, in view of (\ref{CA10}), we find that 
\begin{equation}
\sum\limits_{j}\mathbf{B}_{ij}=\mathbf{I,\quad }\forall i  \label{CA20}
\end{equation}
will be the necessary conditions for (\ref{CA10}) to be G-monotone (cf. 
\cite[p. 1560]{Borisov and Sorek 2004}). Here and in what follows, $\mathbf{I%
}$ denotes the identity operator.

\begin{theorem}
\label{Necessary conditions for GOS-monotonicity}Let an explicit linear
scheme be written in the form (\ref{CA10}), and let any constant be a
solution to (\ref{CA10}). If (\ref{CA10}) is GOS-monotone, then the diagonal
elements, $B_{ij}^{kk}$, of the matrices $\mathbf{B}_{ij}\equiv \left\{
B_{ij}^{kl}\right\} $ are non-negative, i.e. 
\begin{equation}
B_{ij}^{kk}\geq 0,\quad \forall i,j,k,  \label{CA30}
\end{equation}
and 
\begin{equation}
B_{ij}^{kk}\ is\ a\ \mu -function\ of\ i\ for\ all\ j\ and\ k.  \label{CA40}
\end{equation}
\end{theorem}

\begin{proof}
We consider (\ref{CA10}) when $\mathbf{y}_{j}=y_{j}\mathbf{b}_{l}$, $l$ $=$ $%
1$, $2$, ...,$M$, where $y_{j}$ is a scalar value. Let $\left\{
z_{1,i}^{l},z_{2,i}^{l},\ldots ,z_{M,i}^{l}\right\} ^{T}$ be the left-hand
side of (\ref{CA10}) under $\mathbf{y}_{j}=y_{j}\mathbf{b}_{l}$. Then we
obtain from (\ref{CA10}) the following system of decoupled scalar equalities 
\begin{equation}
z_{k,i}^{l}=\sum\limits_{j}B_{ij}^{kl}y_{j},\quad k,l=1,2,\ldots ,M.
\label{CA50}
\end{equation}
In view of Corollary 3.14 in \cite{Borisov and Sorek 2004}, the scheme (\ref
{CA50}) will be S-monotone \emph{iff} 
\begin{equation}
\sum\limits_{j}\left| B_{ij}^{kl}\right| \leq 1,\quad \forall i,k,l.
\label{CA60}
\end{equation}
In view of (\ref{CA20}) we have 
\begin{equation}
\sum\limits_{j}B_{ij}^{kk}=1,\quad \forall i,k,  \label{CA70}
\end{equation}
and hence 
\begin{equation}
\sum\limits_{j}\left| B_{ij}^{kk}\right| \geq 1,\quad \forall i,k.
\label{CA80}
\end{equation}
By virtue of (\ref{CA80}) and using (\ref{CA60}) under $k=l$ we obtain that 
\begin{equation}
\sum\limits_{j}\left| B_{ij}^{kk}\right| =1,\quad \forall i,k.  \label{CA90}
\end{equation}
Thus, (\ref{CA70}) and (\ref{CA90}) must be valid simultaneously. It is
possible \emph{iff} all coefficients $B_{ij}^{kk}$ comply with (\ref{CA30}).

To prove (\ref{CA40}) we consider (\ref{CA50}) under $k=l$. Let $m$ be the
scheme matrix column number and $\delta _{mj}$ denote the Kronecker delta.
Assuming that the scheme will be GO-monotone, it will transform $%
y_{j}=\delta _{mj}$ into a $\mu $-function as $\delta _{mj}$ is a $\wedge $%
-function of $j$. Then we obtain from (\ref{CA50}), under $k=l$, that $%
z_{k,i}^{k}=B_{im}^{kk}$. Hence, $B_{im}^{kk}$ is a $\mu $-function of $i$, $%
\forall k,m$.
\end{proof}

Consider the special case of the scheme (\ref{CA10}), namely, $\mathbf{B}%
_{ij}$ in (\ref{CA10}) depends on a square matrix $\mathbf{A}_{i}$ 
\begin{equation}
\mathbf{B}_{ij}=\varphi _{ij}\left( \mathbf{A}_{i}\right) ,\quad \forall i,j,
\label{CA100}
\end{equation}
where $\mathbf{A}_{i}$ is similar \cite[p. 119]{Mirsky 1990} to a diagonal
matrix $\mathbf{\Lambda }_{i}$, i.e. there exists a non-singular matrix $%
\mathbf{S}_{i}$ such that 
\begin{equation}
\mathbf{S}_{i}^{-1}\cdot \mathbf{A}_{i}\cdot \mathbf{S}_{i}=\mathbf{\Lambda }%
_{i}\mathbf{\equiv }diag\left\{ \lambda _{i}^{1},\lambda _{i}^{2},\ldots
,\lambda _{i}^{M}\right\} .  \label{CA105}
\end{equation}
It is assumed that $\mathbf{B}_{ij}=0$ if $j$ $\notin $ $J_{i}$ $\equiv $ $%
\left\{ j:\ i-k_{L}\leq j\leq i+k_{R}\right\} $, where $k_{L},$ $k_{R}$ $=$ $%
const$ $\geq $ $0$. Notice, it is not assumed here that any constant (i.e., $%
\mathbf{z}_{i}$ $=$ $\mathbf{y}_{j}$ $\equiv $ $\mathbf{c}$ $=const$) is
bound to be a solution to (\ref{CA10}). The following notation is used: 
\begin{equation*}
\mathbf{y}=\left\{ \ldots ,\mathbf{y}_{j}^{T},\ldots \right\} ^{T},\ 
\overline{\mathbf{y}}_{j}=\mathbf{S}_{j}^{-1}\cdot \mathbf{y}_{j},\ 
\overline{\mathbf{y}}=\left\{ \ldots ,\overline{\mathbf{y}}_{j}^{T},\ldots
\right\} ^{T},\ \overline{\mathbf{y}}_{i,j}=\mathbf{S}_{i}^{-1}\cdot \mathbf{%
y}_{j},
\end{equation*}
\begin{equation*}
\widetilde{\mathbf{y}}_{i}=\left\{ \ldots ,\overline{\mathbf{y}}%
_{i,i-k_{L}-1}^{T},\overline{\mathbf{y}}_{i,i-k_{L}}^{T},\ldots ,\overline{%
\mathbf{y}}_{i,i+k_{R}}^{T},\overline{\mathbf{y}}_{i,i+k_{R}+1}^{T},\ldots
\right\} ^{T},
\end{equation*}
\begin{equation}
\mathbf{B=}\left\{ \mathbf{B}_{ij}\right\} ,\quad \overline{\mathbf{B}}_{ij}=%
\mathbf{S}_{i}^{-1}\cdot \mathbf{B}_{ij}\cdot \mathbf{S}_{i},\quad \overline{%
\mathbf{B}}_{i}\mathbf{=}\left\{ \ldots ,\overline{\mathbf{B}}_{ij-1},%
\overline{\mathbf{B}}_{ij},\overline{\mathbf{B}}_{ij+1},\ldots \right\} .
\label{CA109}
\end{equation}
The stability for (\ref{CA10}) provided (\ref{CA100})-(\ref{CA105}) can be
addressed by the following.

\begin{lemma}
\label{MaxPrin01}Consider an explicit scheme that can be written in the form
(\ref{CA10}) under (\ref{CA100}), (\ref{CA105}). Let $s_{i}=s\left( \mathbf{A%
}_{i}\right) $ be the spectrum of $\mathbf{A}_{i}$, and $\varphi _{ij}\left(
\lambda \right) $ be represented by an absolutely convergent power series at
each point $\lambda \in s_{i}.$ Let $\mathbf{B}_{ij}=0$ in (\ref{CA10}) if $%
j $ $\notin $ $J_{i}$ $=$ $\left\{ j:\right. \ i-k_{L}\leq j\leq \left.
i+k_{R}\right\} $. Then the scheme will be stable if 
\begin{equation}
\underset{\lambda \in s_{i}}{\max }\sum\limits_{j}\left| \varphi _{ij}\left(
\lambda \right) \right| \leq 1,\quad \forall i,  \label{CA110}
\end{equation}
\begin{equation}
\left\| \left( \mathbf{S}_{i}^{-1}-\mathbf{S}_{j}^{-1}\right) \cdot \mathbf{S%
}_{j}\right\| _{\infty }\leq \Theta =const,\quad \forall i,\ \forall j\in
J_{i}.  \label{CA115}
\end{equation}
\end{lemma}

\begin{proof}
It is easy to see that 
\begin{equation}
\mathbf{S}_{i}^{-1}\equiv \left\{ \mathbf{I}+\left( \mathbf{S}_{i}^{-1}-%
\mathbf{S}_{j}^{-1}\right) \cdot \mathbf{S}_{j}\right\} \cdot \mathbf{S}%
_{j}^{-1}.  \label{CA116}
\end{equation}
Then, in view of (\ref{CA115}), we find $\forall i$, $\forall j\in J_{i}$ 
\begin{equation*}
\left\| \mathbf{S}_{i}^{-1}\cdot \mathbf{y}_{j}\right\| _{\infty }\leq
\left\| \mathbf{I}+\left( \mathbf{S}_{i}^{-1}-\mathbf{S}_{j}^{-1}\right)
\cdot \mathbf{S}_{j}\right\| _{\infty }\left\| \mathbf{S}_{j}^{-1}\cdot 
\mathbf{y}_{j}\right\| _{\infty }\leq
\end{equation*}
\begin{equation}
\left\{ 1+\left\| \left( \mathbf{S}_{i}^{-1}-\mathbf{S}_{j}^{-1}\right)
\cdot \mathbf{S}_{j}\right\| _{\infty }\right\} \left\| \mathbf{S}%
_{j}^{-1}\cdot \mathbf{y}_{j}\right\| _{\infty }\leq \left( 1+\Theta \right)
\left\| \mathbf{S}_{j}^{-1}\cdot \mathbf{y}_{j}\right\| _{\infty }.
\label{CA117}
\end{equation}
By virtue of (\ref{CA109}), we rewrite (\ref{CA10}) to read 
\begin{equation}
\overline{\mathbf{z}}_{i}\equiv \mathbf{S}_{i}^{-1}\cdot \mathbf{z}%
_{i}=\sum\limits_{j}\overline{\mathbf{B}}_{ij}\cdot \left( \mathbf{S}%
_{i}^{-1}\cdot \mathbf{y}_{j}\right) \equiv \overline{\mathbf{B}}_{i}\cdot 
\widetilde{\mathbf{y}}_{i}\equiv \left\langle \overline{\mathbf{B}}%
_{i}\right\rangle \cdot \left\langle \widetilde{\mathbf{y}}_{i}\right\rangle
,\quad \forall i,  \label{CA120}
\end{equation}
where $\left\langle \overline{\mathbf{B}}_{i}\right\rangle $ and $%
\left\langle \widetilde{\mathbf{y}}_{i}\right\rangle $ denote the ordinary
matrix and vector obtained from $\overline{\mathbf{B}}_{i}$ and $\widetilde{%
\mathbf{y}}_{i}$, respectively, by removing the partitions. Notice, $%
\overline{\mathbf{B}}_{ij}=0$ if $j$ $\notin $ $J_{i}$, since $\mathbf{B}%
_{ij}=0$ for these $j$. Thus, it can be written that $\overline{\mathbf{B}}%
_{ij}=\mathbf{S}_{i}^{-1}\cdot \mathbf{B}_{ij}\cdot \mathbf{S}_{j}$ if $j$ $%
\notin $ $J_{i}$, and hence $\overline{\mathbf{y}}_{i,j}$ $=$ $\overline{%
\mathbf{y}}_{j,j}$ $=$ $\mathbf{S}_{j}^{-1}\cdot \mathbf{y}_{j}$ $=$ $%
\overline{\mathbf{y}}_{j}$ $\left( \forall j\notin J_{i}\right) $. In view
of (\ref{CA120}) we obtain that 
\begin{equation}
\left\| \overline{\mathbf{z}}_{i}\right\| _{\infty }\equiv \left\| \mathbf{S}%
_{i}^{-1}\cdot \mathbf{z}_{i}\right\| _{\infty }\leq \left\| \left\langle 
\overline{\mathbf{B}}_{i}\right\rangle \right\| _{\infty }\left\|
\left\langle \widetilde{\mathbf{y}}_{i}\right\rangle \right\| _{\infty
},\quad \forall i,  \label{CA130}
\end{equation}
The norm $\left\| \left\langle \widetilde{\mathbf{y}}_{i}\right\rangle
\right\| _{\infty }$ in (\ref{CA130}) can be estimated by virtue of (\ref
{IN75}) and (\ref{CA117}): 
\begin{equation}
\left\| \left\langle \widetilde{\mathbf{y}}_{i}\right\rangle \right\|
_{\infty }=\left\| \widetilde{\mathbf{y}}_{i}\right\| _{\infty }\leq \left(
1+\Theta \right) \underset{j}{\max }\left\| \mathbf{S}_{j}^{-1}\cdot \mathbf{%
y}_{j}\right\| _{\infty }=\left( 1+\Theta \right) \left\| \overline{\mathbf{y%
}}\right\| _{\infty },\ \forall i.  \label{CA135}
\end{equation}
Let us estimate $\left\| \left\langle \overline{\mathbf{B}}_{i}\right\rangle
\right\| _{\infty }$ in (\ref{CA130}). In view of (\ref{CA105}) $\mathbf{%
\Lambda }_{i}=\mathbf{S}_{i}^{-1}\cdot \mathbf{A}_{i}\cdot \mathbf{S}_{i}$.
It can be verified, by induction with respect to $n$, that $\left( \mathbf{%
\Lambda }_{i}\right) ^{n}=\mathbf{S}_{i}^{-1}\cdot \left( \mathbf{A}%
_{i}\right) ^{n}\cdot \mathbf{S}_{i}$. Then, in view of Theorem 11.2.2 and
Theorem 11.2.4 in \cite{Mirsky 1990}, we find 
\begin{equation}
\overline{\mathbf{B}}_{ij}\equiv \mathbf{S}_{i}^{-1}\cdot \mathbf{B}%
_{ij}\cdot \mathbf{S}_{i}=\varphi _{ij}\left( \mathbf{S}_{i}^{-1}\cdot 
\mathbf{A}_{i}\cdot \mathbf{S}_{i}\right) =\varphi _{ij}\left( \mathbf{%
\Lambda }_{i}\right) .  \label{CA140}
\end{equation}
Thus, $\overline{\mathbf{B}}_{ij}$ can be written in the form 
\begin{equation}
\overline{\mathbf{B}}_{ij}=diag\left\{ \Lambda _{ij}^{1},\Lambda
_{ij}^{2},\ldots ,\Lambda _{ij}^{M}\right\} ,\quad \Lambda _{ij}^{k}=\varphi
_{ij}\left( \lambda _{j}^{k}\right) ,\quad k=1,2,\ldots ,M.  \label{CA145}
\end{equation}
In view of (\ref{CA145}), we find that 
\begin{equation}
\left\| \left\langle \overline{\mathbf{B}}_{i}\right\rangle \right\|
_{\infty }=\underset{k}{\max }\sum\limits_{j}\left| \Lambda _{ij}^{k}\right|
=\underset{\lambda \in s_{i}}{\max }\sum\limits_{j}\left| \varphi
_{ij}\left( \lambda \right) \right| ,\quad \forall i.  \label{CA150}
\end{equation}
By virtue of (\ref{CA135}), (\ref{CA150}), and (\ref{CA110}) we obtain from (%
\ref{CA130}) that 
\begin{equation}
\left\| \mathbf{S}_{i}^{-1}\cdot \mathbf{z}_{i}\right\| _{\infty }\leq
\left( 1+\Theta \right) \left\| \overline{\mathbf{y}}\right\| _{\infty
},\quad \forall i.  \label{CA160}
\end{equation}
Since 
\begin{equation}
\left\| \overline{\mathbf{z}}\right\| _{\infty }\equiv \underset{i}{\max }%
\left\| \mathbf{S}_{i}^{-1}\cdot \mathbf{z}_{i}\right\| _{\infty },
\label{CA162}
\end{equation}
we obtain, in view of (\ref{CA160}), (\ref{CA162}), that 
\begin{equation}
\left\| \mathbf{z}\right\| _{\ast }\equiv \left\| \overline{\mathbf{z}}%
\right\| _{\infty }\leq \left( 1+\Theta \right) \left\| \overline{\mathbf{y}}%
\right\| _{\infty }\equiv \left( 1+\Theta \right) \left\| \mathbf{y}\right\|
_{\ast }.  \label{CA165}
\end{equation}
The last inequality establishes Lemma \ref{MaxPrin01}.
\end{proof}

We consider the following explicit linear scheme 
\begin{equation}
\mathbf{v}_{i}^{n+1}=\sum\limits_{j}\mathbf{B}_{ij}^{n}\cdot \mathbf{v}%
_{j}^{n},\quad n\geq 0,  \label{CB10}
\end{equation}
where 
\begin{equation}
\mathbf{B}_{ij}^{n}=\left\{ 
\begin{array}{cc}
\varphi _{ij}^{n}\left( \mathbf{A}_{j}^{n}\right) , & j\in J_{i} \\ 
0, & j\notin J_{i}
\end{array}
\right. ,\quad \forall i,j,n,  \label{CB20}
\end{equation}
\begin{equation}
J_{i}=\left\{ j:\right. \ i-k_{L}\leq j\leq \left. i+k_{R}\right\} ,\quad
k_{L},k_{R}=const,\quad \forall i,n,  \label{CB20a}
\end{equation}
$k_{L}$, $k_{R}$, denote the non-negative integer constants being
independent of $t$, $x$, $\Delta x$, and $\Delta t$. It is assumed that the
matrix-valued function $\mathbf{A=A}\left( x,t\right) $ is
Lipschitz-continuous and $\mathbf{A}$\ is diagonizable, i.e. for $\mathbf{A}%
_{i}^{n}\mathbf{=A}\left( x_{i},t_{n}\right) $ there exists a non-singular
matrix $\mathbf{S}_{i}^{n}=\mathbf{S}\left( x_{i},t_{n}\right) $ such that 
\begin{equation}
\left( \mathbf{S}_{i}^{n}\right) ^{-1}\cdot \mathbf{A}_{i}^{n}\cdot \mathbf{S%
}_{i}^{n}=\mathbf{\Lambda }_{i}^{n}\mathbf{\equiv }diag\left\{ \lambda
_{i}^{n,1},\lambda _{i}^{n,2},\ldots ,\lambda _{i}^{n,M}\right\} ,\quad
\forall i,n.  \label{CB30}
\end{equation}

Let us note that even if $\mathbf{B}_{ij}^{n}=\varphi _{ij}^{n}\left( 
\mathbf{A}_{i}^{n}\right) $ in (\ref{CB20}) and Lemma \ref{MaxPrin01} be
valid for the linear scheme (\ref{CB10}) with $\Theta =O\left( \Delta
t\right) $ at every time step, the scheme (\ref{CB10}) will be ``locally
stable'' only, i.e. any growth in error is, at most, order $O\left( \Delta
t\right) $ in one time step. However, we cannot, in general, show on the
basis of (\ref{CA165}) that 
\begin{equation}
\left\| \mathbf{v}^{N_{T}}\right\| _{\ast \ast }\leq C_{T}\left\| \mathbf{v}%
^{0}\right\| _{\ast },\quad C_{T}=const,  \label{CB90}
\end{equation}
where $\left\| \mathbf{\cdot }\right\| _{\ast \ast }$ and $\left\| \mathbf{%
\cdot }\right\| _{\ast }$ denote some norms, $\mathbf{v}^{n}=\left\{ \ldots
,\left( \mathbf{v}_{i}^{n}\right) ^{T},\ldots \right\} ^{T}$, $N_{T}$
denotes the time level corresponding to time $T=N_{T}\Delta t$ over which we
wish to compute. The reason is that the vector norm in (\ref{CA165}) depends
on the time level $t_{n}$, and hence we maynot apply (\ref{CA165})
recursively to obtain (\ref{CB90}). The stability of the system (\ref{CB10}%
), can be addressed by the following.

\begin{theorem}
\label{StabilitySufficient}Consider an explicit scheme that can be written
in the form (\ref{CB10}) under (\ref{CB20})-(\ref{CB30}), where the
functions $\varphi _{ij}^{n}\left( \mathbf{A}\right) $ and $\mathbf{A}\left(
x,t\right) $ are both Lipschitz-continuous. Let there exist $\Delta x_{0}>0$
such that the function $\varphi _{ij}^{n}\left( \lambda \right) $ in (\ref
{CB20}) can be represented by an absolutely convergent power series at each
point of the spectrum $s_{i}^{n}=s\left( \mathbf{A}_{i}^{n}\right) $ $%
\forall i,n$, $\forall j\in J_{i},$ $\forall \Delta x$ $\leq $ $\Delta x_{0}$%
, and let the matrix-valued functions $\mathbf{S}\left( x,t\right) $, and $%
\mathbf{S}^{-1}\left( x,t\right) $ in (\ref{CB30}) can be taken such that
the matrix-valued functions $\left[ \left( \mathbf{S}_{i}^{n}\right) ^{-1}%
\mathbf{-}\left( \mathbf{S}^{n}\right) ^{-1}\left( x\right) \right] \cdot 
\mathbf{S}^{n}\left( x\right) $ and $\left[ \left( \mathbf{S}_{i}\right)
^{-1}\left( t\right) -\left( \mathbf{S}_{i}^{n}\right) ^{-1}\right] \cdot 
\mathbf{S}_{i}^{n}$ will be Lipschitz-continuous in space and, respectively,
time $\forall i,n$. Let 
\begin{equation}
\left\| \left( \mathbf{S}_{j}^{n}\right) ^{-1}\right\| _{\infty }\leq \beta
_{-1}=const,\quad \left\| \mathbf{S}_{j}^{n}\right\| _{\infty }\leq \beta
_{0}=const,\quad \forall j,n\text{.}  \label{CB60}
\end{equation}
Then the scheme (\ref{CB10}) will be stable, i.e. (\ref{CB90}) will be
valid, if 
\begin{equation}
\underset{\lambda \in s_{i}^{n}}{\max }\sum\limits_{j}\left| \varphi
_{ij}^{n}\left( \lambda \right) \right| \leq 1,\quad \forall i,n.
\label{CB65}
\end{equation}
\end{theorem}

\begin{proof}
Let $\mathbf{\check{B}}_{ij}^{n}=\varphi _{ij}^{n}\left( \mathbf{A}%
_{i}^{n}\right) $, and let us rewrite (\ref{CB10}) to read 
\begin{equation}
\mathbf{v}_{i}^{n+1}=\mathbf{\check{v}}_{i}^{n}+\mathbf{\hat{v}}%
_{i}^{n},\quad \forall i,n,  \label{CP10}
\end{equation}
where 
\begin{equation}
\mathbf{\check{v}}_{i}^{n}=\sum\limits_{j}\mathbf{\check{B}}_{ij}^{n}\cdot 
\mathbf{v}_{j}^{n},\quad \forall i,n,\ \forall j\in J_{i},  \label{CP20}
\end{equation}
\begin{equation}
\mathbf{\hat{v}}_{i}^{n}=\sum\limits_{j}\left( \mathbf{B}_{ij}^{n}-\mathbf{%
\check{B}}_{ij}^{n}\right) \cdot \mathbf{v}_{j}^{n},\quad \forall i,n,\
\forall j\in J_{i}.  \label{CP30}
\end{equation}

First, let us estimate the norm, $h_{n}\left( \cdot \right) $, of $\mathbf{%
\check{v}}^{n}:$

\begin{equation}
h_{n}\left( \mathbf{\check{v}}^{n}\right) \equiv \left\| \overline{\mathbf{%
\check{v}}}^{n}\right\| _{\infty }\equiv \underset{i}{\max }\left\| \left( 
\mathbf{S}_{i}^{n}\right) ^{-1}\cdot \mathbf{\check{v}}_{i}^{n}\right\|
_{\infty }.  \label{CA35}
\end{equation}
Since $\left[ \left( \mathbf{S}_{i}^{n}\right) ^{-1}\mathbf{-}\left( \mathbf{%
S}^{n}\right) ^{-1}\left( x\right) \right] \cdot \mathbf{S}^{n}\left(
x\right) $ and $\left[ \left( \mathbf{S}_{i}\right) ^{-1}\left( t\right)
-\left( \mathbf{S}_{i}^{n}\right) ^{-1}\right] \cdot \mathbf{S}_{i}^{n}$ are
Lipschitz-continuous in space and time, respectively, we may write 
\begin{equation}
\left\| \left[ \left( \mathbf{S}_{i}^{n}\right) ^{-1}\mathbf{-}\left( 
\mathbf{S}_{i+1}^{n}\right) ^{-1}\right] \cdot \mathbf{S}_{i+1}^{n}\right\|
_{\infty }\leq \beta _{1}\Delta x,\ \mathbf{\beta }_{1}\mathbf{=}const,\quad
\forall i,n,  \label{CB40}
\end{equation}
\begin{equation}
\left\| \left[ \left( \mathbf{S}_{i}^{n+1}\right) ^{-1}-\left( \mathbf{S}%
_{i}^{n}\right) ^{-1}\right] \cdot \mathbf{S}_{i}^{n}\right\| _{\infty }\leq
\beta _{2}\Delta t,\ \beta _{2}=const,\quad \forall i,n.  \label{CB45}
\end{equation}
By virtue of (\ref{CB40}), we find 
\begin{equation}
\left\| \left[ \left( \mathbf{S}_{i}^{n}\right) ^{-1}-\left( \mathbf{S}%
_{j}^{n}\right) ^{-1}\right] \cdot \mathbf{S}_{j}^{n}\right\| _{\infty }\leq
\beta _{3}\Delta x,\ \beta _{3}=const,\quad \forall i,n,\ \forall j\in J_{i},
\label{CB50}
\end{equation}
where $\beta _{3}=\beta _{1}\max (k_{L},k_{R})$. We assume for the explicit
scheme (\ref{CB10}), that $\Delta x=O\left( \Delta t\right) $ (i.e. $\exists 
$ $\Delta t_{0}>0$, $\exists $ $\alpha _{0}>0$ such that $\Delta x$ $\leq $ $%
\alpha _{0}\Delta t$ $\forall \Delta t\leq \Delta t_{0}$). Then we find by
virtue of (\ref{CB50}) that 
\begin{equation}
\left\| \left\{ \left( \mathbf{S}_{i}^{n}\right) ^{-1}-\left( \mathbf{S}%
_{j}^{n}\right) ^{-1}\right\} \cdot \mathbf{S}_{j}^{n}\right\| _{\infty
}\leq \beta _{4}\Delta t,\ \beta _{4}=\alpha _{0}\beta _{3},\ \forall i,n,\
\forall j\in J_{i}.  \label{CB70}
\end{equation}
The inequality (\ref{CB70}) coincides with the assumption (\ref{CA115}) in
Lemma \ref{MaxPrin01} under $\Theta =\beta _{4}\Delta t$. Then, in view of
Lemma \ref{MaxPrin01}, we obtain for the scheme (\ref{CP20}), that 
\begin{equation*}
h_{n}\left( \mathbf{\check{v}}^{n}\right) \equiv \left\| \overline{\mathbf{%
\check{v}}}^{n}\right\| _{\infty }\equiv \underset{i}{\max }\left\| \left( 
\mathbf{S}_{i}^{n}\right) ^{-1}\cdot \mathbf{\check{v}}_{i}^{n}\right\|
_{\infty }\leq
\end{equation*}
\begin{equation}
\left[ 1+\beta _{4}\Delta t\right] \underset{i}{\max }\left\| \left( \mathbf{%
S}_{i}^{n}\right) ^{-1}\cdot \mathbf{v}_{i}^{n}\right\| _{\infty }\equiv %
\left[ 1+\beta _{4}\Delta t\right] h_{n}\left( \mathbf{v}^{n}\right) .
\label{CB80}
\end{equation}

Let us now estimate the norm $h_{n}\left( \mathbf{\hat{v}}^{n}\right) $.
Since $\varphi _{ij}^{n}\left( \mathbf{A}\right) $, $\mathbf{A}\left(
x,t\right) $ are both Lipschitz continuous, we may write 
\begin{equation}
\left\| \varphi _{ij}^{n}\left( \mathbf{A}_{j}^{n}\right) -\varphi
_{ij}^{n}\left( \mathbf{A}_{i}^{n}\right) \right\| _{\infty }\leq \alpha
_{1}\left\| \mathbf{A}_{i}^{n}-\mathbf{A}_{j}^{n}\right\| _{\infty },\quad
\forall i,n,\ \forall j\in J_{i},  \label{CB83}
\end{equation}
\begin{equation}
\left\| \mathbf{A}_{i}^{n}-\mathbf{A}_{j}^{n}\right\| _{\infty }\leq \alpha
_{2}\left| x_{j}-x_{i}\right| \leq \alpha _{3}\Delta x,\quad \forall i,n,\
\forall j\in J_{i},  \label{CB85}
\end{equation}
where $\alpha _{3}$ $=$ $\alpha _{2}\max (k_{L},k_{R})$, $\alpha _{1}$, $%
\alpha _{2}$ $=$ $const$. By virtue of (\ref{CB83}), (\ref{CB85}), and
assuming that $\Delta x=O\left( \Delta t\right) $, we obtain 
\begin{equation}
\left\| \mathbf{B}_{ij}^{n}-\mathbf{\check{B}}_{ij}^{n}\right\| _{\infty
}\equiv \left\| \varphi _{ij}^{n}\left( \mathbf{A}_{j}^{n}\right) -\varphi
_{ij}^{n}\left( \mathbf{A}_{i}^{n}\right) \right\| _{\infty }\leq \alpha
_{4}\Delta t,\quad \forall i,n,\ \forall j\in J_{i},  \label{CB85a}
\end{equation}
where $\alpha _{4}$ $=$ $\alpha _{0}\alpha _{1}\alpha _{3}$ $=$ $const$. We
obtain from (\ref{CP30}) that 
\begin{equation}
\left( \mathbf{S}_{i}^{n}\right) ^{-1}\cdot \mathbf{\hat{v}}%
_{i}^{n}=\sum\limits_{j}\left( \mathbf{S}_{i}^{n}\right) ^{-1}\cdot \left( 
\mathbf{B}_{ij}^{n}-\mathbf{\check{B}}_{ij}^{n}\right) \cdot \mathbf{S}%
_{j}^{n}\cdot \left( \mathbf{S}_{j}^{n}\right) ^{-1}\cdot \mathbf{v}_{j}^{n}.
\label{CB86}
\end{equation}
Whence, by virtue of (\ref{CB85a}) and (\ref{CB60}), we obtain 
\begin{equation}
\left\| \left( \mathbf{S}_{i}^{n}\right) ^{-1}\cdot \mathbf{\hat{v}}%
_{i}^{n}\right\| _{\infty }\leq \alpha _{5}\Delta t\underset{j}{\max }%
\left\| \left( \mathbf{S}_{j}^{n}\right) ^{-1}\cdot \mathbf{v}%
_{j}^{n}\right\| _{\infty }\equiv \alpha _{5}\Delta th_{n}\left( \mathbf{v}%
^{n}\right) ,\quad \forall i,n,  \label{CB87}
\end{equation}
where $\alpha _{5}=\beta _{-1}\beta _{0}\alpha _{4}\max (k_{L},k_{R})=const$%
. By virtue of (\ref{CP10}), (\ref{CB80}), and (\ref{CB87}), we obtain 
\begin{equation}
h_{n}\left( \mathbf{v}^{n+1}\right) \leq \left[ 1+\beta \Delta t\right]
h_{n}\left( \mathbf{v}^{n}\right) ,\ \beta =\beta _{4}+\alpha
_{5}=const,\quad \forall n.  \label{CB88}
\end{equation}

It is easy to see that 
\begin{equation}
\left( \mathbf{S}_{i}^{n+1}\right) ^{-1}\equiv \left\{ \mathbf{I}+\left(
\left( \mathbf{S}_{i}^{n+1}\right) ^{-1}-\left( \mathbf{S}_{i}^{n}\right)
^{-1}\right) \cdot \mathbf{S}_{i}^{n}\right\} \cdot \left( \mathbf{S}%
_{i}^{n}\right) ^{-1},  \label{CB100}
\end{equation}
whence, by virtue of (\ref{CB45}), we find 
\begin{equation*}
h_{n+1}\left( \mathbf{v}^{n+1}\right) =\underset{i}{\max }\left\| \left( 
\mathbf{S}_{i}^{n+1}\right) ^{-1}\cdot \mathbf{v}_{i}^{n+1}\right\| _{\infty
}\leq
\end{equation*}
\begin{equation*}
\underset{i}{\max }\left\| \mathbf{I}+\left( \left( \mathbf{S}%
_{i}^{n+1}\right) ^{-1}-\left( \mathbf{S}_{i}^{n}\right) ^{-1}\right) \cdot 
\mathbf{S}_{i}^{n}\right\| _{\infty }\left\| \left( \mathbf{S}%
_{i}^{n}\right) ^{-1}\cdot \mathbf{v}_{i}^{n+1}\right\| _{\infty }\leq
\end{equation*}
\begin{equation}
\left( 1+\beta _{2}\Delta t\right) \underset{i}{\max }\left\| \left( \mathbf{%
S}_{i}^{n}\right) ^{-1}\cdot \mathbf{v}_{i}^{n+1}\right\| _{\infty }=\left(
1+\beta _{2}\Delta t\right) h_{n}\left( \mathbf{v}^{n+1}\right) .
\label{CB110}
\end{equation}
In view of (\ref{CB88}) and (\ref{CB110}), we find 
\begin{equation}
h_{n+1}\left( \mathbf{v}^{n+1}\right) \leq \left[ 1+\gamma \Delta t\right]
^{2}h_{n}\left( \mathbf{v}^{n}\right) ,\ \gamma =\max \left( \alpha ,\beta
_{2}\right) ,\quad \forall n.  \label{CB120}
\end{equation}
Applying (\ref{CB120}) recursively gives 
\begin{equation}
h_{N_{T}}\left( \mathbf{v}^{N_{T}}\right) \leq \left( 1+\gamma \Delta
t\right) ^{2N_{T}}h_{0}\left( \mathbf{v}^{0}\right) \leq C_{T}h_{0}\left( 
\mathbf{v}^{0}\right) ,\quad C_{T}=\exp \left( 2\gamma T\right) .
\label{CB130}
\end{equation}
The inequalities in (\ref{CB130}) prove the theorem.
\end{proof}

Let us consider the case when the operator $\mathbf{B}_{ij}^{n}$ in (\ref
{CB10}) depends on a matrix $\mathbf{A}_{j}^{n}$ belonging to a set of
pairwise commutative diagonizable matrices: 
\begin{equation}
\mathbf{B}_{ij}^{n}=\varphi _{ij}^{n}\left( \mathbf{A}_{j}^{n}\right) ,\ 
\mathbf{A}_{j}^{n}\cdot \mathbf{A}_{k}^{m}=\mathbf{A}_{k}^{m}\cdot \mathbf{A}%
_{j}^{n},\quad \forall i,j,n,k,m.  \label{CC10}
\end{equation}
In such a case, the S-monotonicity of the system (\ref{CB10}), can be
addressed by the following.

\begin{theorem}
\label{S-monotoneIFF}Consider an explicit scheme that can be written in the
form (\ref{CB10}) provided (\ref{CC10}). Let $\varphi _{ij}^{n}\left(
\lambda \right) $ in (\ref{CC10}) can be represented by an absolutely
convergent power series at each point of the spectrum $s_{j}^{n}=s\left( 
\mathbf{A}_{j}^{n}\right) $ $\forall i,j,n$. Then the scheme (\ref{CB10})
will be S-monotone \emph{iff} 
\begin{equation}
\underset{\lambda \in s_{j}^{n}}{\max }\sum\limits_{i}\left| \varphi
_{ij}^{n}\left( \lambda \right) \right| \leq 1,\quad \forall j,n.
\label{CC20}
\end{equation}
\end{theorem}

\begin{proof}
As $\mathbf{A}_{j}^{n}$ belongs to the set of pair-wise permutable
diagonizable matrices, the matrices of the set are simultaneously similar to
diagonal matrices \cite[p. 318]{Mirsky 1990}, i.e., there exists a
non-singular matrix $\mathbf{S}$ such that 
\begin{equation}
\mathbf{S}^{-1}\cdot \mathbf{A}_{j}^{n}\cdot \mathbf{S}=\mathbf{\Lambda }%
_{j}^{n}\mathbf{\equiv }diag\left\{ \lambda _{j}^{n,1},\lambda
_{j}^{n,2},\ldots ,\lambda _{j}^{n,M}\right\} ,\quad \forall j,n.
\label{CC30}
\end{equation}
where $\lambda _{j}^{n,m}$ denotes the $m$-th eigenvalue of $\mathbf{A}%
_{j}^{n}$. The following notation is used: 
\begin{equation}
\overline{\mathbf{v}}_{j}^{n}=\mathbf{S}^{-1}\cdot \mathbf{v}_{j}^{n},\ 
\overline{\mathbf{B}}_{ij}^{n}=\mathbf{S}^{-1}\cdot \mathbf{B}_{ij}^{n}\cdot 
\mathbf{S},\ \overline{\mathbf{B}}^{n}=\left\{ \overline{\mathbf{B}}%
_{ij}^{n}\right\} ,\ \mathbf{B}^{n}=\left\{ \mathbf{B}_{ij}^{n}\right\} .
\label{CC40}
\end{equation}
By virtue of (\ref{CC40}), we rewrite (\ref{CB10}) to read 
\begin{equation}
\overline{\mathbf{v}}_{i}^{n+1}\equiv \mathbf{S}^{-1}\cdot \mathbf{v}%
_{i}^{n+1}=\sum\limits_{j}\overline{\mathbf{B}}_{ij}^{n}\cdot \overline{%
\mathbf{v}}_{j}^{n}.  \label{CC50}
\end{equation}
Using $\overline{\mathbf{v}}^{n}\equiv \left\{ \ldots ,\left( \overline{%
\mathbf{v}}_{j}^{n}\right) ^{T},\ldots \right\} ^{T}$, we rewrite (\ref{CC50}%
) to read 
\begin{equation}
\overline{\mathbf{v}}^{n+1}=\overline{\mathbf{B}}^{n}\cdot \overline{\mathbf{%
v}}^{n},\quad \overline{\mathbf{v}}^{n+1}\equiv \left\{ \ldots ,\left( 
\overline{\mathbf{v}}_{i}^{n+1}\right) ^{T},\ldots \right\} ^{T}.
\label{CC52}
\end{equation}
In view of (\ref{CC52}) we obtain that 
\begin{equation}
h\left( \mathbf{v}^{n+1}\right) \equiv \left\| \left\langle \overline{%
\mathbf{v}}^{n+1}\right\rangle \right\| _{1}\leq \left\| \left\langle 
\overline{\mathbf{B}}^{n}\right\rangle \right\| _{1}\left\| \left\langle 
\overline{\mathbf{v}}^{n}\right\rangle \right\| _{1}\equiv h\left( \mathbf{v}%
^{n}\right) ,  \label{CC55}
\end{equation}
where $\left\langle \overline{\mathbf{B}}^{n}\right\rangle $ and $%
\left\langle \overline{\mathbf{v}}^{n}\right\rangle $ denote the ordinary
matrix and vector obtained from $\overline{\mathbf{B}}^{n}$ and $\overline{%
\mathbf{v}}^{n}$, respectively, by removing the partitions. Let us estimate
the norm of $\left\langle \overline{\mathbf{B}}^{n}\right\rangle $ in (\ref
{CC55}). Since $\varphi _{ij}^{n}\left( \lambda \right) $ can be represented
by an absolutely convergent power series at each point $\lambda \in
s_{j}^{n}=s\left( \mathbf{A}_{j}^{n}\right) $, we find, in view of Theorem
11.2.2 and Theorem 11.2.4 in \cite{Mirsky 1990}, that 
\begin{equation}
\overline{\mathbf{B}}_{ij}^{n}\equiv \mathbf{S}^{-1}\cdot \mathbf{B}%
_{ij}^{n}\cdot \mathbf{S}=\varphi _{ij}^{n}\left( \mathbf{S}^{-1}\cdot 
\mathbf{A}_{j}^{n}\cdot \mathbf{S}\right) =\varphi _{ij}^{n}\left( \mathbf{%
\Lambda }_{j}^{n}\right) .  \label{CC60}
\end{equation}
Thus, $\overline{\mathbf{B}}_{ij}^{n}$ can be written in the form 
\begin{equation}
\overline{\mathbf{B}}_{ij}=diag\left\{ \Lambda _{ij}^{n,1},\Lambda
_{ij}^{n,2},\ldots ,\Lambda _{ij}^{n,M}\right\} ,\ \Lambda
_{ij}^{n,k}=\varphi _{ij}^{n}\left( \lambda _{j}^{n,k}\right) ,\
k=1,2,\ldots ,M.  \label{CC70}
\end{equation}
In view of (\ref{CC70}), we obtain that 
\begin{equation}
\left\| \left\langle \overline{\mathbf{B}}^{n}\right\rangle \right\| _{1}=%
\underset{j}{\max }\left( \underset{k=1,\ldots ,M}{\max }\sum\limits_{i}%
\left| \Lambda _{ij}^{n,k}\right| \right) =\underset{j}{\max }\left( 
\underset{\lambda \in s_{j}^{n}}{\max }\sum\limits_{i}\left| \varphi
_{ij}^{n}\left( \lambda \right) \right| \right) .  \label{CC80}
\end{equation}
Whence, in view of (\ref{CC20}), we find 
\begin{equation}
\left\| \left\langle \overline{\mathbf{B}}^{n}\right\rangle \right\|
_{1}\leq 1,\quad \forall n.  \label{CC90}
\end{equation}
The vector norm $h\left( \cdot \right) $ in (\ref{CC55}) does not depend on
time level. Then, in view of Proposition 3.2 in \cite{Borisov V.S. 2003},
the inequality (\ref{CC90}) proves Theorem \ref{S-monotoneIFF}
\end{proof}

\begin{proposition}
\label{Proposition01}If (\ref{CA10}) is a variational scheme, then (\ref
{CA20}) is, in general, not valid. Notice, Lemma \ref{MaxPrin01} and Theorem 
\ref{StabilitySufficient} are proven without assumption (\ref{CA20}).
However, in addition to the Lipschitz-continuity of $\mathbf{A}\left(
x,t\right) $ (see (\ref{CA100}) and (\ref{CB20})), it is assumed in Lemma 
\ref{MaxPrin01} and Theorem \ref{StabilitySufficient} that some functions of 
$\mathbf{S}\left( x,t\right) $ (see (\ref{CA105}), (\ref{CB30})) are also
Lipschitz-continuous. Let us note that the stability of a linear scheme can
often be proven without assumption (\ref{CA20}) as well as without
additional assumptions on the continuity. To demonstrate it, let us
generalize the theorem of Friedrichs (1954) (see, e.g., \cite[p. 120]
{Richtmyer and Morton 1967}, \cite[p. 374]{Samarskiy and Gulin 1973}) to be
applicable to variational schemes. We consider the following difference
scheme 
\begin{equation}
\mathbf{y}^{n+1}\left( x\right) =\sum\limits_{k=-m}^{m}\mathbf{B}%
_{k}^{n}\left( x\right) \cdot \mathbf{y}^{n}\left( x+k\Delta x\right) ,\ 
\mathbf{y}^{n}\in \mathbb{R}^{M},\ \mathbf{B}_{k}^{n}\in \mathbb{R}^{M\times
M},  \label{PROP10}
\end{equation}
where $x\in \left( -\infty ,\infty \right) $, $\mathbf{B}_{k}^{n}$ is a
symmetric and non-negative matrix. It is assumed that $\mathbf{y}^{n}\left(
x\right) $ and $\mathbf{B}_{k}^{n}\left( x\right) $ are periodic (with the
period equal to 1) functions of $x$. Let 
\begin{equation}
\mathbf{y}_{j}^{n}\equiv \mathbf{y}^{n}\left( x+j\Delta x\right) ,\ \mathbf{B%
}_{k,j}^{n}\equiv \mathbf{B}_{k}^{n}\left( x+j\Delta x\right) ,
\label{PROP15}
\end{equation}
and let 
\begin{equation}
\left( \mathbf{u,v}\right) \equiv \int\limits_{0}^{1}\left[ \mathbf{u}%
^{T}\left( x\right) \mathbf{\cdot v}\left( x\right) \right] dx,\quad \left\| 
\mathbf{u}\right\| \equiv \sqrt{\left( \mathbf{u,u}\right) }.  \label{PROP20}
\end{equation}
If there exist $c_{1}$, $c_{2}$ $=$ $const$ such that 
\begin{equation}
\left\| \sum\limits_{k}\mathbf{B}_{k}^{n}\right\| \leq 1+c_{1}\Delta x,\quad
\left\| \mathbf{B}_{k,k}^{n}-\mathbf{B}_{k}^{n}\right\| \leq \frac{c_{2}}{%
2m+1}\Delta x,  \label{PROP30}
\end{equation}
then the scheme is stable. Notice, it is not assumed that $\sum_{k}\mathbf{B}%
_{k}^{n}\left( x\right) =\mathbf{I}$.

The proof is very little different from the proof when $\sum_{k}\mathbf{B}%
_{k}^{n}\left( x\right) =\mathbf{I}$. Actually, in view of (\ref{PROP10})
and the first inequality in (\ref{PROP30}), we obtain 
\begin{equation*}
\left\| \mathbf{y}^{n+1}\right\| ^{2}=\sum\limits_{k}\left( \mathbf{B}%
_{k}^{n}\cdot \mathbf{y}_{k}^{n},\mathbf{y}^{n+1}\right) \leq
0.5\sum\limits_{k}\left[ \left( \mathbf{B}_{k}^{n}\cdot \mathbf{y}_{k}^{n},%
\mathbf{y}_{k}^{n}\right) +\left( \mathbf{B}_{k}^{n}\cdot \mathbf{y}^{n+1},%
\mathbf{y}^{n+1}\right) \right]
\end{equation*}
\begin{equation}
\leq 0.5\sum\limits_{k}\left( \mathbf{B}_{k}^{n}\cdot \mathbf{y}_{k}^{n},%
\mathbf{y}_{k}^{n}\right) +0.5\left( 1+c_{1}\Delta x\right) \left\| \mathbf{y%
}^{n+1}\right\| ^{2}.  \label{PROP40}
\end{equation}
Since $\mathbf{y}^{n}\left( x\right) $ and $\mathbf{B}_{k}^{n}\left(
x\right) $ are periodic functions, we obtain, by virtue of (\ref{CA05}), (%
\ref{PROP30}), and (\ref{PROP40}), that 
\begin{equation*}
\left( 1-\mu _{0}c_{1}\Delta t\right) \left\| \mathbf{y}^{n+1}\right\|
^{2}\leq \sum\limits_{k}\left( \mathbf{B}_{k,k}^{n}\cdot \mathbf{y}_{k}^{n},%
\mathbf{y}_{k}^{n}\right) +\sum\limits_{k}\left( \left( \mathbf{B}_{k}^{n}-%
\mathbf{B}_{k,k}^{n}\right) \cdot \mathbf{y}_{k}^{n},\mathbf{y}%
_{k}^{n}\right) \leq
\end{equation*}
\begin{equation}
\sum\limits_{k}\left( \mathbf{B}_{k}^{n}\cdot \mathbf{y}^{n},\mathbf{y}%
^{n}\right) +\frac{c_{2}}{2m+1}\Delta x\sum\limits_{k}\left( \mathbf{y}%
_{k}^{n},\mathbf{y}_{k}^{n}\right) =\left( 1+\mu _{0}\left(
c_{1}+c_{2}\right) \Delta t\right) \left\| \mathbf{y}^{n}\right\| ^{2}.
\label{PROP45}
\end{equation}
It follows from (\ref{PROP45}) that 
\begin{equation}
\left\| \mathbf{y}^{n+1}\right\| ^{2}\leq \frac{1+\mu _{0}\left(
c_{1}+c_{2}\right) \Delta t}{1-\mu _{0}c_{1}\Delta t}\left\| \mathbf{y}%
^{n}\right\| ^{2}.  \label{PROP60}
\end{equation}
Let $\Delta t_{0}=const$ such that $1-\mu _{0}c_{1}\Delta t_{0}>0$. In
particular, let $\Delta t_{0}=0.5\diagup \left( \mu _{0}c_{1}\right) $.
Then, for all $\Delta t<\Delta t_{0}$ the following inequality will be valid 
\begin{equation}
\left\| \mathbf{y}^{n+1}\right\| ^{2}\leq \left( 1+c_{3}\Delta t\right)
\left\| \mathbf{y}^{n}\right\| ^{2},\quad c_{3}=2\mu _{0}\left(
2c_{1}+c_{2}\right) =const.  \label{PROP70}
\end{equation}
The inequality in (\ref{PROP70}) proves Proposition \ref{Proposition01}.
\end{proposition}

Notice, in practice, we often deal with systems for which the matrices $%
\mathbf{B}_{k}^{n}$ in Scheme (\ref{PROP10}) are not symmetric. Let us
generalize the theorem of Friedrichs, \cite[p. 120]{Richtmyer and Morton
1967}, \cite[p. 374]{Samarskiy and Gulin 1973}, to be applicable to
variational schemes with non-symmetric matrices $\mathbf{B}_{k}^{n}$.

It is assumed that there exist matrix-valued functions $\mathbf{W}\left(
x,t\right) $ and $\mathbf{A}_{k}\left( x,t\right) $ such that $\mathbf{W=W}%
^{T}$, $\mathbf{W\cdot A}_{k}=\left( \mathbf{W\cdot A}_{k}\right) ^{T},$%
\begin{equation}
\lambda _{0}\left( \mathbf{u,u}\right) \leq \left( \mathbf{W\cdot u,u}%
\right) \leq \Lambda _{0}\left( \mathbf{u,u}\right) ,\quad \lambda
_{0},\Lambda _{0}=const,\ 0<\lambda _{0}\leq \Lambda _{0},  \label{Fr10}
\end{equation}
\begin{equation}
\left\| \mathbf{W}^{n+1}-\mathbf{W}^{n}\right\| \leq \alpha _{0}\Delta
t,\quad \alpha _{0}=const,\quad \mathbf{W}^{n}=\mathbf{W}\left(
x,t_{n}\right) ,  \label{Fr15}
\end{equation}
\begin{equation}
\left\| \mathbf{A}_{k}^{n}-\mathbf{B}_{k}^{n}\right\| _{\mathbf{W}^{n}}\leq 
\frac{\beta _{0}}{2m+1}\Delta x,\quad \beta _{0}=const,\ \mathbf{A}_{k}^{n}=%
\mathbf{A}_{k}\left( x,t_{n}\right) ,  \label{Fr20}
\end{equation}
\begin{equation}
\left\| \mathbf{A}_{k}^{n}-\mathbf{A}_{k,j}^{n}\right\| _{\mathbf{W}%
^{n}}\leq b_{0}\Delta x,\quad b_{0}=const,\ \mathbf{A}_{k,j}^{n}=\mathbf{A}%
_{k}\left( x+j\Delta x,t_{n}\right) ,  \label{Fr30}
\end{equation}
where $\left\| \mathbf{A}\right\| $ is the norm of an operator $\mathbf{A}$
on a Hilbert space equipped by the scalar product (\ref{PROP20}), while $%
\left\| \mathbf{A}\right\| _{\mathbf{W}}$ is the norm of an operator $%
\mathbf{A}$ on a Hilbert space equipped by the following scalar product 
\begin{equation}
\left( \mathbf{u,v}\right) _{\mathbf{W}}\equiv \left( \mathbf{W\cdot u,v}%
\right) ,\quad \left\| \mathbf{u}\right\| _{\mathbf{W}}\equiv \sqrt{\left( 
\mathbf{u,u}\right) _{\mathbf{W}}}.  \label{Fr40}
\end{equation}

The generalization can be addressed by the following theorem, where the
notation (\ref{PROP15}), (\ref{PROP20}), and (\ref{Fr40}) will be used.

\begin{theorem}
\label{Generalized Friedrichs}Consider an explicit scheme that can be
written in the form (\ref{PROP10}), where $\mathbf{y}^{n}\in \mathbb{R}^{M}$%
, $\mathbf{B}_{k}^{n}\in \mathbb{R}^{M\times M}$ are periodic (with the
period equal to 1) functions of $x$. Let $\mathbf{W}\left( x,t\right) $ be
symmetric, i.e. $\mathbf{W=W}^{T}$, and positive ($\mathbf{W}>0$ in terms of
(\ref{Fr10})) matrix-valued function, and let (\ref{Fr20}), (\ref{Fr30}) be
valid for the case of symmetrizable, i.e. $\mathbf{W}^{n}\mathbf{\cdot A}%
_{k}^{n}=\left( \mathbf{W}^{n}\mathbf{\cdot A}_{k}^{n}\right) ^{T}$, and
periodic matrix-valued function $\mathbf{A}_{k}^{n}\left( x\right) $. If $%
\sum_{k}\mathbf{A}_{k}\left( x,t\right) =\mathbf{I}$ and 
\begin{equation}
\lambda _{1}\left( \mathbf{u,u}\right) \leq \left( \left( \mathbf{W\cdot A}%
_{k}\right) \mathbf{\cdot u,u}\right) \leq \Lambda _{1}\left( \mathbf{u,u}%
\right) ,\ \lambda _{1},\Lambda _{1}=const,\ 0<\lambda _{1}\leq \Lambda _{1},
\label{Fr50}
\end{equation}
then Scheme (\ref{PROP10}) will be stable
\end{theorem}

\begin{proof}
Multiplying (\ref{PROP10}) by $\mathbf{W}^{n}$, and by $\mathbf{y}^{n+1}$,
we obtain, after integrating both sides, that 
\begin{equation}
\left\| \mathbf{y}^{n+1}\right\| _{\mathbf{W}^{n}}^{2}\leq
\sum\limits_{k}\left( \mathbf{A}_{k}^{n}\cdot \mathbf{y}_{k}^{n},\mathbf{y}%
^{n+1}\right) _{\mathbf{W}^{n}}+\sum\limits_{k}\left( \left( \mathbf{B}%
_{k}^{n}-\mathbf{A}_{k}^{n}\right) \cdot \mathbf{y}_{k}^{n},\mathbf{y}%
^{n+1}\right) _{\mathbf{W}^{n}}.  \label{Fr60}
\end{equation}
Since $\mathbf{W}^{n}\cdot \mathbf{A}_{k}^{n}$ is symmetric and, in view of (%
\ref{Fr50}), positive, and since $\sum_{k}\mathbf{A}_{k}^{n}=\mathbf{I}$, we
write: 
\begin{equation}
\sum\limits_{k}\left( \mathbf{A}_{k}^{n}\cdot \mathbf{y}_{k}^{n},\mathbf{y}%
^{n+1}\right) _{\mathbf{W}^{n}}\leq 0.5\sum\limits_{k}\left( \mathbf{A}%
_{k}^{n}\cdot \mathbf{y}_{k}^{n},\mathbf{y}_{k}^{n}\right) _{\mathbf{W}%
^{n}}+0.5\left\| \mathbf{y}^{n+1}\right\| _{\mathbf{W}^{n}}^{2}.
\label{Fr70}
\end{equation}
By virtue of (\ref{Fr30}), periodicity of $\mathbf{A}_{k}^{n}$ and $\mathbf{y%
}_{k}^{n}$, and since $\sum_{k}\mathbf{A}_{k}^{n}=\mathbf{I}$, we obtain: 
\begin{equation*}
\sum\limits_{k}\left( \mathbf{A}_{k}^{n}\cdot \mathbf{y}_{k}^{n},\mathbf{y}%
_{k}^{n}\right) _{\mathbf{W}^{n}}=\sum\limits_{k}\left( \mathbf{A}%
_{k,k}^{n}\cdot \mathbf{y}_{k}^{n},\mathbf{y}_{k}^{n}\right) _{\mathbf{W}%
^{n}}+\sum\limits_{k}\left( \left( \mathbf{A}_{k}^{n}-\mathbf{A}%
_{k,k}^{n}\right) \cdot \mathbf{y}_{k}^{n},\mathbf{y}_{k}^{n}\right) _{%
\mathbf{W}^{n}}\leq 
\end{equation*}
\begin{equation}
\sum\limits_{k}\left( \mathbf{A}_{k}^{n}\cdot \mathbf{y}^{n},\mathbf{y}%
^{n}\right) _{\mathbf{W}^{n}}+b_{0}\Delta x\sum\limits_{k}\left( \mathbf{y}%
_{k}^{n},\mathbf{y}_{k}^{n}\right) _{\mathbf{W}^{n}}\leq \left(
1+b_{1}\Delta x\right) \left\| \mathbf{y}^{n}\right\| _{\mathbf{W}^{n}}^{2},
\label{Fr80}
\end{equation}
where $b_{1}=\left( 2m+1\right) b_{0}=const$. In view of (\ref{Fr20}) and
the assumption of periodicity, we obtain for the last term in the right-hand
side of (\ref{Fr60}) that 
\begin{equation*}
\sum\limits_{k}\left( \left( \mathbf{B}_{k}^{n}-\mathbf{A}_{k}^{n}\right)
\cdot \mathbf{y}_{k}^{n},\mathbf{y}^{n+1}\right) _{\mathbf{W}^{n}}\leq
\sum\limits_{k}\left\| \left( \mathbf{B}_{k}^{n}-\mathbf{A}_{k}^{n}\right)
\cdot \mathbf{y}_{k}^{n}\right\| _{\mathbf{W}^{n}}\left\| \mathbf{y}%
^{n+1}\right\| _{\mathbf{W}^{n}}\leq 
\end{equation*}
\begin{equation}
\beta _{0}\Delta x\left\| \mathbf{y}^{n}\right\| _{\mathbf{W}^{n}}\left\| 
\mathbf{y}^{n+1}\right\| _{\mathbf{W}^{n}}\leq 0.5\beta _{0}\Delta x\left\| 
\mathbf{y}^{n}\right\| _{\mathbf{W}^{n}}^{2}+0.5\beta _{0}\Delta x\left\| 
\mathbf{y}^{n+1}\right\| _{\mathbf{W}^{n}}^{2}.  \label{Fr90}
\end{equation}
After elementary transformations we find from (\ref{Fr60})-(\ref{Fr90}) that 
\begin{equation}
\left\| \mathbf{y}^{n+1}\right\| _{\mathbf{W}^{n}}^{2}\leq \frac{1+\left(
\beta _{0}+b_{1}\right) \Delta x}{1-\beta _{0}\Delta x}\left\| \mathbf{y}%
^{n}\right\| _{\mathbf{W}^{n}}^{2}.  \label{Fr100}
\end{equation}
Let $\Delta x_{0}=const$ such that $1-\beta _{0}\Delta x_{0}>0$. In
particular, let $\Delta x_{0}=0.5\diagup \beta _{0}$. Then, in view of (\ref
{CA05}),\ for all $\Delta x<\Delta x_{0}$ the following inequality will be
valid 
\begin{equation}
\left\| \mathbf{y}^{n+1}\right\| _{\mathbf{W}^{n}}^{2}\leq \left(
1+b_{2}\Delta t\right) \left\| \mathbf{y}^{n}\right\| _{\mathbf{W}%
^{n}}^{2},\ b_{2}=2\mu _{0}\left( 2\beta _{0}+b_{1}\right) =const.
\label{Fr110}
\end{equation}
By virtue of (\ref{Fr10}) and (\ref{Fr15}), we find that 
\begin{equation*}
\left\| \mathbf{y}^{n+1}\right\| _{\mathbf{W}^{n+1}}^{2}=\left( \mathbf{W}%
^{n+1}\cdot \mathbf{y}^{n+1}\mathbf{,y}^{n+1}\right) =\left( \mathbf{W}%
^{n}\cdot \mathbf{y}^{n+1}\mathbf{,y}^{n+1}\right) +
\end{equation*}
\begin{equation*}
\left( \left( \mathbf{W}^{n+1}-\mathbf{W}^{n}\right) \cdot \mathbf{y}^{n+1}%
\mathbf{,y}^{n+1}\right) \leq \left\| \mathbf{y}^{n+1}\right\| _{\mathbf{W}%
^{n}}^{2}+\alpha _{0}\Delta t\left( \mathbf{y}^{n+1}\mathbf{,y}^{n+1}\right)
\leq 
\end{equation*}
\begin{equation}
\left( 1+b_{3}\Delta t\right) \left\| \mathbf{y}^{n+1}\right\| _{\mathbf{W}%
^{n}}^{2},\quad b_{3}=\frac{\alpha _{0}}{\lambda _{0}}=const.  \label{Fr120}
\end{equation}
Then, in view of (\ref{Fr110}) and (\ref{Fr120}), we write 
\begin{equation}
\left\| \mathbf{y}^{n+1}\right\| _{\mathbf{W}^{n+1}}\leq \left( 1+c\Delta
t\right) \left\| \mathbf{y}^{n}\right\| _{\mathbf{W}^{n}},\quad c=\max
\left( b_{2},b_{3}\right) =const.  \label{Fr130}
\end{equation}
The inequality in (\ref{Fr130}) proves Theorem \ref{Generalized Friedrichs}.
\end{proof}

\section{Monotone $C^{1}$ piecewise cubics in construction of central schemes%
\label{COS}}

In this section we consider some theoretical aspects for high-order
interpolation and employment of monotone $C^{1}$ piecewise cubics (e.g., 
\cite{Fritsch and Carlson 1980}, \cite{Kocic and Milovanovic 1997}) in
construction of monotone central schemes. By virtue of the
operator-splitting idea (see also LOS in \cite{Samarskii 2001}), the
following chain of equations corresponds to the problem (\ref{INA10}) 
\begin{equation}
\frac{1}{2}\frac{\partial \mathbf{u}}{\partial t}+\frac{\partial }{\partial x%
}\mathbf{f}\left( \mathbf{u}\right) =0,\quad t_{n}<t\leq t_{n+0.5},\quad 
\mathbf{u}\left( x,t_{n}\right) =\mathbf{u}^{n}\left( x\right) ,  \label{C10}
\end{equation}
\begin{equation}
\frac{1}{2}\frac{\partial \mathbf{u}}{\partial t}=\frac{1}{\tau }\mathbf{q}%
\left( \mathbf{u}\right) ,\quad t_{n+0.5}<t\leq t_{n+1},\quad \mathbf{u}%
\left( x,t_{n+0.5}\right) =\mathbf{u}^{n+0.5}\left( x\right) .  \label{C20}
\end{equation}
Using the central differencing, we write 
\begin{equation}
\left. \frac{\partial \mathbf{u}}{\partial t}\right| _{t=t_{n+0.125},\
x=x_{i+0.5}}=\frac{\mathbf{u}_{i+0.5}^{n+0.25}-\mathbf{u}_{i+0.5}^{n}}{%
0.25\Delta t}+O\left( \left( \Delta t\right) ^{2}\right) ,  \label{C24}
\end{equation}
\begin{equation}
\left. \frac{\partial \mathbf{f}}{\partial x}\right| _{t=t_{n+0.125},\
x=x_{i+0.5}}=\frac{\mathbf{f}_{i+1}^{n+0.125}-\mathbf{f}_{i}^{n+0.125}}{%
\Delta x}+O\left( \left( \Delta x\right) ^{2}\right) .  \label{C25}
\end{equation}
By virtue of (\ref{C24})-(\ref{C25}) we approximate (\ref{C10}) on the cell $%
\left[ x_{i},x_{i+1}\right] \times \left[ t_{n},t_{n+0.25}\right] $ by the
following difference equation 
\begin{equation}
\mathbf{v}_{i+0.5}^{n+0.25}=\mathbf{v}_{i+0.5}^{n}-\frac{\Delta t}{2\Delta x}%
\left( \mathbf{g}_{i+1}^{n+0.125}-\mathbf{g}_{i}^{n+0.125}\right) ,
\label{C30}
\end{equation}
where $\mathbf{v}_{i}^{n+\beta }$, $\mathbf{g}_{i}^{n+\beta }$ are the grid
functions. As usually, the mathematical treatment for the second step (i.e.,
on the cell $\left[ x_{i-0.5},x_{i+0.5}\right] \times \left[
t_{n+0.25},t_{n+0.5}\right] $) of a staggered scheme will, in general, not
be included in the text, because it is quite similar to the one for the
first step.

\begin{figure}[h]
\centerline{\includegraphics[width=11.50cm,height=4.50cm]{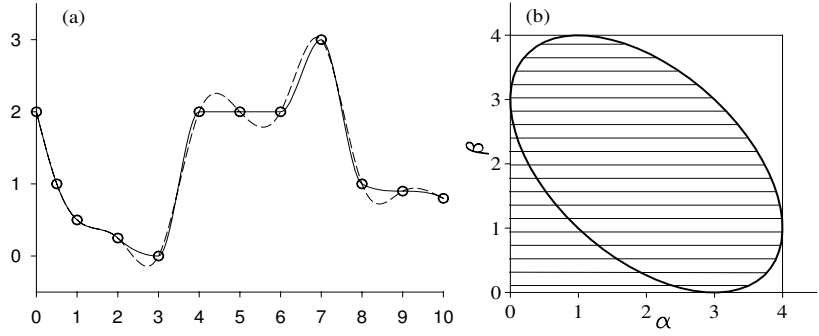}}
\caption{Monotone piecewise cubic interpolation. (a) Interpolation of a 1-D
tabulated function. Circles: prescribed tabulated values; Dashed line:
classic cubic spline; Solid line: monotone piecewise cubic. (b) Necessary
and sufficient conditions for monotonicity. Horizontal hatching: region of
monotonicity; Unshaded: cubic is non-monotone. }
\label{Fritsch}
\end{figure}

Considering that (\ref{C30}) approximate (\ref{C10}) with the accuracy $%
O(\left( \Delta x\right) ^{2}+\left( \Delta t\right) ^{2})$, the next
problem is to approximate $\mathbf{v}_{i+0.5}^{n}$ and $\mathbf{g}%
_{i}^{n+0.125}$ in such a way as to retain the accuracy of the
approximation. For instance, the following approximations 
\begin{equation}
\mathbf{v}_{i+0.5}^{n}=0.5\left( \mathbf{v}_{i}^{n}+\mathbf{v}%
_{i+1}^{n}\right) +O\left( \left( \Delta x\right) ^{2}\right) ,\quad \mathbf{%
g}_{i}^{n+0.125}=\mathbf{f}\left( \mathbf{v}_{i}^{n}\right) +O\left( \Delta
t\right) ,  \label{C50}
\end{equation}
leads to the staggered form of the famed LxF scheme that is of the
first-order approximation (see, e.g., \cite[p. 170]{Godlewski and Raviart
1996}). One way to obtain a higher-order scheme is to use a higher order
interpolation. At the same time it is required of the interpolant to be
monotonicity preserving. Notice, the classic cubic spline does not possess
such a property (see Figure \ref{Fritsch}a). Let us consider the problem of
high-order interpolation of $\mathbf{v}_{i+0.5}^{n}$ in (\ref{C30}) with
closer inspection

Let $\mathbf{p}=\mathbf{p}\left( x\right) \equiv \left\{ p^{1}\left(
x\right) ,\ldots ,p^{k}\left( x\right) ,\ldots ,p^{m}\left( x\right)
\right\} ^{T}$ be a component-wise monotone $C^{1}$ piecewise cubic
interpolant (e.g., \cite{Fritsch and Carlson 1980}, \cite{Kocic and
Milovanovic 1997}), and let 
\begin{equation*}
\mathbf{p}_{i}=\mathbf{p}\left( x_{i}\right) ,\quad \mathbf{p}_{i}^{\prime }=%
\mathbf{p}^{\prime }\left( x_{i}\right) ,\quad \Delta \mathbf{p}_{i}=\mathbf{%
p}_{i+1}-\mathbf{p}_{i},
\end{equation*}
\begin{equation}
\mathbf{p}_{i}^{\prime }=\mathbb{A}_{i}\cdot \frac{\Delta \mathbf{p}_{i}}{%
\Delta x},\quad \mathbf{p}_{i+1}^{\prime }=\mathbb{B}_{i}\cdot \frac{\Delta 
\mathbf{p}_{i}}{\Delta x},  \label{C80}
\end{equation}
where $\mathbf{p}_{i}^{\prime }$ denotes the derivative of the interpolant
at $x=x_{i}$. The diagonal matrices $\mathbb{A}_{i}$ and $\mathbb{B}_{i}$\
in (\ref{C80})\ are defined as follows 
\begin{equation}
\mathbb{A}_{i}=diag\left\{ \alpha _{i}^{1},\alpha _{i}^{2},\ldots ,\alpha
_{i}^{m}\right\} ,\ \mathbb{B}_{i}=diag\left\{ \beta _{i}^{1},\beta
_{i}^{2},\ldots ,\beta _{i}^{m}\right\} .  \label{C85}
\end{equation}
The cubic interpolant, $\mathbf{p}=\mathbf{p}\left( x\right) $, is
component-wise monotone on $\left[ x_{i},x_{i+1}\right] $ \emph{iff} one of
the following conditions (e.g., \cite{Fritsch and Carlson 1980}, \cite{Kocic
and Milovanovic 1997}) is satisfied: 
\begin{equation}
\left( \alpha _{i}^{k}-1\right) ^{2}+\left( \alpha _{i}^{k}-1\right) \left(
\beta _{i}^{k}-1\right) +\left( \beta _{i}^{k}-1\right) ^{2}-3\left( \alpha
_{i}^{k}+\beta _{i}^{k}-2\right) \leq 0,  \label{C90}
\end{equation}
\begin{equation}
\alpha _{i}^{k}+\beta _{i}^{k}\leq 3,\quad \alpha _{i}^{k}\geq 0,\ \beta
_{i}^{k}\geq 0,\quad \forall i,k.  \label{C100}
\end{equation}
As reported in \cite{Kocic and Milovanovic 1997}, the necessary and
sufficient conditions for monotonicity of a $C^{1}$ piecewise cubic
interpolant originally given by Ferguson and Miller (1969), and
independently, by Fritsch and Carlson \cite{Fritsch and Carlson 1980}. The
region of monotonicity is shown in Figure \ref{Fritsch}b. The results of
implementing a monotone $C^{1}$ piecewise cubic interpolation when compared
with the classic cubic spline interpolation, are depicted in Figure \ref
{Fritsch}a. We note (Figure \ref{Fritsch}a) that the constructed function
produces monotone interpolation and this function coincides with the classic
cubic spline at some sections where the classic cubic spline is monotone.

Using the cubic segment of the $C^{1}$ piecewise cubic interpolant, $\mathbf{%
p}=\mathbf{p}\left( x\right) $, (see, e.g., \cite{Fritsch and Carlson 1980}, 
\cite{Kocic and Milovanovic 1997}) for $x\in \left[ x_{i},x_{i+1}\right] $,
we obtain the following interpolation formula 
\begin{equation}
\mathbf{p}_{i+0.5}=0.5\left( \mathbf{p}_{i}+\mathbf{p}_{i+1}\right) -\frac{%
\Delta x}{8}\left( \mathbf{p}_{i+1}^{\prime }-\mathbf{p}_{i}^{\prime
}\right) +O\left( \left( \Delta x\right) ^{r}\right) .  \label{C110}
\end{equation}
If $\mathbf{p}\left( x\right) $ has a continuous fourth derivative, then $%
r=4 $ in (\ref{C110}), see e.g. \cite[p. 111]{Kahaner et al. 1989}. However,
the exact value of $\mathbf{p}_{i}^{\prime }$ in (\ref{C110}) is, in
general, unknown, and hence to construct numerical schemes, employing
formulae similar to (\ref{C110}), the value of derivatives $\mathbf{p}%
_{i}^{\prime }$ must be estimated.

Using (\ref{C110}) and the second formula in (\ref{C50}) we obtain from (\ref
{C30}) the following scheme 
\begin{equation}
\mathbf{v}_{i+0.5}^{n+0.25}=0.5\left( \mathbf{v}_{i}^{n}+\mathbf{v}%
_{i+1}^{n}\right) -\frac{\Delta x}{8}\left( \mathbf{d}_{i+1}^{n}-\mathbf{d}%
_{i}^{n}\right) -\frac{\Delta t}{2}\frac{\mathbf{f}\left( \mathbf{v}%
_{i+1}^{n}\right) -\mathbf{f}\left( \mathbf{v}_{i}^{n}\right) }{\Delta x},
\label{C120}
\end{equation}
where $\mathbf{d}_{i}^{n}$ denotes the derivative of the interpolant at $%
x=x_{i}$. In view of (\ref{C110}) and the second formula in (\ref{C50}), the
local truncation error \cite[p. 142]{LeVeque 2002}, $\psi $, on a
sufficiently smooth solution $\mathbf{u}(x,t)$ to (\ref{C10}) is found to be 
\begin{equation}
\psi =O\left( \Delta t\right) +O\left( \frac{\left( \Delta x\right) ^{r}}{%
\Delta t}\right) +O\left( \left( \Delta t\right) ^{2}+\left( \Delta x\right)
^{2}\right) .  \label{C130}
\end{equation}
In view of (\ref{C130}) we conclude that the scheme (\ref{C120}) generates a
conditional approximation, because it approximates (\ref{C10}) only if $%
\left( \Delta x\right) ^{r}\diagup \Delta t\rightarrow 0$ as $\Delta
x\rightarrow 0$ and $\Delta t\rightarrow 0$. Let $\mathbf{d}_{i}^{n}$\ be
approximated with the accuracy $O\left( \left( \Delta x\right) ^{s}\right) $%
, then the value of $r$ in (\ref{C130}) can be calculated (see Section \ref
{Appendix1}, Proposition \ref{Approximate derivative}) by the following
formula 
\begin{equation}
r=\min \left( 4,s+1\right) .  \label{C140}
\end{equation}
Interestingly, since (\ref{C120}) provides the conditional approximation,
the order of accuracy depends on the pathway taken by $\Delta x$ and $\Delta
t$ as $\Delta x\rightarrow 0$ and $\Delta t\rightarrow 0$. Actually, there
exists a pathway such that $\Delta t$ is proportional to $\left( \Delta
x\right) ^{\mu }$ and the CFL condition is fulfilled provided $\mu \geq 1$
and $\Delta x\leq \Delta x_{0}$, where $\Delta x_{0}$ is a positive value.
If we take $\mu =1$ and $s\geq 1$, then we obtain from (\ref{C130}) that the
scheme (\ref{C120}) is of the first-order. If $\mu =2$ and $s\geq 3$, then (%
\ref{C120}) is of the second-order. However, if $\mu =2$ and $s=2$, then, in
view of (\ref{C130}) and (\ref{C140}), the scheme (\ref{C120}) is of the
first-order. Moreover, under $\mu =2$ and $s=2$, the scheme will be of the
first-order even if $\mathbf{g}_{i}^{n+0.125}$ in (\ref{C30}) will be
approximated with the accuracy$\ O(\left( \Delta t\right) ^{2})$. It seems
likely that Example 6 in \cite{Kurganov and Tadmor 2000} can be seen as an
illustration of the last assertion. The Nessyahu-Tadmor (NT) scheme with the
second-order approximation of $\mathbf{d}_{i}^{n}$ is used \cite{Kurganov
and Tadmor 2000} to solve a Burgers-type equation. Since $\Delta t$ $=$ $%
O(\left( \Delta x\right) ^{2})$ \cite{Kurganov and Tadmor 2000}, the NT
scheme is of the first-order, and hence it can be the main reason for the
scheme to exhibit the smeared discontinuity computed in \cite[Fig. 6.22]
{Kurganov and Tadmor 2000}.

The approximation of derivatives $\mathbf{p}_{i}^{\prime }$ can be done by
the following three steps \cite{Fritsch and Carlson 1980}: (i) an
initialization of the derivatives $\mathbf{p}_{i}^{\prime }$; (ii) the
choice of subregion of monotonicity; (iii) modification of the initialized
derivatives $\mathbf{p}_{i}^{\prime }$ to produce a monotone interpolant.

The matter of initialization of the derivatives is the most subtle issue of
this algorithm. Actually, the approximation of $\mathbf{p}_{i}^{\prime }$
must, in general, be done with accuracy $O(\left( \Delta x\right) ^{3})$ to
obtain the second-order scheme when $\Delta t$ is proportional to $\left(
\Delta x\right) ^{2}$, inasmuch as central schemes generate a conditional
approximation. Thus, using the two-point or the three-point (centered)
difference formula (e.g. \cite{Kocic and Milovanovic 1997}, \cite{Pareschi
Lorenzo 2001}) we obtain, in general, the first-order scheme. The so called
limiter functions \cite{Kocic and Milovanovic 1997} lead, in general, to a
low-order scheme as these limiters are often $O(\Delta x)$ or $O(\left(
\Delta x\right) ^{2})$ accurate. Performing the initialization of the
derivatives $\mathbf{p}_{i}^{\prime }$ in the interpolation formula (\ref
{C110}) by the classic cubic spline interpolation \cite{Press William 1988},
we obtain the approximation, which is $O(\left( \Delta x\right) ^{3})$
accurate (e.g., \cite{Kahaner et al. 1989}, \cite{Kocic and Milovanovic 1997}%
), and hence, in general, the second-order scheme. The same accuracy, $%
O(\left( \Delta x\right) ^{3})$, can be achieved by using the four-point
approximation \cite{Kocic and Milovanovic 1997}. However, the efficiency of
the algorithm based on the classic cubic spline interpolation is comparable
with the one based on the four-point approximation, as the number of
multiplications and divisions (as well as additions and subtractions) per
one node is approximately the same for both algorithms. We will use the
classic cubic spline interpolation for the initialization of the derivatives 
$\mathbf{P}_{i}^{\prime }$ in the interpolation formula (\ref{C110}), as it
is based on the tridiagonal algorithm, which is `the rare case of an
algorithm that, in practice, is more robust than theory says it should be' 
\cite{Press William 1988}.

Obviously, for each interval $\left[ x_{i},x_{i+1}\right] $ in which the
initialized derivatives $\mathbf{p}_{i}^{\prime }$, $\mathbf{p}%
_{i+1}^{\prime }$ such that at least one point ($\alpha _{i}^{k}$, $\beta
_{i}^{k}$) does not belong to the region of monotonicity (\ref{C90})-(\ref
{C100}), the derivatives $\mathbf{p}_{i}^{\prime }$, $\mathbf{p}%
_{i+1}^{\prime }$ must be modified to $\widetilde{\mathbf{p}}_{i}^{\prime }$%
, $\widetilde{\mathbf{p}}_{i+1}^{\prime }$ such that the point ($\widetilde{%
\alpha }_{i}^{k}$, $\widetilde{\beta }_{i}^{k}$) will be in the region of
monotonicity. 
The modification of the initialized derivatives, would be much simplified if
we take a square as a subregion of monotonicity. In connection with this, we
will make use the subregions of monotonicity represented in the following
form: 
\begin{equation}
0\leq \alpha _{i}^{k}\leq 4\aleph ,\quad 0\leq \beta _{i}^{k}\leq 4\aleph
,\quad \forall i,k,  \label{CA180}
\end{equation}
where $\aleph $ is a monotonicity parameter. 
Obviously, the condition (\ref{CA180}) is sufficient for the monotonicity
(see Figure \ref{Fritsch}b) provided that $0\leq $ $\aleph $ $\leq 0.75$.

\label{aaa1}Let us now find necessary and sufficient conditions for (\ref
{C110}) to be G-monotone. By virtue of (\ref{C80}), the interpolation
formula (\ref{C110}) can be rewritten to read 
\begin{equation}
\mathbf{p}_{i+0.5}=\left( 0.5\mathbf{I}+\frac{\mathbb{B}_{i}-\mathbb{A}_{i}}{%
8}\right) \cdot \mathbf{p}_{i}+\left( 0.5\mathbf{I}-\frac{\mathbb{B}_{i}-%
\mathbb{A}_{i}}{8}\right) \cdot \mathbf{p}_{i+1}.  \label{CA185}
\end{equation}
The coefficients of (\ref{CA185}) will be non-negative \emph{iff} $\left|
\beta _{i}-\alpha _{i}\right| \leq 4$. Hence (\ref{C110}) will be G-monotone 
\emph{iff} (\ref{CA180}) will be valid provided $0\leq $ $\aleph $ $\leq 1$.
Notice, there is no any contradiction between the sufficient conditions, (%
\ref{CA180}) provided $0\leq $ $\aleph $ $\leq 0.75$, for the interpolant, $%
\mathbf{p}=\mathbf{p}\left( x\right) $,\ to be monotone through the interval 
$\left[ x_{i},x_{i+1}\right] $, and the necessary and sufficient conditions,
(\ref{CA180}) provided $0\leq $ $\aleph $ $\leq 1$, for the scheme (\ref
{CA185}) to be G-monotone. In the latter case the interpolant, $\mathbf{p}=%
\mathbf{p}\left( x\right) $,\ may, in general, be non-monotone, however at
the point $i+0.5$ the value of an arbitrary component of $\mathbf{p}_{i+0.5}$
will be between the corresponding components of $\mathbf{p}_{i}$ and $%
\mathbf{p}_{i+1}$.

To fulfill the conditions of monotonicity (\ref{CA180}), the modification of
derivatives $\mathbf{p}_{i}^{\prime }=\left\{ p_{i}^{\prime 1},p_{i}^{\prime
2},\ldots ,p_{i}^{\prime m}\right\} $ can be done by the following algorithm
suggested, in fact, by Fritsch and Carlson \cite{Fritsch and Carlson 1980}
(see also \cite{Kocic and Milovanovic 1997}): 
\begin{equation}
S_{i}^{k}:=4\aleph \min \func{mod}(\Delta _{i-1}^{k},\Delta _{i}^{k}),\quad 
\widetilde{p}_{i}^{\prime k}:=\min \func{mod}(p_{i}^{\prime
k},S_{i}^{k}),\quad \aleph =const,  \label{REMa10}
\end{equation}
where $\Delta _{i}^{k}=\left( p_{i+1}^{k}-p_{i}^{k}\right) \diagup \Delta x$%
, the function $\min \func{mod}(x,y)$ is defined (e.g., \cite{Kocic and
Milovanovic 1997}, \cite{Kurganov and Tadmor 2000}, \cite{Morton 2001}, \cite
{Pareschi Lorenzo 2001}, \cite{Serna and Marquina 2005}) as follows 
\begin{equation}
\min \func{mod}(x,y)\equiv \frac{1}{2}\left[ sgn(x)+sgn(y)\right] \min
\left( \left| x\right| ,\left| y\right| \right) .  \label{REMa20}
\end{equation}

\section{Construction of first- and second-order central schemes\label{COSN}}

Central difference schemes with first- and second-order accuracy are
introduced in this section. The construction of the central schemes is based
on: (i) Variational GOS-monotonicity notion, (ii) Monotone piecewise cubic
interpolation (e.g., \cite{Fritsch and Carlson 1980}, \cite{Kocic and
Milovanovic 1997}), (iii)\ Operator-splitting techniques (see also LOS in 
\cite{Samarskii 2001}).

\subsection{First-order central schemes\label{COS1}}

Let us note that instead of point values, $\mathbf{v}_{i+0.5}^{n}$, employed
in the construction of the scheme (\ref{C30}), it can be used the cell
averages (e.g., \cite{Balaguer and Conde 2005}, \cite{Kurganov and Tadmor
2000}, \cite{LeVeque 2002}) calculated on the basis of the monotone $C^{1}$
piecewise cubics. In such a case we obtain, instead of (\ref{C110}), the
following interpolation formula 
\begin{equation}
\mathbf{p}_{i+0.5}=0.5\left( \mathbf{p}_{i}+\mathbf{p}_{i+1}\right)
-\varkappa \frac{\Delta x}{8}\left( \mathbf{p}_{i+1}^{\prime }-\mathbf{p}%
_{i}^{\prime }\right) ,  \label{CA200}
\end{equation}
where $\varkappa =2\diagup 3$. The region of monotonicity in this case will
also be

\begin{equation}
0\leq \mathbb{A}_{i}\leq 4\aleph \mathbf{I},\ 0\leq \mathbb{B}_{i}\leq
4\aleph \mathbf{I},\quad 0\leq \aleph \leq 1,\quad \forall i.  \label{CA203}
\end{equation}
Notice, the interpolation formula (\ref{CA200}) coincides with (\ref{C110})
under $\varkappa =1$. Thus, in view of the interpolation formula (\ref{CA200}%
), the staggered scheme (\ref{C30}) is written to read 
\begin{equation}
\mathbf{v}_{i+0.5}^{n+0.25}=0.5\left( \mathbf{v}_{i+1}^{n}+\mathbf{v}%
_{i}^{n}\right) -\varkappa \frac{\Delta x}{8}\left( \mathbf{d}_{i+1}^{n}-%
\mathbf{d}_{i}^{n}\right) -\frac{\Delta t}{2}\frac{\mathbf{f}\left( \mathbf{v%
}_{i+1}^{n}\right) -\mathbf{f}\left( \mathbf{v}_{i}^{n}\right) }{\Delta x},
\label{CA210}
\end{equation}
where $\mathbf{d}_{i}^{n}$ denotes the derivative of the interpolant at $%
x=x_{i}$, the range of values for the parameter $\varkappa $ is the segment $%
0\leq \varkappa \leq 1$. As usually, the mathematical treatments for the
second step of the staggered scheme will, in general, not be included in the
text, because the second step is quite similar to (\ref{CA210}). If $%
\varkappa =1$ (or $\varkappa =0$), then Scheme (\ref{CA210}) coincides with
the scheme (\ref{C120}) (or with the LxF scheme, respectively). As it was
shown above, the scheme (\ref{CA210}) is of the first order provided $\Delta
t$ $=$ $O\left( \Delta x\right) $. In such a case, since the source terms
can be, in general, stiff (i.e., $\tau \ll 1$), it is natural to use the
following first-order implicit scheme for (\ref{C20}). 
\begin{equation}
\mathbf{v}_{i}^{n+1}=\mathbf{v}_{i}^{n+0.5}+\frac{\Delta t}{\tau }\mathbf{q}%
\left( \mathbf{v}_{i}^{n+1}\right) .  \label{C55}
\end{equation}
The first order central scheme (\ref{CA210})-(\ref{C55}) based on
operator-splitting techniques will be abbreviated to as COS1.

Let us investigate the stability of Scheme (\ref{CA210}). It is assumed that
the vector-valued function $\mathbf{u=u}\left( x,t\right) $ is G\^{a}teaux-
(or Fr\'{e}chet-) differentiable on the convex set $\Omega _{xt}$ $\subset $ 
$\mathbb{R}$ $\times $ $[0,+\infty )$, and its derivative is bounded on $%
\Omega _{xt}$. It is also assumed that the operator $\mathbf{A}$ $(=\partial 
\mathbf{f}\left( \mathbf{u}\right) \diagup \partial \mathbf{u)}$ is
Fr\'{e}chet-differentiable on the convex set $\Omega _{\mathbf{u}}$ $\subset 
$ $\mathbb{R}^{M}$, and its derivative is bounded on $\Omega _{\mathbf{u}}$.
Hence $\mathbf{A}$ $\mathbf{=A}\left( x,t\right) $ will be G\^{a}teaux- (or,
respectively, Fr\'{e}chet-) differentiable \cite[p. 62]{Ortega and
Rheinboldt 1970} and its derivative will be bounded on $\Omega _{xt}$, and
hence $\mathbf{A}\left( x,t\right) $ will be Lipschitz-continuous on $\Omega
_{xt}$ \cite[p. 70]{Ortega and Rheinboldt 1970}. Since the Jacobian matrix $%
\mathbf{A}$ $\mathbf{=A}\left( x,t\right) $ in (\ref{IN80}) possesses $M$
linearly independent eigenvectors, $\mathbf{A}_{i}^{n}$ ($\mathbf{=A}\left(
x_{i},t_{n}\right) $) is similar to a diagonal matrix \cite{Mirsky 1990},
i.e. there exists a non-singular matrix $\mathbf{S}_{i}^{n}=\mathbf{S}\left(
x_{i},t_{n}\right) $ such that 
\begin{equation}
\left( \mathbf{S}_{i}^{n}\right) ^{-1}\cdot \mathbf{A}_{i}^{n}\cdot \mathbf{S%
}_{i}^{n}=diag\left\{ \lambda _{i}^{n,1},\lambda _{i}^{n,2},\ldots ,\lambda
_{i}^{n,M}\right\} ,\quad \forall i,n.  \label{CAB40}
\end{equation}
The right and left eigenvectors of $\mathbf{A}_{i}^{n}\mathbf{=A}\left(
x_{i},t_{n}\right) $ can be defined in such a way that $\mathbf{S}_{i}^{n}=%
\mathbf{S}\left( x_{i},t_{n}\right) $ will be the matrix having the right
eigenvectors as its columns, and the rows of $\left( \mathbf{S}%
_{i}^{n}\right) ^{-1}=\mathbf{S}^{-1}\left( x_{i},t_{n}\right) $ will be the
left eigenvectors \cite[p. 62]{Lancaster 1969}. For Theorem \ref
{StabilitySufficient} to be used, it must be proven that $\left[ \left( 
\mathbf{S}_{i}^{n}\right) ^{-1}\right. $ $-$ $\left. \left( \mathbf{S}%
^{n}\right) ^{-1}\left( x\right) \right] \cdot \mathbf{S}^{n}\left( x\right) 
$ and $\left[ \left( \mathbf{S}_{i}\right) ^{-1}\left( t\right) \right. $ $-$
$\left. \left( \mathbf{S}_{i}^{n}\right) ^{-1}\right] \cdot \mathbf{S}%
_{i}^{n}$ will be Lipschitz-continuous in space and, respectively, time $%
\forall i,n$. Since each of the above functions depends on one parameter ($x$
or $t$) only, it can be done with ease for strictly hyperbolic systems, i.e.
when the eigenvalues of the operator $\mathbf{A}$ $(=\partial \mathbf{f}%
\left( \mathbf{u}\right) \diagup \partial \mathbf{u)}$ in (\ref{IN80}) will
be all distinct \cite[p. 2]{LeVeque 2002}. In this case the proof follows
from the long-known results of perturbation theory for simple eigenvalues 
\cite[p. 67]{Wilkinson 1969} (see also Theorem 2.1 in \cite{Andrew et al.
1993}). For a detailed discussion on the derivatives (sensitivities) of
eigenvectors of matrix-valued functions depending on several parameters when
the eigenvalues are multiple see \cite{Andrew et al. 1993}, \cite{Xie and
Dai 2003}.

By virtue of (\ref{C80}), the second term in right-hand side of (\ref{CA210}%
) can be written in the form 
\begin{equation}
\varkappa \frac{\Delta x}{8}\left( \mathbf{d}_{i+1}^{n}-\mathbf{d}%
_{i}^{n}\right) =\frac{\varkappa }{8}\left( \mathbb{B}_{i}^{n}-\mathbb{A}%
_{i}^{n}\right) \cdot \left( \mathbf{v}_{i+1}^{n}-\mathbf{v}_{i}^{n}\right) .
\label{DD10}
\end{equation}
Then, the variational scheme corresponding to (\ref{CA210}) is the following 
\begin{equation*}
\delta \mathbf{v}_{i+0.5}^{n+0.25}=0.5\left( \delta \mathbf{v}%
_{i}^{n}+\delta \mathbf{v}_{i+1}^{n}\right) +\frac{\varkappa }{8}\left[
\left( \mathbf{v}_{i}^{n}-\mathbf{v}_{i+1}^{n}\right) ^{T}\cdot \delta 
\mathbb{D}_{i}^{n}\right] ^{T}+
\end{equation*}
\begin{equation}
\frac{\varkappa }{8}\mathbb{D}_{i}^{n}\cdot \left( \delta \mathbf{v}%
_{i}^{n}-\delta \mathbf{v}_{i+1}^{n}\right) +\frac{\Delta t}{2\Delta x}%
\left( \mathbf{A}_{i}^{n}\cdot \delta \mathbf{v}_{i}^{n}-\mathbf{A}%
_{i+1}^{n}\cdot \delta \mathbf{v}_{i+1}^{n}\right) ,  \label{DD20}
\end{equation}
where $\mathbb{D}_{i}^{n}=diag\left\{ D_{i,1}^{n},D_{i,2}^{n},\ldots
,D_{i,M}^{n}\right\} \equiv \mathbb{B}_{i}^{n}-\mathbb{A}_{i}^{n}$. By
virtue of (\ref{CA203}), we find that $-4\aleph \mathbf{I}\leq \mathbb{D}%
_{i}^{n}\leq 4\aleph \mathbf{I}$, and hence $-8\aleph \mathbf{I}\leq \delta 
\mathbb{D}_{i}^{n}\leq 8\aleph \mathbf{I}$. Thus, we may write that 
\begin{equation}
\left\| \delta \mathbb{D}_{i}^{n}\right\| _{2}\leq 8\aleph .  \label{DD30}
\end{equation}
Considering that $\mathbf{v}_{i}^{n}$ in (\ref{CA210}) is
Lipschitz-continuous on $\Omega _{xt}$, we write 
\begin{equation}
\left\| \mathbf{v}_{i}^{n}-\mathbf{v}_{i+1}^{n}\right\| _{2}\leq C_{v}\Delta
x,\quad C_{v}=const.  \label{DD40}
\end{equation}
It is assumed, see (\ref{CA05}), that there exists $\alpha _{0}$ $=$ $const$
such that $\Delta x$ $\leq $ $\alpha _{0}\Delta t$ for a sufficiently small $%
\Delta t$. Then, by virtue of (\ref{DD30}) and (\ref{DD40}), and since $%
0\leq $ $\varkappa ,\aleph $ $\leq 1$, we find the following estimation for
the second term in right-hand side of (\ref{DD20}): 
\begin{equation}
\left\| \frac{\varkappa }{8}\left[ \left( \mathbf{v}_{i}^{n}-\mathbf{v}%
_{i+1}^{n}\right) ^{T}\cdot \delta \mathbb{D}_{i}^{n}\right] ^{T}\right\|
_{2}\leq \frac{\varkappa }{8}\left\| \mathbf{v}_{i}^{n}-\mathbf{v}%
_{i+1}^{n}\right\| _{2}\left\| \delta \mathbb{D}_{i}^{n}\right\| _{2}\leq
\alpha _{0}C_{v}\Delta t.  \label{DD50}
\end{equation}
In view of (\ref{DD50}), the scheme (\ref{DD20}) will be stable if the
following scheme will be stable (e.g., \cite[pp. 390-392]{Samarskii 2001}) 
\begin{equation*}
\delta \mathbf{v}_{i+0.5}^{n+0.25}=0.5\left( \mathbf{I+E}_{i}^{n}\right)
\cdot \delta \mathbf{v}_{i}^{n}+0.5\left( \mathbf{I-E}_{i+1}^{n}\right)
\cdot \delta \mathbf{v}_{i+1}^{n}\equiv
\end{equation*}
\begin{equation}
0.5\left( \mathbf{I+}\frac{\varkappa }{4}\mathbb{D}_{i}^{n}+\frac{\Delta t}{%
\Delta x}\mathbf{A}_{i}^{n}\right) \cdot \delta \mathbf{v}_{i}^{n}+0.5\left( 
\mathbf{I-}\frac{\varkappa }{4}\mathbb{D}_{i}^{n}-\frac{\Delta t}{\Delta x}%
\mathbf{A}_{i+1}^{n}\right) \cdot \delta \mathbf{v}_{i+1}^{n}.  \label{DD60}
\end{equation}
We write, in view of (\ref{IN80}) and (\ref{CA203}), that 
\begin{equation}
s\left( \mathbf{E}_{i}^{n}\right) \subset \left[ -\varkappa \aleph -\frac{%
\Delta t}{\Delta x}\lambda _{\max },\ \varkappa \aleph +\frac{\Delta t}{%
\Delta x}\lambda _{\max }\right] ,\quad \forall i,n.  \label{DD70}
\end{equation}
Hence, by virtue of Theorem \ref{StabilitySufficient} we find that the
scheme (\ref{DD60}) will be stable if 
\begin{equation}
\underset{\lambda \in s\left( \mathbf{E}_{i}^{n}\right) }{\max }0.5\left(
\left| {1+\lambda }\right| +\left| {1-\lambda }\right| \right) \leqslant
1,\quad \forall i,n.  \label{DD80}
\end{equation}
We obtain from (\ref{DD80}) the following condition for the stability of the
variational scheme (\ref{DD60}): 
\begin{equation}
\varkappa \aleph +C_{r}\leqslant 1,\quad C_{r}=\frac{\Delta t\lambda _{\max }%
}{\Delta x},  \label{DD90}
\end{equation}
where $C_{r}$ denotes the Courant number. Thus, in view of Theorem \ref
{Linear-nonlinear stability} the scheme (\ref{CA210}) will be stable if (\ref
{DD90}) will be valid.

Notice, (\ref{DD90}) was obtained on the basis that $\mathbf{E}%
_{i}^{n}\equiv \frac{\varkappa }{4}\mathbb{D}_{i}^{n}+\frac{\Delta t}{\Delta
x}\mathbf{A}_{i}^{n}$ is diagonizable. Such an assumption should be verified
for a concrete problem. It might be well to point out that this assumption
has a rigorous basis if $\mathbf{A}_{i}^{n}$ is diagonizable and $\mathbb{D}%
_{i}^{n}$ is a scalar matrix, i.e. $\mathbb{D}_{i}^{n}=4\theta ^{n}\mathbf{I}
$, $\theta ^{n}=const$ ($-\aleph \leq \theta ^{n}\leq \aleph $). In such a
case $\mathbf{E}_{i}^{n}$ will be diagonizable.\ Thus, for instance, LxF
scheme will be stable if $C_{r}\leqslant 1$. In the case, when $\mathbb{D}%
_{i}^{n}$ is not a scalar matrix, whereas the Jacobian matrix $\mathbf{A}$
in (\ref{IN80}) is symmetric, the condition (\ref{DD90}) for stability of (%
\ref{CA210}) can be found with ease by virtue of Proposition \ref
{Proposition01}. In a more general case, when $\mathbf{E}_{i}^{n}$ is not
diagonizable as well as$\ \mathbf{A}$ is not symmetric, the stability of (%
\ref{CA210}) can be investigated by virtue of Theorem \ref{Generalized
Friedrichs}.

Let us find a necessary condition for the variational S-monotonicity (see
Definition \ref{Variational monotonicity}) of the COS1 scheme, (\ref{CA210}%
)-(\ref{C55}). Considering (\ref{DD20}) on a constant ($\mathbf{v}_{i}^{n}$ $%
=$ $\mathbf{C}$ $=$ $const$, $\forall i,n$), we obtain 
\begin{equation*}
\delta \mathbf{v}_{i+0.5}^{n+0.25}=0.5\left( \delta \mathbf{v}%
_{i}^{n}+\delta \mathbf{v}_{i+1}^{n}\right) +\frac{\varkappa }{8}\mathbb{D}%
\cdot \left( \delta \mathbf{v}_{i}^{n}-\delta \mathbf{v}_{i+1}^{n}\right) +
\end{equation*}
\begin{equation}
+\frac{\Delta t}{2\Delta x}\mathbf{A}_{c}\cdot \left( \delta \mathbf{v}%
_{i}^{n}-\delta \mathbf{v}_{i+1}^{n}\right) ,\quad \mathbf{A}_{c}=\mathbf{A}%
\left( \mathbf{C}\right) ,\quad \mathbf{A}\left( \mathbf{u}\right) =\frac{%
\partial \mathbf{f}\left( \mathbf{u}\right) }{\partial \mathbf{u}}.
\label{CAB240}
\end{equation}
Here, in (\ref{CAB240}), $\mathbf{A}_{c}$\ is a diagonizable matrix such
that its spectrum $s\left( \mathbf{A}_{c}\right) $ $\subset $ $\left[
-\lambda _{\max },\right. $ $\left. \lambda _{\max }\right] $, see (\ref
{IN80}). In view of (\ref{DD10}), $\mathbb{D}$ in (\ref{CAB240}) can be
taken at will, as $\mathbf{v}_{i}^{n}$ $=$ $\mathbf{C}$ $=$ $const$, $%
\mathbf{v}_{i+1}^{n}$ $-$ $\mathbf{v}_{i}^{n}$ $=$ $0$ and $\mathbf{d}%
_{i}^{n}$ $=$ $\mathbf{d}_{i+1}^{n}$ $=$ $0$. Aiming to obtain the necessary
condition for (\ref{DD20}) to be S-monotone, it is assumed that $\mathbb{D}%
=\zeta \mathbf{I}$\ is a scalar matrix with $\mathbb{\zeta }$ $\in $ $\left[
-4\aleph ,4\aleph \right] $. In view of Theorem \ref{S-monotoneIFF}, the
scheme (\ref{CAB240}) will be S-monotone \emph{iff} 
\begin{equation}
\underset{\left| \zeta \right| \leq 4\aleph ,\ \lambda \in s\left( \mathbf{A}%
_{c}\right) }{\max }\left| {\frac{1}{2}+\frac{\varkappa \zeta }{8}+\frac{%
\Delta t}{2\Delta x}\lambda }\right| +\left| {\frac{1}{2}-\frac{\varkappa
\zeta }{8}-\frac{\Delta t}{2\Delta x}\lambda }\right| \leqslant 1.
\label{REM40}
\end{equation}
By virtue of (\ref{REM40}), we find that (\ref{DD90}) is the necessary
condition for the variational S-monotonicity of (\ref{CA210}).

The variational scheme corresponding to (\ref{C55}) reads 
\begin{equation}
\delta \mathbf{v}_{i}^{n+1}=\delta \mathbf{v}_{i}^{n+0.5}+\frac{\Delta t}{%
\tau }\mathbf{G}\left( \mathbf{v}_{i}^{n+1}\right) \cdot \delta \mathbf{v}%
_{i}^{n+1},  \label{CAB210}
\end{equation}
where $\mathbf{G}\left( \mathbf{V}\right) =\partial \mathbf{q}\left( \mathbf{%
V}\right) \diagup \partial \mathbf{V}$. Let us rewrite (\ref{CAB210}) to
read 
\begin{equation}
\delta \mathbf{v}_{i}^{n+1}=\left\{ \mathbf{I}-\frac{\Delta t}{\tau }\mathbf{%
G}\left( \mathbf{v}_{i}^{n+1}\right) \right\} ^{-1}\cdot \delta \mathbf{v}%
_{i}^{n+0.5}.  \label{CAB220}
\end{equation}
Let $\mathbf{G}$ will be a normal matrix. In such a case, in view of 
\cite[Theorem 3.3]{Borisov and Sorek 2004} the variational scheme (\ref
{CAB220}) will be S-monotone if 
\begin{equation}
\underset{k}{\max }\frac{1}{\left| 1-\frac{\Delta t}{\tau }\xi _{k}\left( 
\mathbf{v}_{i}^{n+1}\right) \right| }\leq 1,\quad \forall i,n\mathbf{,}
\label{CAB230}
\end{equation}
where $\xi _{k}$ denotes the $k-th$ eigenvalue of the matrix $\mathbf{G}$.
In view of (\ref{INA20}), the inequality (\ref{CAB230}) is valid, and hence
the variational scheme (\ref{CAB210}) will be unconditionally S-monotone.

Thus, the COS1 scheme, (\ref{CA210})-(\ref{C55}), will be variationally
S-monotone only if (\ref{DD90}) will be valid.

\subsection{Second-order central scheme\label{Second-order central schemes}}

In this section, the second-order scheme for (\ref{C10}) is developed by
approximating $\mathbf{v}_{i+0.5}^{n}$\ and $\mathbf{g}_{i}^{n+0.125}$ in (%
\ref{C30}) with the accuracy $O(\left( \Delta x\right) ^{2}+\left( \Delta
t\right) ^{2})$. The sufficient conditions for stability as well as the
necessary condition for S-monotonicity of this scheme are found.

Using Taylor series expansion, we write 
\begin{equation}
\mathbf{g}_{i}^{n+0.125}=\mathbf{f}\left( \mathbf{v}_{i}^{n}\right) +\left. 
\frac{\partial \mathbf{f}\left( \mathbf{v}_{i}^{n}\right) }{\partial t}%
\right| _{t=t_{n}}\frac{\Delta t}{8}+O\left( \Delta t^{2}\right) .
\label{SA10}
\end{equation}
By virtue of the PDE system, (\ref{C10}), we find 
\begin{equation}
\frac{\partial \mathbf{f}}{\partial t}=\frac{\partial \mathbf{f}}{\partial 
\mathbf{u}}\cdot \frac{\partial \mathbf{u}}{\partial t}=-2\frac{\partial 
\mathbf{f}}{\partial \mathbf{u}}\cdot \frac{\partial \mathbf{f}}{\partial 
\mathbf{u}}\cdot \frac{\partial \mathbf{u}}{\partial x}=-2\left( \frac{%
\partial \mathbf{f}}{\partial \mathbf{u}}\right) ^{2}\cdot \frac{\partial 
\mathbf{u}}{\partial x}.  \label{SA20}
\end{equation}
Using the interpolation formula (\ref{CA200}) and the formulae (\ref{SA10})-(%
\ref{SA20}), we obtain from (\ref{C30}) the following second order central
scheme 
\begin{equation*}
\mathbf{v}_{i+0.5}^{n+0.25}=0.5\left( \mathbf{v}_{i+1}^{n}+\mathbf{v}%
_{i}^{n}\right) -\varkappa \frac{\Delta x}{8}\left( \mathbf{d}_{i+1}^{n}-%
\mathbf{d}_{i}^{n}\right) +
\end{equation*}
\begin{equation}
\frac{\left( \Delta t\right) ^{2}}{8\Delta x}\left[ \left( \mathbf{A}%
_{i+1}^{n}\right) ^{2}\cdot \mathbf{d}_{i+1}^{n}-\left( \mathbf{A}%
_{i}^{n}\right) ^{2}\cdot \mathbf{d}_{i}^{n}\right] -\frac{\Delta t}{2}\frac{%
\mathbf{f}\left( \mathbf{v}_{i+1}^{n}\right) -\mathbf{f}\left( \mathbf{v}%
_{i}^{n}\right) }{\Delta x}.  \label{SA30}
\end{equation}
Since $\mathbf{d}_{i}^{n}$ is the derivative of the interpolant at $x=x_{i}$%
, the third term in the right-hand side of (\ref{SA30}) can be seen as the
non-negative numerical viscosity introduced into the first order scheme (\ref
{CA210}). Owing to this term, the scheme (\ref{SA30}) is $O(\left( \Delta
x\right) ^{2}+\left( \Delta t\right) ^{2})$ accurate. Since the source terms
in (\ref{INA10}) can be, in general, stiff (i.e., $\tau \ll 1$), it is
natural to use the following second-order implicit Runge-Kutta scheme for (%
\ref{C20}), since this scheme possesses a discrete analogy to the continuous
asymptotic limit, 
\begin{equation*}
\mathbf{v}_{i}^{n+0.75}=\mathbf{v}_{i}^{n+1}-\frac{\gamma }{2}\mathbf{q}%
\left( \mathbf{v}_{i}^{n+1}\right) ,
\end{equation*}
\begin{equation}
\mathbf{v}_{i}^{n+1}=\mathbf{v}_{i}^{n+0.5}+\gamma \mathbf{q}\left( \mathbf{v%
}_{i}^{n+0.75}\right) ,\quad \gamma \equiv \frac{\Delta t}{\tau }.
\label{SA45}
\end{equation}
Scheme (\ref{SA30})-(\ref{SA45}), based on operator-splitting techniques,
will be abbreviated to as COS2.

Let us find the sufficient conditions for stability of Scheme (\ref{SA30}).
It is assumed that the following inequalities are valid in a suitable norm 
\begin{equation}
\left\| \frac{\partial \mathbf{A}_{i}^{n}}{\partial \mathbf{v}_{i}^{n}}\cdot
\delta \mathbf{v}_{i}^{n}\right\| \leq \beta _{A}\left\| \delta \mathbf{v}%
_{i}^{n}\right\| ,\quad \alpha _{A},\beta _{A}=const.  \label{SS10}
\end{equation}
In view of (\ref{C80}), the variational scheme corresponding to (\ref{SA30})
is the following 
\begin{equation*}
\delta \mathbf{v}_{i+0.5}^{n+0.25}=0.5\left( \delta \mathbf{v}%
_{i}^{n}+\delta \mathbf{v}_{i+1}^{n}\right) +\frac{\varkappa }{8}\left[
\left( \mathbf{v}_{i}^{n}-\mathbf{v}_{i+1}^{n}\right) ^{T}\cdot \delta 
\mathbb{D}_{i}^{n}\right] ^{T}+\frac{\varkappa }{8}\mathbb{D}_{i}^{n}\cdot
\left( \delta \mathbf{v}_{i}^{n}-\delta \mathbf{v}_{i+1}^{n}\right) 
\end{equation*}
\begin{equation*}
+\frac{\left( \Delta t\right) ^{2}}{8\Delta x^{2}}\left\{ \left[ \delta
\left( \left( \mathbf{A}_{i+1}^{n}\right) ^{2}\cdot \mathbb{B}_{i}\right) %
\right] \cdot \left( \mathbf{v}_{i+1}^{n}-\mathbf{v}_{i}^{n}\right) -\left[
\delta \left( \left( \mathbf{A}_{i}^{n}\right) ^{2}\cdot \mathbb{A}%
_{i}\right) \right] \cdot \left( \mathbf{v}_{i+1}^{n}-\mathbf{v}%
_{i}^{n}\right) \right\} +
\end{equation*}
\begin{equation*}
\frac{\left( \Delta t\right) ^{2}}{8\Delta x^{2}}\left[ \left( \mathbf{A}%
_{i+1}^{n}\right) ^{2}\cdot \mathbb{B}_{i}-\left( \mathbf{A}_{i}^{n}\right)
^{2}\cdot \mathbb{A}_{i}\right] \cdot \left( \delta \mathbf{v}%
_{i+1}^{n}-\delta \mathbf{v}_{i}^{n}\right) +
\end{equation*}
\begin{equation}
\frac{\Delta t}{2\Delta x}\left( \mathbf{A}_{i}^{n}\cdot \delta \mathbf{v}%
_{i}^{n}-\mathbf{A}_{i+1}^{n}\cdot \delta \mathbf{v}_{i+1}^{n}\right) ,
\label{SS20}
\end{equation}
In view of (\ref{IN80}), (\ref{DD30}), (\ref{DD40}), and (\ref{SS10}),
Scheme (\ref{SS20}) will be stable if the following scheme will be stable. 
\begin{equation*}
\delta \mathbf{v}_{i+0.5}^{n+0.25}=0.5\left( \delta \mathbf{v}%
_{i}^{n}+\delta \mathbf{v}_{i+1}^{n}\right) +\frac{\varkappa }{8}\mathbb{D}%
_{i}^{n}\cdot \left( \delta \mathbf{v}_{i}^{n}-\delta \mathbf{v}%
_{i+1}^{n}\right) +
\end{equation*}
\begin{equation*}
\frac{\left( \Delta t\right) ^{2}}{8\Delta x^{2}}\left[ \left( \mathbf{A}%
_{i+1}^{n}\right) ^{2}\cdot \mathbb{B}_{i}-\left( \mathbf{A}_{i}^{n}\right)
^{2}\cdot \mathbb{A}_{i}\right] \cdot \left( \delta \mathbf{v}%
_{i+1}^{n}-\delta \mathbf{v}_{i}^{n}\right) +
\end{equation*}
\begin{equation}
\frac{\Delta t}{2\Delta x}\left( \mathbf{A}_{i}^{n}\cdot \delta \mathbf{v}%
_{i}^{n}-\mathbf{A}_{i+1}^{n}\cdot \delta \mathbf{v}_{i+1}^{n}\right) .
\label{SS30}
\end{equation}
We rewrite (\ref{SS30}) to read 
\begin{equation}
\delta \mathbf{v}_{i+0.5}^{n+0.25}=0.5\left( \mathbf{I+E}_{i}^{n}\right)
\cdot \delta \mathbf{v}_{i}^{n}+0.5\left( \mathbf{I-E}_{i+1}^{n}\right)
\cdot \delta \mathbf{v}_{i+1}^{n},  \label{SS40}
\end{equation}
where 
\begin{equation}
\mathbf{E}_{i}^{n}=\frac{\varkappa }{4}\mathbb{D}_{i}^{n}-\frac{\left(
\Delta t\right) ^{2}}{4\left( \Delta x\right) ^{2}}\left[ \left( \mathbf{A}%
_{i+1}^{n}\right) ^{2}\cdot \mathbb{B}_{i}-\left( \mathbf{A}_{i}^{n}\right)
^{2}\cdot \mathbb{A}_{i}\right] +\frac{\Delta t}{\Delta x}\mathbf{A}_{i}^{n},
\label{SS50}
\end{equation}
\begin{equation}
\mathbf{E}_{i+1}^{n}=\frac{\varkappa }{4}\mathbb{D}_{i}^{n}-\frac{\left(
\Delta t\right) ^{2}}{4\left( \Delta x\right) ^{2}}\left[ \left( \mathbf{A}%
_{i+1}^{n}\right) ^{2}\cdot \mathbb{B}_{i}-\left( \mathbf{A}_{i}^{n}\right)
^{2}\cdot \mathbb{A}_{i}\right] +\frac{\Delta t}{\Delta x}\mathbf{A}%
_{i+1}^{n}.  \label{SS60}
\end{equation}
We write, in view of (\ref{IN80}) and (\ref{CA203}), that $s\left( \mathbf{E}%
_{i}^{n}\right) \subset \left[ -\lambda _{E},\lambda _{E}\right] $, where 
\begin{equation}
\lambda _{E}=\varkappa \aleph -\left( \frac{\Delta t\lambda _{\max }}{\Delta
x}\right) ^{2}\aleph +\frac{\Delta t}{\Delta x}\lambda _{\max },\quad
\forall i,n.  \label{SS70}
\end{equation}
Hence, by virtue of Theorem \ref{StabilitySufficient} we find that the
scheme (\ref{SS40}) will be stable if 
\begin{equation}
\left( \varkappa -C_{r}^{2}\right) \aleph +C_{r}\leq 1,\quad C_{r}=\frac{%
\Delta t\lambda _{\max }}{\Delta x}.  \label{SS80}
\end{equation}
Thus, in view of Theorem \ref{Linear-nonlinear stability}, the scheme (\ref
{SA30}) will be stable if (\ref{SS80}) will be valid.

By analogy with the COS1 scheme, we find the necessary conditions for the
variational S-monotonicity (see Definition \ref{Variational monotonicity})
of the COS2 scheme, (\ref{SA30})-(\ref{SA45}), assuming that $\mathbf{v}%
_{i}^{n}$ $=$ $\mathbf{C}$ $=$ $const$. The variational scheme corresponding
to (\ref{SA30}) under (\ref{C80}), (\ref{DD10}) is the following 
\begin{equation*}
\delta \mathbf{v}_{i+0.5}^{n+0.25}=0.5\left( \delta \mathbf{v}%
_{i}^{n}+\delta \mathbf{v}_{i+1}^{n}\right) +\frac{\varkappa }{8}\mathbb{D}%
\cdot \left( \delta \mathbf{v}_{i}^{n}-\delta \mathbf{v}_{i+1}^{n}\right) -
\end{equation*}
\begin{equation}
\frac{\left( \Delta t\right) ^{2}}{8\left( \Delta x\right) ^{2}}\left( 
\mathbf{A}_{c}\right) ^{2}\cdot \mathbb{D}\cdot \left( \delta \mathbf{v}%
_{i}^{n}-\delta \mathbf{v}_{i+1}^{n}\right) +\frac{\Delta t}{2\Delta x}%
\mathbf{A}_{c}\cdot \left( \delta \mathbf{v}_{i}^{n}-\delta \mathbf{v}%
_{i+1}^{n}\right) ,  \label{SA100}
\end{equation}
where $\mathbf{A}_{c}$ $=$ $\mathbf{A}\left( \mathbf{C}\right) $, $\mathbf{A}%
\left( \mathbf{u}\right) $ $=$ $\partial \mathbf{f}\left( \mathbf{u}\right)
\diagup \partial \mathbf{u}$, $\mathbb{D}=\zeta \mathbf{I}$. In view of
Theorem \ref{S-monotoneIFF}, the scheme (\ref{SA100}) will be S-monotone 
\emph{iff} 
\begin{equation*}
\underset{\left| \zeta \right| \leq 4\aleph ,\ \lambda \in s\left( \mathbf{A}%
_{c}\right) }{\max }\left( \left| {\frac{1}{2}+\frac{\varkappa \zeta }{8}-%
\frac{\left( \Delta t\right) ^{2}}{8\left( \Delta x\right) ^{2}}}\zeta
\lambda ^{2}{+\frac{\Delta t}{2\Delta x}\lambda }\right| \right. +
\end{equation*}
\begin{equation}
\left. \left| {\frac{1}{2}-\frac{\varkappa \zeta }{8}+{\frac{\left( \Delta
t\right) ^{2}}{8\left( \Delta x\right) ^{2}}}\zeta \lambda ^{2}-\frac{\Delta
t}{2\Delta x}\lambda }\right| \right) \leqslant 1.  \label{SA110}
\end{equation}
By virtue of (\ref{SA110}) we find that (\ref{SS80}) will be the necessary
condition for the variational S-monotonicity (see Definition \ref
{Variational monotonicity}) of the non-linear scheme (\ref{SA30}).

The variational scheme corresponding to (\ref{SA45}) (under $\mathbf{v}%
_{i}^{n}$ $=$ $\mathbf{C}$ $=$ $const$) reads 
\begin{equation*}
\delta \mathbf{v}_{i}^{n+0.75}=\delta \mathbf{v}_{i}^{n+1}-\frac{\gamma }{2}%
\mathbf{G}_{c}\cdot \delta \mathbf{v}_{i}^{n+1},\quad \gamma \equiv \frac{%
\Delta t}{\tau },
\end{equation*}
\begin{equation}
\delta \mathbf{v}_{i}^{n+1}=\delta \mathbf{v}_{i}^{n+0.5}+\gamma \mathbf{G}%
_{c}\cdot \delta \mathbf{v}_{i}^{n+0.75},\quad \mathbf{G}_{c}=\mathbf{G}%
\left( C\right) ,  \label{SA130}
\end{equation}
where $\mathbf{G}\left( \mathbf{u}\right) $ $=$ $\partial \mathbf{q}\left( 
\mathbf{u}\right) \diagup \partial \mathbf{u}$. Let $\mathbf{G}$ will be a
normal matrix. In such a case, since all eigenvalues of $\mathbf{G}_{c}$
have non-positive real parts, see (\ref{INA20}), the first step of the
scheme (\ref{SA45}) will be S-monotone (see \cite[Theorem 3.3]{Borisov and
Sorek 2004}, Theorem \ref{Monotonicity due to variation}). It remains to
prove that the scheme (\ref{SA130}), taken as a whole, will be S-monotone
under the same condition. Eliminating $\delta \mathbf{v}_{i}^{n+0.75}$ we
obtain that 
\begin{equation}
\delta \mathbf{v}_{i}^{n+1}=\left[ \mathbf{I}-\gamma \mathbf{G}_{c}\cdot
\left( \mathbf{I}-\frac{\gamma }{2}\mathbf{G}_{c}\right) \right] ^{-1}\cdot
\delta \mathbf{v}_{i}^{n+0.5}.  \label{SA140}
\end{equation}
In view of \cite[Theorem 3.3]{Borisov and Sorek 2004}, the scheme (\ref
{SA140}) will be S-monotone if 
\begin{equation}
\underset{k}{\max }\left| \frac{1}{1-\gamma \xi _{k}\left( \mathbf{C}\right)
\left( 1-\frac{\gamma }{2}\xi _{k}\left( \mathbf{C}\right) \right) }\right|
\leq 1,\quad \forall \mathbf{C\in }\Omega _{\mathbf{u}},  \label{SA150}
\end{equation}
where $\xi _{k}\left( \mathbf{C}\right) $ denotes the $k$-th eigenvalue of
the matrix $\mathbf{G}_{c}$. Since all eigenvalues of $\mathbf{G}_{c}$ have
non-positive real parts, the variational scheme (\ref{SA140}) will be
unconditionally S-monotone.

Thus, the COS2 scheme, (\ref{SA30})-(\ref{SA45}), will be variationally
S-monotone only if (\ref{SS80}) will be valid.

\subsection{Second order schemes based on operator splitting techniques\label%
{SOSOST}}

Using the first order schemes (\ref{CA210}) and (\ref{C55}), it can be
constructed (see Section \ref{Appendix1}, Proposition \ref{First to Second
order scheme}) a scheme approximating (\ref{INA10}) with the accuracy $%
O(\left( \Delta x\right) ^{2}+\left( \Delta t\right) ^{2})$. Actually, since 
\begin{equation}
\frac{\left( \Delta t\right) ^{2}}{32}\left( \frac{\partial ^{2}\mathbf{u}}{%
\partial t^{2}}\right) _{i+0.5}^{n+0.25}=\frac{\left( \Delta t\right) ^{2}}{%
32}\left( \frac{\partial ^{2}\mathbf{u}}{\partial t^{2}}\right)
_{i}^{n+0.25}+O\left( \Delta x\left( \Delta t\right) ^{2}\right) ,
\label{CA205}
\end{equation}
the scheme will be as follows 
\begin{equation}
\mathbf{v}_{i}^{n+0.25}=\mathbf{v}_{i}^{n}+\frac{\Delta t}{2\tau }\mathbf{q}%
\left( \mathbf{v}_{i}^{n+0.25}\right) ,  \label{CA215}
\end{equation}
\begin{equation*}
\mathbf{v}_{i+0.5}^{n+0.5}=0.5\left( \mathbf{v}_{i+1}^{n+0.25}+\mathbf{v}%
_{i}^{n+0.25}\right) -
\end{equation*}
\begin{equation}
\varkappa \frac{\Delta x}{8}\left( \mathbf{d}_{i+1}^{n+0.25}-\mathbf{d}%
_{i}^{n+0.25}\right) -\frac{\Delta t}{2}\frac{\mathbf{f}\left( \mathbf{v}%
_{i+1}^{n+0.25}\right) -\mathbf{f}\left( \mathbf{v}_{i}^{n+0.25}\right) }{%
\Delta x}.  \label{CA230}
\end{equation}
Hence, the stability as well as monotonicity behavior of (\ref{CA230}) and (%
\ref{CA215}) can be interesting itself. As demonstrated in Section \ref
{Exemplification and discussion}, the behavior of Scheme (\ref{CA230}) is
similar to the one of the NT scheme, i.e. it can produce spurious
oscillations if CFL number is not sufficiently small. Let us develop another
scheme approximating (\ref{INA10}) with the accuracy $O(\left( \Delta
x\right) ^{2}+\left( \Delta t\right) ^{2})$ and such that its components
(after operator splitting) will be of the second order. It can be done on
the basis of the second order scheme (\ref{CA215})-(\ref{CA230}) with ease.
Actually, adding to and subtracting from Equation (\ref{FS15}) (see Section 
\ref{Appendix1}, Proposition \ref{First to Second order scheme}) the same
quantity (\ref{CA205}), we obtain (after operator splitting) the following
scheme, instead of (\ref{CA215})-(\ref{CA230}), 
\begin{equation}
\mathbf{v}_{i}^{n+0.25}=\mathbf{v}_{i}^{n}+\frac{\Delta t}{2\tau }\mathbf{q}%
_{i}^{n+0.25}-\frac{\left( \Delta t\right) ^{2}}{32}\left( \frac{\partial
^{2}\mathbf{u}}{\partial t^{2}}\right) _{i}^{n+0.25},  \label{SOS10}
\end{equation}
\begin{equation*}
\mathbf{v}_{i+0.5}^{n+0.5}=0.5\left( \mathbf{v}_{i+1}^{n+0.25}+\mathbf{v}%
_{i}^{n+0.25}\right) -\varkappa \frac{\Delta x}{8}\left( \mathbf{d}%
_{i+1}^{n+0.25}-\mathbf{d}_{i}^{n+0.25}\right) +
\end{equation*}
\begin{equation}
\frac{\left( \Delta t\right) ^{2}}{32}\left( \frac{\partial ^{2}\mathbf{u}}{%
\partial t^{2}}\right) _{i+0.5}^{n+0.25}-\frac{\Delta t}{2}\frac{\mathbf{f}%
\left( \mathbf{v}_{i+1}^{n+0.25}\right) -\mathbf{f}\left( \mathbf{v}%
_{i}^{n+0.25}\right) }{\Delta x}.  \label{SOS20}
\end{equation}
Thus, Scheme (\ref{SOS10}) as well as Scheme (\ref{SOS20}) are of the second
order, and Scheme (\ref{SOS10})-(\ref{SOS20}), taken as a whole, is of the
second order as well.

Using Taylor series expansion, and central differencing, we find 
\begin{equation}
\mathbf{v}_{i}^{n+0.125}=\mathbf{v}_{i}^{n+0.25}-\frac{\Delta t}{8}\left( 
\frac{\partial \mathbf{u}}{\partial t}\right) _{i}^{n+0.25}+\frac{1}{2}%
\left( \frac{\Delta t}{8}\right) ^{2}\left( \frac{\partial ^{2}\mathbf{u}}{%
\partial t^{2}}\right) _{i}^{n+0.25},  \label{SOS25}
\end{equation}
\begin{equation}
\mathbf{v}_{i}^{n+0.25}=\mathbf{v}_{i}^{n}+\frac{\Delta t}{4}\left( \frac{%
\partial \mathbf{u}}{\partial t}\right) _{i}^{n+0.125}+O\left( \left( \Delta
t\right) ^{3}\right) .  \label{SOS27}
\end{equation}
Considering the equation in (\ref{C10}) at $t_{n+0.25}<t\leq t_{n+0.5}$, we
obtain, in view of (\ref{SA20}), that 
\begin{equation}
\frac{\partial ^{2}\mathbf{u}}{\partial t^{2}}=-2\frac{\partial }{\partial t}%
\left( \frac{\partial \mathbf{f}}{\partial x}\right) =-2\frac{\partial }{%
\partial x}\left( \frac{\partial \mathbf{f}}{\partial t}\right) =4\frac{%
\partial }{\partial x}\left( \mathbf{A}^{2}\cdot \frac{\partial \mathbf{u}}{%
\partial x}\right) ,  \label{SOS40}
\end{equation}
where $\mathbf{A=}\partial \mathbf{f\diagup }\partial \mathbf{u}$. Then 
\begin{equation*}
\left[ \frac{\partial }{\partial x}\left( \mathbf{A}^{2}\cdot \frac{\partial 
\mathbf{u}}{\partial x}\right) \right] _{i+0.5}^{n+0.25}=
\end{equation*}
\begin{equation}
\frac{1}{\Delta x}\left[ \left( \mathbf{A}_{i+1}^{n+0.25}\right) ^{2}\cdot 
\mathbf{d}_{i+1}^{n}-\left( \mathbf{A}_{i}^{n+0.25}\right) ^{2}\cdot \mathbf{%
d}_{i}^{n}\right] +O\left( \left( \Delta x\right) ^{2}\right) .
\label{SOS50}
\end{equation}
By virtue of (\ref{C20}), (\ref{SOS25})-(\ref{SOS50}), we rewrite Scheme (%
\ref{SOS10})-(\ref{SOS20}) to read 
\begin{equation*}
\mathbf{v}_{i}^{n+0.125}=\mathbf{v}_{i}^{n+0.25}-\frac{\Delta t}{8\tau }%
\left( \mathbf{q}_{i}^{n+0.125}+\mathbf{q}_{i}^{n+0.25}\right) ,
\end{equation*}
\begin{equation}
\mathbf{v}_{i}^{n+0.25}=\mathbf{v}_{i}^{n}+\frac{\Delta t}{2\tau }\mathbf{q}%
_{i}^{n+0.125},  \label{SOS60}
\end{equation}
\begin{equation*}
\mathbf{v}_{i+0.5}^{n+0.5}=0.5\left( \mathbf{v}_{i+1}^{n+0.25}+\mathbf{v}%
_{i}^{n+0.25}\right) -\varkappa \frac{\Delta x}{8}\left( \mathbf{d}%
_{i+1}^{n+0.25}-\mathbf{d}_{i}^{n+0.25}\right) +
\end{equation*}
\begin{equation*}
\frac{\left( \Delta t\right) ^{2}}{8\Delta x}\left[ \left( \mathbf{A}%
_{i+1}^{n+0.25}\right) ^{2}\cdot \mathbf{d}_{i+1}^{n+0.25}-\left( \mathbf{A}%
_{i}^{n+0.25}\right) ^{2}\cdot \mathbf{d}_{i}^{n+0.25}\right]
\end{equation*}
\begin{equation}
-\frac{\Delta t}{2}\frac{\mathbf{f}\left( \mathbf{v}_{i+1}^{n+0.25}\right) -%
\mathbf{f}\left( \mathbf{v}_{i}^{n+0.25}\right) }{\Delta x}.  \label{SOS70}
\end{equation}
Notice, Scheme (\ref{SOS70}) coincides, in fact, with Scheme (\ref{SA30}),
however there is a difference between (\ref{SOS60}) and (\ref{SA45}). In
spite of the fact that both of these implicit Runge-Kutta schemes, (\ref
{SOS60}) and (\ref{SA45}), are of the second order, only the above
combination, i.e. Scheme (\ref{SOS60})-(\ref{SOS70}), is of the second order
as a whole. Thus, even if a higher order implicit Runge-Kutta scheme will be
used to approximate (\ref{C20}), nevertheless, there is a risk that Equation
(\ref{INA10}) will be approximated with the first order in time.

It can be proven, by analogy with the proof in Section \ref{Second-order
central schemes}, that (\ref{SS80}) is the sufficient condition for
stability as well as the necessary condition for S-monotonicity of Scheme (%
\ref{SOS60})-(\ref{SOS70}).

There exists one more problem associated with the stiffness, i.e. $\tau \ll 1
$, of (\ref{C20}). To demonstrate it, let us consider the following system
of ordinary differential equations (ODEs) with a relaxation operator in the
right-hand side: 
\begin{equation}
\frac{dX\left( t\right) }{dt}=-\frac{\varphi \left( X,Y\right) }{\tau }%
\left( X-Y\right) ,\quad X\left( 0\right) =X_{0},  \label{SOS80}
\end{equation}
\begin{equation}
\frac{dY\left( t\right) }{dt}=-\frac{\varphi \left( X,Y\right) }{\tau }%
\left( Y-X\right) ,\quad Y\left( 0\right) =Y_{0}.  \label{SOS90}
\end{equation}
The system will be considered at the interval $0<t\leq \Delta t$. The
following notation will be used: $X_{\nu }=X\left( \nu \Delta t\right) $. If 
$\varphi \left( X,Y\right) \equiv 1$, then the approximation, $X_{1}^{\ast }$%
, to $X_{1}$, in the case when the first order scheme (\ref{C55}) is used,
will be 
\begin{equation}
X_{1}^{\ast }=Z_{0}+\frac{X_{0}-Z_{0}}{1+\lambda },  \label{SOS100}
\end{equation}
where $Z_{0}=0.5(X_{0}+Y_{0})$, $\lambda =2\Delta t\diagup \tau $. In the
case of the second order scheme, (\ref{SA45}), the approximation to $X_{1}$
will be 
\begin{equation}
X_{1}^{\ast \ast }=Z_{0}+\frac{X_{0}-Z_{0}}{1+\lambda +0.5\lambda ^{2}}.
\label{SOS110}
\end{equation}
The analytic solution, $X_{1}$, will be the following 
\begin{equation}
X_{1}=Z_{0}+\frac{X_{0}-Z_{0}}{\exp \left( \lambda \right) }.  \label{SOS120}
\end{equation}

Notice, in the case of underresolved numerical scheme we obtain that $%
\lambda \gg 1$. A comparison between (\ref{SOS120}) and (\ref{SOS110}) shows
that there is a need to develop an implicit Runge-Kutta scheme of higher
order accuracy, and such that the scheme (based on the operator splitting
techniques), taken as a whole, will be, at least, of the second order.
However, the higher order will be the implicit Runge-Kutta scheme, the more
time-consuming procedure will be obtained. To tide over this problem, a
semi-analytic approach could be useful. Let us demonstrate one, as applied
to System (\ref{SOS80})-(\ref{SOS90}). Let $\theta _{\nu }\equiv \varphi
\left( X_{\nu },Y_{\nu }\right) $. Integrating (\ref{SOS80})-(\ref{SOS90}),
and using the midpoint rule, we obtain 
\begin{equation}
X_{1}-X_{0}=-\frac{\theta _{0.5}}{\tau }\int\limits_{0}^{\Delta t}\left(
X-Y\right) dt+O\left( \left( \Delta t\right) ^{2}\right) ,  \label{SOS140}
\end{equation}
\begin{equation}
Y_{1}-Y_{0}=-\frac{\theta _{0.5}}{\tau }\int\limits_{0}^{\Delta t}\left(
Y-X\right) dt+O\left( \left( \Delta t\right) ^{2}\right) .  \label{SOS150}
\end{equation}
To resolve (\ref{SOS140})-(\ref{SOS150}), a linearized version of System (%
\ref{SOS80})-(\ref{SOS90}) can be used, namely, instead of $\varphi \left(
X,Y\right) $ in (\ref{SOS80})-(\ref{SOS90}), it can be taken $\theta
_{0.5}=const$. The analytic solution will be the following 
\begin{equation}
X_{1}=Z_{0}+\frac{X_{0}-Z_{0}}{\exp \left( 2\theta _{0.5}\Delta t\diagup
\tau \right) },\quad Y_{1}=Z_{0}+\frac{Y_{0}-Z_{0}}{\exp \left( 2\theta
_{0.5}\Delta t\diagup \tau \right) }.  \label{SOS160}
\end{equation}
It remains to estimate the value of $\theta _{0.5}$. Since $\theta _{0.5}$ $=
$ $0.5(\varphi \left( X_{0},Y_{0}\right) +\varphi \left( X_{1},Y_{1}\right) )
$ $+$ $O(\left( \Delta t\right) ^{2})$, we obtain the following, in general,
non-linear equation in $\theta _{0.5}$%
\begin{equation}
\theta _{0.5}=0.5(\varphi \left( X_{0},Y_{0}\right) +\varphi \left(
X_{1},Y_{1}\right) ),  \label{SOS170}
\end{equation}
where $X_{1}$ and $Y_{1}$ are defined via (\ref{SOS160}).

\section{Exemplification and discussion\label{Exemplification and discussion}%
}

In this section we are mainly concerned with verification of the second
order central scheme, (\ref{SA30}). To demonstrate the superiority of this
scheme over Scheme (\ref{CA210}) being $O(\left( \Delta x\right) ^{2}+\Delta
t)$ accurate, we will use the inviscid Burgers equation. The COS2 scheme, (%
\ref{SA30})-(\ref{SA45}), is verified using Pember's rarefaction test
problem \cite{Pember 1993a}. Euler equations of gas dynamics, namely Sod's
as well as Lax problems, are also used for the verification of the second
order central scheme, (\ref{SA30}).

\subsection{Scalar non-linear equation}

As the first stage in the verification, we will focus on the following
scalar version of the problem (\ref{INA10}): 
\begin{equation}
\frac{\partial u}{\partial t}+\frac{\partial }{\partial x}f\left( u\right)
=0,\quad x\in \mathbb{R},\ 0<t\leq T_{\max };\quad \left. u\left( x,t\right)
\right| _{t=0}=u^{0}\left( x\right) .  \label{VS10}
\end{equation}
To test the first order scheme, (\ref{CA210}), we will solve the inviscid
Burgers equation (i.e. $f\left( u\right) \equiv u^{2}\diagup 2$) with the
following initial condition 
\begin{equation}
u\left( x,0\right) =\left\{ 
\begin{array}{cc}
u_{0}, & x\in \left( h_{L},h_{R}\right)  \\ 
0, & x\notin \left( h_{L},h_{R}\right) 
\end{array}
\right. ,\quad h_{R}>h_{L},\ u_{0}=const\neq 0.  \label{VS70}
\end{equation}
The exact solution to (\ref{VS10}), (\ref{VS70}) is given by 
\begin{equation}
u\left( x,t\right) =\left\{ 
\begin{array}{cc}
u_{1}\left( x,t\right) , & 0<t\leq T \\ 
u_{2}\left( x,t\right) , & t>T
\end{array}
\right. ,  \label{VS80}
\end{equation}
where $T=2S\diagup u_{0}$, $S=h_{R}-h_{L}$, 
\begin{equation}
u_{1}\left( x,t\right) =\left\{ 
\begin{array}{cc}
\frac{x-h_{L}}{b-h_{L}}u_{0}, & h_{L}<x\leq b,\ b=u_{0}t+h_{L} \\ 
u_{0}, & b<x\leq 0.5u_{0}t+h_{R} \\ 
0, & x\leq h_{L}\ or\ x>0.5u_{0}t+h_{R}
\end{array}
\right. ,  \label{VS90}
\end{equation}
\begin{equation}
u_{2}\left( x,t\right) =\left\{ 
\begin{array}{cc}
\frac{2S\left( x-h_{L}\right) }{\left( L-h_{L}\right) ^{2}}u_{0}, & 
h_{L}<x\leq L \\ 
0, & x\leq h_{L}\ or\ x>L
\end{array}
\right. ,  \label{VS100}
\end{equation}
\begin{equation}
L=2\sqrt{S^{2}+0.5u_{0}S\left( t-T\right) }+h_{L}.  \label{VS110}
\end{equation}

\begin{figure}[h]
\centerline{\includegraphics[width=11.50cm,height=3.8cm]{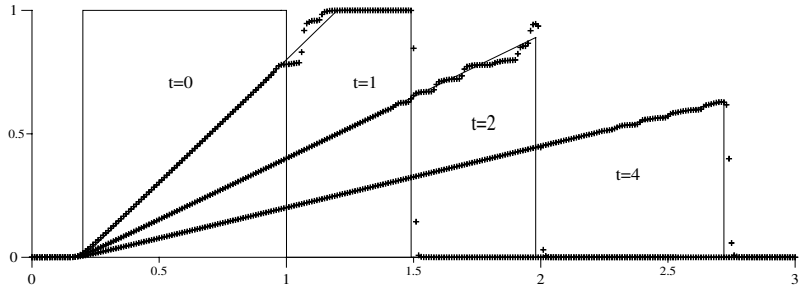}}
\caption{Inviscid Burgers equation. The scheme COS1 ($\varkappa =1$) versus
the analytical solution. Crosses: numerical solution; Solid line: analytical
solution and initial data. $C_{r}=\aleph =0.5$, $\Delta x=0.01$. }
\label{K2COS01}
\end{figure}

The numerical solutions were computed on a uniform grid with spatial
increments of $\Delta x=0.01$, the velocity $u_{0}=1$ in (\ref{VS70}), $%
h_{L}=0.2$, $h_{R}=1$, the monotonicity parameter $\aleph =0.5$, the Courant
number $Cr$ $\equiv $ $u_{0}\Delta t\diagup \Delta x=$ $0.5$, and the
parameter $\varkappa =1$ in (\ref{CA210}). The results of simulation are
depicted with the exact solution in Figure \ref{K2COS01}. 
We note (Figure \ref{K2COS01}) that the first order scheme, (\ref{CA210}),
exhibits a typical second-order nature, however spurious solutions are
produced by the scheme. To obtain necessary conditions for the variational
GOS-monotonicity (see Definition \ref{Variational monotonicity}) of the COS1
scheme,\ (\ref{CA210}), we consider the scheme in variations (\ref{DD20}),
corresponding to (\ref{CA210}), on a constant ($v_{i}^{n}$ $=$ $C$ $=$ $const
$, $\forall i,n$), i.e. (\ref{CAB240}). By virtue of Theorem \ref
{S-monotoneIFF} and Theorem \ref{Necessary conditions for GOS-monotonicity},
we find that, in the case of the Burgers equation, Scheme (\ref{CAB240})
will be GOS-monotone only if (\ref{DD90}) will be fulfilled. Since all of
the coefficients of (\ref{CAB240}) will be non-negative (the scalar version
is considered) under (\ref{DD90}), the scheme in question, (\ref{CA210}),
will be H-monotone (and hence TVD and G-monotone) only if (\ref{DD90}) will
be valid. Notice, the numerical simulations were performed with such values
of the parameter $\varkappa $, Courant number, $Cr$, and monotonicity
parameter, $\aleph $, that (\ref{DD90}) was not violated. As it can be seen
in Figure \ref{K2COS01}, the boundary maximum principle is not violated by
the scheme, i.e., the maximum positive values of the dependent variable, $v$%
, occur at the boundary $t=0$. It is interesting that the spurious solution
(see Figure \ref{K2COS01}) produced by the scheme COS1 has the monotonicity
property \cite{Harten 1983}, since no new local extrema in $x$ are created
as well as the value of a local minimum is non-decreasing and the value of a
local maximum is non-increasing.

Let us note that the problem of building free-of-spurious-oscillations
schemes is, in general, unsettled up to the present. Even the best modern
high-resolution schemes can produce spurious oscillations, and these
oscillations are often of ENO type (see, e.g., \cite{Pareschi et al. 2005}
and references therein). We found that the oscillations produced by the COS1
scheme, (\ref{CA210}), are of ENO type, namely their amplitude decreases
rapidly with decreasing the time-increment $\Delta t$, and the oscillations
virtually disappear under a relatively low Courant number, $Cr$ $\leq $ $%
0.15 $. However, the reduction of the Courant number causes some smearing of
the solution. The spurious oscillations (see Figure \ref{K2COS01}) can be
eradicated without reduction Courant number, but decreasing the parameter $%
\varkappa $. Particularly, the spurious oscillations disappear if $\varkappa
=2\diagup 3$, $C_{r}=0.5$, however, this introduces more numerical smearing
than in the case of the Courant number reduction. Satisfactory results are
obtained under $\varkappa =0.82$ ($C_{r}=0.5$). The results of simulations
are not depicted here.

To gain insight to why the scheme COS1, (\ref{CA210}), can exhibit spurious
solutions, let us consider the, so called, \emph{first differential
approximation} of this scheme (\cite[p. 45]{Ganzha and Vorozhtsov 1996}, 
\cite[p. 376]{Samarskiy and Gulin 1973}; see also `modified equations' in 
\cite[p. 45]{Ganzha and Vorozhtsov 1996}, \cite{LeVeque 2002}, \cite{Morton
1996}). As reported in \cite{Ganzha and Vorozhtsov 1996}, \cite{Samarskiy
and Gulin 1973}, this heuristic method was originally presented by Hirt
(1968) (see \cite[p. 45]{Ganzha and Vorozhtsov 1996}) as well as by Shokin
and Yanenko (1968) (see \cite[p. 376]{Samarskiy and Gulin 1973}), and has
since been widely employed in the development of stable difference schemes
for PDEs.

We found that the local truncation error, $\psi $, for the scheme COS1 can
be written in the following form 
\begin{equation*}
\psi =\frac{\left( 1-\varkappa \right) \left( \Delta x\right) ^{2}}{4\Delta t%
}\frac{\partial ^{2}u\left( x,t\right) }{\partial x^{2}}+\frac{\Delta t}{4}%
\frac{\partial ^{2}f\left( u\right) }{\partial t\partial x}+
\end{equation*}
\begin{equation}
O\left( \frac{\left( \Delta x\right) ^{4}}{\Delta t}+\left( \Delta t\right)
^{2}+\left( \Delta x\right) ^{2}\right) .  \label{VS190}
\end{equation}
By virtue of (\ref{VS190}), we find the first differential approximation of
the scheme COS1 
\begin{equation}
\frac{\partial u}{\partial t}+\frac{\partial f\left( u\right) }{\partial x}=%
\frac{\Delta t}{4}\frac{\partial }{\partial x}\left( B\frac{\partial u\left(
x,t\right) }{\partial x}\right) ,  \label{VS200}
\end{equation}
where $B=\left( 1-\varkappa \right) \left( \Delta x\diagup \Delta t\right)
^{2}-A^{2}$. The term in right-hand side of (\ref{VS200}) will be
dissipative if 
\begin{equation}
\left( 1-\varkappa \right) \left( \frac{\Delta x}{\Delta t}\right)
^{2}-A^{2}>0,\ \Longrightarrow \ C_{r}^{2}<1-\varkappa .  \label{VS210}
\end{equation}
Thus, the scheme COS1, (\ref{CA210}), is non-dissipative under $\varkappa =1$%
, and hence can produce spurious oscillations. Notice, if $\varkappa =0.82$,
then we obtain from (\ref{VS210}) that $C_{r}<0.42$. Nevertheless, as it is
reported above, satisfactory results can be obtained under $C_{r}=0.5$ as
well.

\begin{figure}[h]
\centerline{\includegraphics[width=11.50cm,height=3.8cm]{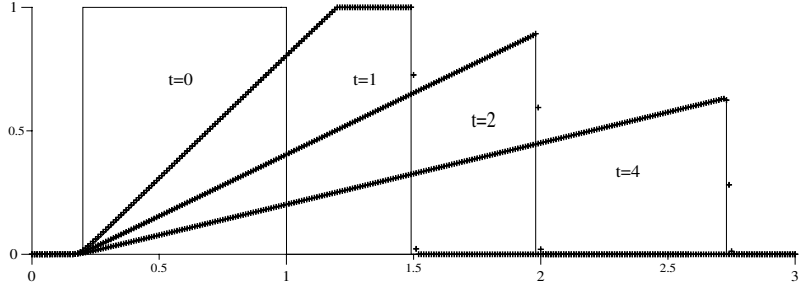}}
\caption{Inviscid Burgers equation. The scheme COS2 ($\varkappa =1$) versus
the analytical solution. Crosses: numerical solution; Solid line: analytical
solution and initial data. $C_{r}=1$, $\aleph =0.5$, $\Delta x=0.01$. }
\label{K2COS02}
\end{figure}

So then, the notion of first differential approximation has enabled us to
understand that the spurious solutions exhibited by the scheme COS1, (\ref
{CA210}), are mainly associated with the negative numerical viscosity
introduced to obtain the scheme of the second order in space, i.e. $O(\left(
\Delta x\right) ^{2}+\Delta t)$. Let us consider the scheme COS2, (\ref{SA30}%
), approximating (\ref{VS10}) with the accuracy $O(\left( \Delta x\right)
^{2}+\left( \Delta t\right) ^{2})$. Notice, the second order scheme COS2, (%
\ref{SA30}), is nothing more than the scheme COS1, (\ref{CA210}), with the
additional non-negative numerical viscosity. To test the scheme COS2, (\ref
{SA30}), the inviscid Burgers equation was solved under the initial
condition (\ref{VS70}). The numerical solutions were computed under the same
values of parameters as in the case of the scheme COS1, but $C_{r}=1$. The
results of simulation are depicted with the exact solution in Figure \ref
{K2COS02}.

We note (Figure \ref{K2COS02}) that the scheme COS2, (\ref{SA30}), exhibits
a typical second-order nature without any spurious oscillations. Increasing
the value of $\aleph $ (up to $\aleph =1$) leads to a minor improvement of
the numerical solutions, whereas decreasing the value of $Cr$ leads to a
mild smearing of the solutions. The results of simulations are not depicted
here.

\subsection{Hyperbolic conservation laws with relaxation}

Let us consider the model system of hyperbolic conservation laws with
relaxation developed in \cite{Pember 1993a}: 
\begin{equation}
\frac{\partial w}{\partial t}+\frac{\partial }{\partial x}\left( \frac{1}{2}%
u^{2}+aw\right) =0,  \label{VS240}
\end{equation}
\begin{equation}
\frac{\partial z}{\partial t}+\frac{\partial }{\partial x}az=\frac{1}{\tau }%
Q(w,z),  \label{VS250}
\end{equation}
where 
\begin{equation}
Q(w,z)=z-m(u-u_{0}),\quad u=w-q_{0}z,  \label{VS260}
\end{equation}
$\tau $ denotes the relaxation time of the system, $q_{0}$, $m$, $a$, and $%
u_{0}$ are constants. The Jacobian, $\mathbf{A}$, can be written in the form 
\begin{equation}
\mathbf{A=}\left\{ 
\begin{array}{cc}
w-q_{0}z+a & -q_{0}\left( w-q_{0}z\right) \\ 
0 & a
\end{array}
\right\} .  \label{VS270}
\end{equation}
The system (\ref{VS240})-(\ref{VS250}) has the following frozen \cite{Pember
1993a} characteristic speeds $\lambda _{1}$ $=$ $a$, $\lambda _{2}$ $=$ $u+a$%
. The equilibrium equation for (\ref{VS240})-(\ref{VS250}) is 
\begin{equation}
\frac{\partial w}{\partial t}+\frac{\partial }{\partial x}\left( \frac{1}{2}%
u_{\ast }^{2}+aw\right) =0,  \label{VS280}
\end{equation}
where 
\begin{equation}
u_{\ast }=w-q_{0}z_{\ast },\quad z_{\ast }=\frac{m}{1+mq_{0}}\left(
w-u_{0}\right) .  \label{VS290}
\end{equation}
The equilibrium characteristic speed $\lambda _{\ast }$ can be written in
the form 
\begin{equation}
\lambda _{\ast }\left( w\right) =\frac{u_{\ast }\left( w\right) }{1+mq_{0}}%
+a.  \label{VS300}
\end{equation}

\begin{figure}[tbph]
\centerline{\includegraphics[width=14.cm,height=6.cm]{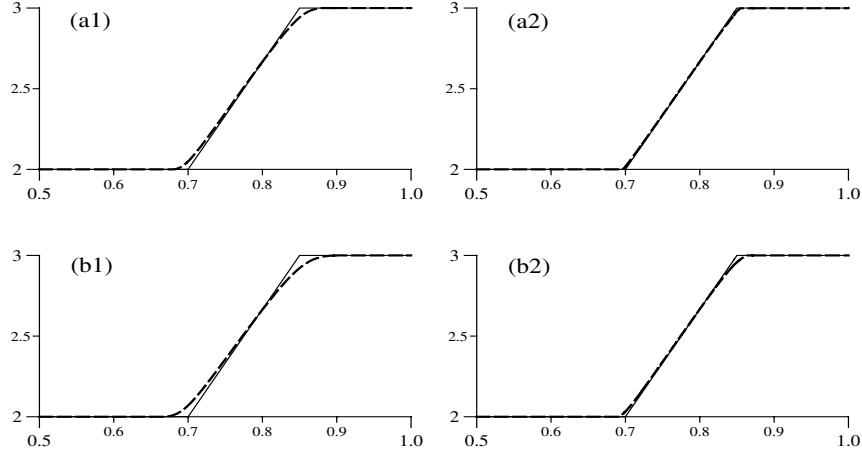}}
\caption{Pember's rarefaction test problem. The second-order scheme COS2 ($%
\varkappa =1$) versus the analytical solution for $u$. Dashed line:
numerical solution; Solid line: analytical solution. Time $t=0.3$, Courant
number $C_{r}=1$, monotonicity parameter $\aleph =1$. (a1): $\Delta
x=10^{-3} $, $a=1$; (a2): $\Delta x=2.5\times 10^{-4}$, $a=1$; (b1): $\Delta
x=10^{-3}$, $a=-1$; (b2): $\Delta x=2.5\times 10^{-4}$, $a=-1$.}
\label{RAREF1}
\end{figure}

Pember's rarefaction test problem is to find the solution $\left\{
w,z\right\} $ to (\ref{VS240})-(\ref{VS250}), and hence the function $%
u=u\left( x,t\right) $, under $\tau \rightarrow 0$, and where 
\begin{equation}
\left\{ w,z\right\} =\left\{ 
\begin{array}{cc}
\left\{ w_{L},z_{\ast }\left( w_{L}\right) \right\} , & x<x_{0} \\ 
\left\{ w_{R},z_{\ast }\left( w_{R}\right) \right\} , & x>x_{0}
\end{array}
\right. ,  \label{VS310}
\end{equation}
\begin{equation}
0<u_{L}=w_{L}-q_{0}z_{\ast }\left( w_{L}\right) <u_{R}=w_{R}-q_{0}z_{\ast
}\left( w_{R}\right) .  \label{VS320}
\end{equation}
The analytical solution of this problem can be found in \cite{Pember 1993a}.
The parameters of the model system are assumed as follows: $q_{0}=-1$, $m=-1$%
, $u_{0}=3$, $a=\pm 1$, $\tau =10^{-8}$. The initial conditions of the
rarefaction problem are defined by 
\begin{equation}
u_{L}=2,\ \Longrightarrow \ z_{L}=m\left( u_{L}-u_{0}\right) =1,\
w_{L}=u_{L}+q_{0}z_{L}=1,  \label{VS330}
\end{equation}
\begin{equation}
u_{R}=3,\ \Longrightarrow \ z_{R}=m\left( u_{R}-u_{0}\right) =0,\
w_{R}=u_{R}+q_{0}z_{R}=3.  \label{VS340}
\end{equation}
The position of the initial discontinuity, $x_{0}$, is set according to the
value of $a$ so that the solutions of all the rarefaction problems are
identical \cite{Pember 1993a}. Let a position, $x_{R}^{t}$, of leading edge
or a position, $x_{L}^{t}$, of trailing edge of the rarefaction be known
(e.g., $x_{R}^{t}=0.85$, $x_{L}^{t}=0.7$ in \cite{Pember 1993a}), then 
\begin{equation}
x_{0}=x_{R}^{t}-\left( \frac{u_{R}}{1+mq_{0}}+a\right) t=x_{L}^{t}-\left( 
\frac{u_{L}}{1+mq_{0}}+a\right) t.  \label{VS350}
\end{equation}
At $t=0.3$, under (\ref{VS330})-(\ref{VS340}) we have \cite{Pember 1993a} 
\begin{equation}
u=\left\{ 
\begin{array}{cc}
2, & x\leq 0.7 \\ 
2+\frac{x-0.7}{0.85-0.7}, & 0.7<x<0.85 \\ 
3, & x\geq 0.85
\end{array}
\right. .  \label{VS360}
\end{equation}
The results of simulations, based upon the scheme COS2, (\ref{SA30})-(\ref
{SA45}), under different values of the parameter $a$ ($a=1$, $a=-1$) and
different values of a grid spacing, $\Delta x$, are depicted in Figure \ref
{RAREF1}.

One can clearly see (Figure \ref{RAREF1}) that the scheme COS2 is free from
spurious oscillations. Let us also note that the results generated by the
scheme COS2 are less accurate in the case of negative value of $a$ than
those in the case of positive value of $a$. Specifically, in the numerical
solutions produced under $a=-1$, the representations of the trailing and
leading edges of the rarefaction are more smeared than those in the
solutions produced under $a=1$. Notice, under some negative value of $a$,
the frozen and the equilibrium characteristic speeds do not all have the
same sign.

\subsection{Euler equations of gas dynamics}

In this subsection we apply the second order scheme COS2, (\ref{SA30}), to
the Euler equations of gamma-law gas: 
\begin{equation}
\frac{\partial \mathbf{u}\left( x,t\right) }{\partial t}+\frac{\partial }{%
\partial x}\mathbf{F}\left( \mathbf{u}\right) =0,\quad x\in \mathbb{R},\
t>0;\quad \mathbf{u}\left( x,0\right) =\mathbf{u}^{0}\left( x\right) ,
\label{VE10}
\end{equation}
\begin{equation}
\mathbf{u\equiv }\left\{ u_{1},u_{2},u_{3}\right\} ^{T}=\left\{ \rho ,\rho
v,e\right\} ^{T},\quad \mathbf{F}\left( \mathbf{u}\right) =\left\{ \rho
v,\rho v^{2}+p,\left( e+p\right) v\right\} ^{T},  \label{VE20}
\end{equation}
\begin{equation}
e=\frac{p}{\gamma -1}+\frac{1}{2}\rho v^{2},\quad \gamma =const,
\label{VE30}
\end{equation}
where $\rho $, $v$, $p$, $e$ denote the density, velocity, pressure, and
total energy respectively. We consider the Riemann problem subject to
Riemann initial data 
\begin{equation}
\mathbf{u}^{0}\left( x\right) =\left\{ 
\begin{array}{cc}
\mathbf{u}_{L} & x<x_{0} \\ 
\mathbf{u}_{R} & x>x_{0}
\end{array}
\right. ,\quad \mathbf{u}_{L},\mathbf{u}_{R}=const.  \label{VE40}
\end{equation}
The analytic solution to the Riemann problem can be found in \cite[Sec. 14]
{LeVeque 2002}.

Aiming to suppress possible spurious oscillations, the scheme COS2, (\ref
{SA30}), is modified in such a way as to prevent a violation of the
necessary conditions (see Theorem \ref{Necessary conditions for
GOS-monotonicity}) for variational monotonicity of the scheme COS2, (\ref
{SA30}). The modification is to use $\aleph =\aleph (x,t)$ instead of $%
\aleph =const$. Leaving out the terms equal to zero under $\mathbf{v}_{i}^{n}
$ $=$ $\mathbf{C}$ $=$ $const$ in (\ref{SA30}), we write the reduced
variational scheme corresponding to the scheme COS2 in the following form 
\begin{equation}
\delta \mathbf{v}_{i+0.5}^{n+0.5}=\mathbf{C}_{i}^{n}\cdot \delta \mathbf{v}%
_{i}^{n}+\mathbf{D}_{i}^{n}\cdot \delta \mathbf{v}_{i+1}^{n},  \label{VE60}
\end{equation}
\begin{equation}
\delta \mathbf{v}_{i}^{n+1}=\mathbf{C}_{i-0.5}^{n+0.5}\cdot \delta \mathbf{v}%
_{i-0.5}^{n+0.5}+\mathbf{D}_{i-0.5}^{n+0.5}\cdot \delta \mathbf{v}%
_{i+0.5}^{n+0.5},  \label{VE70}
\end{equation}
where 
\begin{equation}
\mathbf{C}_{i}^{n}=\frac{1}{2}\left( \mathbf{I}+\frac{\Delta t}{\Delta x}%
\mathbf{A}_{i}^{n}+\mathbf{B}_{i}^{n}\right) ,\ \mathbf{D}_{i}^{n}=\frac{1}{2%
}\left( \mathbf{I}-\frac{\Delta t}{\Delta x}\mathbf{A}_{i+1}^{n}-\mathbf{B}%
_{i}^{n}\right) ,  \label{VE80}
\end{equation}
\begin{equation}
\mathbf{B}_{i}^{n}=\frac{1}{4}\left( \mathbb{B}_{i}^{n}-\mathbb{A}%
_{i}^{n}\right) -\frac{\Delta t^{2}}{4\Delta x^{2}}\left[ \left( \mathbf{A}%
_{i+1}^{n}\right) ^{2}\cdot \mathbb{B}_{i}^{n}-\left( \mathbf{A}%
_{i}^{n}\right) ^{2}\cdot \mathbb{A}_{i}^{n}\right] ,  \label{VE90}
\end{equation}
\begin{equation}
\mathbf{C}_{i-0.5}^{n+0.5}=\frac{1}{2}\left( \mathbf{I}+\frac{\Delta t}{%
\Delta x}\mathbf{A}_{i-0.5}^{n+0.5}+\mathbf{B}_{i-0.5}^{n+0.5}\right) ,
\label{VE100}
\end{equation}
\begin{equation}
\mathbf{D}_{i-0.5}^{n+0.5}=\frac{1}{2}\left( \mathbf{I}-\frac{\Delta t}{%
\Delta x}\mathbf{A}_{i+0.5}^{n+0.5}-\mathbf{B}_{i-0.5}^{n+0.5}\right) ,
\label{VE110}
\end{equation}
\begin{equation*}
\mathbf{B}_{i-0.5}^{n+0.5}=\frac{1}{4}\left( \mathbb{B}_{i-0.5}^{n+0.5}-%
\mathbb{A}_{i-0.5}^{n+0.5}\right) -
\end{equation*}
\begin{equation}
\frac{\Delta t^{2}}{4\Delta x^{2}}\left[ \left( \mathbf{A}%
_{i+0.5}^{n+0.5}\right) ^{2}\cdot \mathbb{B}_{i-0.5}^{n+0.5}-\left( \mathbf{A%
}_{i-0.5}^{n+0.5}\right) ^{2}\cdot \mathbb{A}_{i-0.5}^{n+0.5}\right] .
\label{VE120}
\end{equation}
Notice, if the diagonal elements of the matrices $\mathbf{C}_{i}^{n}$ and $%
\mathbf{D}_{i}^{n}$ are non-negative, i.e. $C_{i}^{n,kk}\geq 0$ and $%
D_{i}^{n,kk}\geq 0$, then any matrix column of the scheme (\ref{VE60})
consists of zeros, but two entries. In such a case this column is a $\mu $%
-function. Hence, at the first stage of modification, it is sufficient to
verify $C_{i}^{n,kk}$ and $D_{i}^{n,kk}$. If $C_{i}^{n,kk}<0$ or $%
D_{i}^{n,kk}<0$ for, at least, one value of $k$, then the value of $\aleph $
at the node $i$ (and, if it is necessary, at the neighbor nodes) is reduced
up to $\aleph _{\min }\geq 0$. Then, the modified values of $\aleph $, i.e. $%
\aleph _{i}^{n}$, are used in the scheme COS2, (\ref{SA30}). After that we
turn to the second stage of modifications. Eliminating $\delta \mathbf{v}%
_{i+0.5}^{n+0.5}$ from (\ref{VE60})-(\ref{VE70}) we convert the variational
scheme based on staggered spatial grid into the following non-staggered form 
\begin{equation}
\delta \mathbf{v}_{i}^{n+1}=\mathbf{F}_{i}^{n}\cdot \delta \mathbf{v}%
_{i-1}^{n}+\mathbf{G}_{i}^{n}\cdot \delta \mathbf{v}_{i}^{n}+\mathbf{H}%
_{i}^{n}\cdot \delta \mathbf{v}_{i+1}^{n},  \label{VE130}
\end{equation}
where 
\begin{equation}
\mathbf{F}_{i}^{n}=\mathbf{C}_{i-0.5}^{n+0.5}\cdot \mathbf{C}_{i-1}^{n},\ 
\mathbf{G}_{i}^{n}=\mathbf{C}_{i-0.5}^{n+0.5}\cdot \mathbf{D}_{i-1}^{n}+%
\mathbf{D}_{i-0.5}^{n+0.5}\cdot \mathbf{C}_{i}^{n},\ \mathbf{H}_{i}^{n}=%
\mathbf{D}_{i-0.5}^{n+0.5}\cdot \mathbf{D}_{i}^{n}.  \label{VE140}
\end{equation}
In view of Theorem \ref{Necessary conditions for GOS-monotonicity}, we
obtain that the following inequalities must be valid 
\begin{equation}
F_{i}^{n,kk},\ G_{i}^{n,kk},\ H_{i}^{n,kk}\geq 0,\quad G_{i}^{n,kk}\geq \min
\left\{ F_{i+1}^{n,kk},H_{i-1}^{n,kk}\right\} \quad \forall i,\ \forall k.
\label{VE150}
\end{equation}
where $F_{i}^{n,kk}$, $G_{i}^{n,kk}$, $H_{i}^{n,kk}$ denote the diagonal
elements of the matrices $\mathbf{F}_{i}^{n}$,\ $\mathbf{G}_{i}^{n}$, and $%
\mathbf{H}_{i}^{n}$, respectively. By analogy with the first stage, the
value of $\aleph _{i}^{n}$ is modified at every node, where at least one
inequality in (\ref{VE150}) is violated.

\begin{figure}[tbph]
\centerline{\includegraphics[width=11.9cm,height=15.4cm]{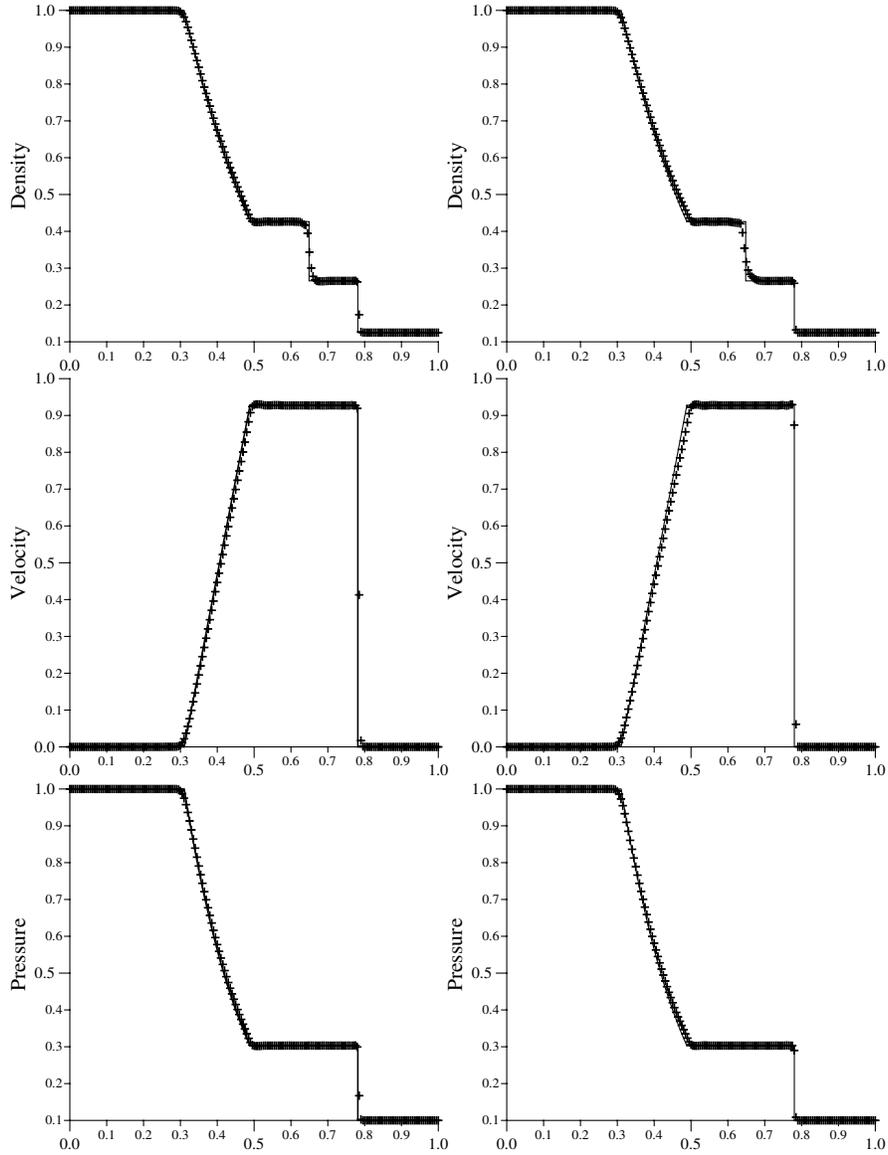}}
\caption{Sod's problem. The scheme COS2 with two-stage modification under $%
C_{r}=0.9$, $\aleph \in \left[ 0.3,1\right] $ (left column) and $C_{r}=1$, $%
\aleph \in \left[ 0,1\right] $ (right column) versus the analytical
solution. Time $t=0.16$, spatial increment $\Delta x=0.005$.}
\label{C09A03}
\end{figure}

First we solve the shock tube problem (see, e.g., \cite{Balaguer and Conde
2005}, \cite{LeVeque 2002}, \cite{Liu and Tadmor 1998}) with Sod's initial
data: 
\begin{equation}
\mathbf{u}_{L}=\left\{ 
\begin{array}{c}
1 \\ 
0 \\ 
2.5
\end{array}
\right\} ,\quad \mathbf{u}_{R}=\left\{ 
\begin{array}{c}
0.125 \\ 
0 \\ 
0.25
\end{array}
\right\} .  \label{VE50}
\end{equation}
Following Balaguer and Conde \cite{Balaguer and Conde 2005} as well as Liu
and Tadmor \cite{Liu and Tadmor 1998} we assume that the computational
domain is $0\leq x\leq 1$; the point $x_{0}$ is located at the middle of the
interval $\left[ 0,1\right] $, i.e. $x_{0}=0.5$; the equations (\ref{VE10})
are integrated up to $t=0.16$ on a spatial grid with 200 nodes as in \cite
{Balaguer and Conde 2005} and in \cite{Liu and Tadmor 1998}. The Courant
number is taken to be $Cr=1$ (or $Cr=0.9$) in contrast to \cite{Balaguer and
Conde 2005} and \cite{Liu and Tadmor 1998}, where the simulations were done
under $\Delta t=0.1\Delta x$ (i.e. $0.13\lesssim Cr\lesssim 0.22$). The
results of simulations with the two-stage modification of the monotonicity
parameter $\aleph $ are depicted in Figure \ref{C09A03}. 

\begin{figure}[tbph]
\centerline{\includegraphics[width=12.75cm,height=16.5cm]{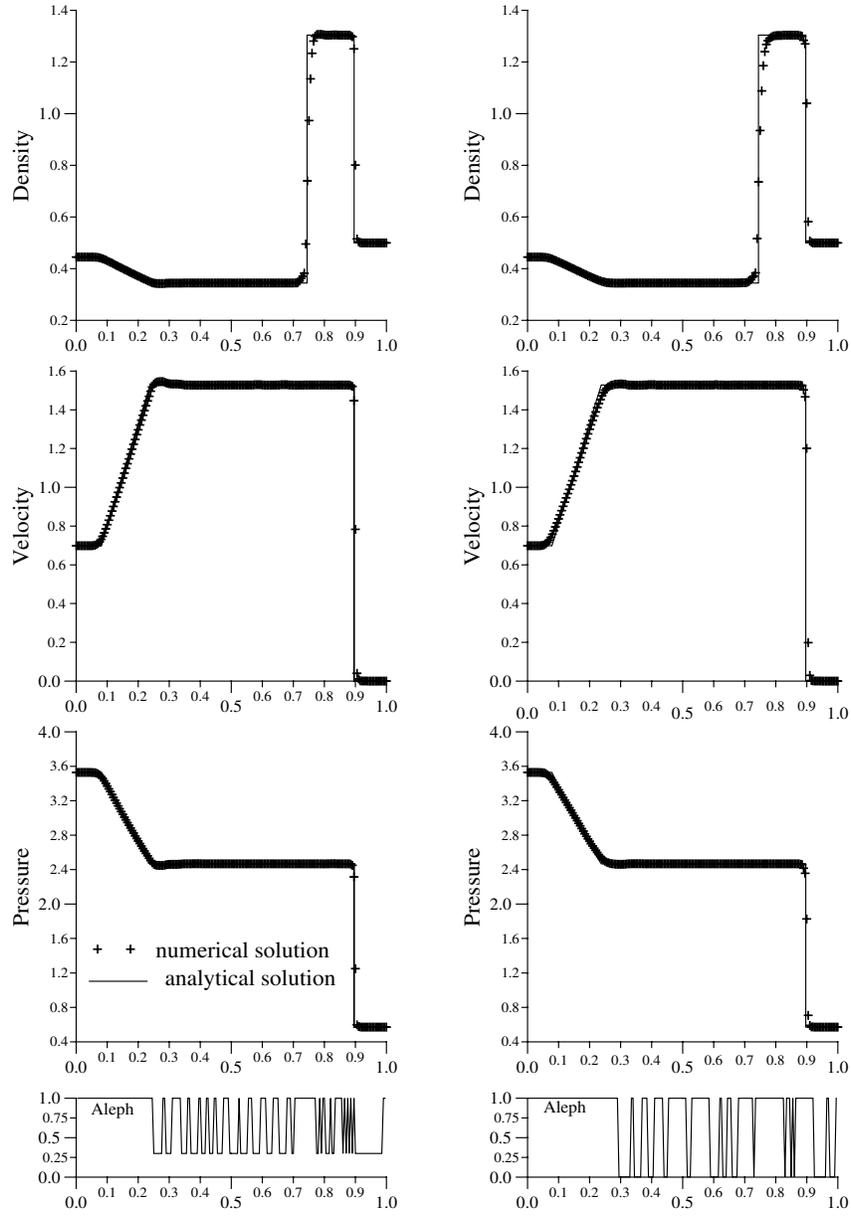}}
\caption{Lax problem. The scheme COS2 with two-stage modification under $%
C_{r}=0.9$, $\aleph \in \left[ 0.3,1\right] $ (left column) and $\aleph \in 
\left[ 0.001,1\right] $ (right column) versus the analytical solution. Time $%
t=0.16$, spatial increment $\Delta x=0.005$, Aleph: the distribution of $%
\aleph $ after the second stage of modifications}
\label{C09LA010}
\end{figure}

The results depicted in Figure \ref{C09A03} are not worse in comparison to
the corresponding third-order central results of \cite[p. 418]{Liu and
Tadmor 1998} as well as to the results obtained by the fourth-order
non-oscillatory scheme in \cite[p. 472]{Balaguer and Conde 2005}. Moreover,
the scheme COS2, (\ref{SA30}), does not produce any spurious oscillations.
We continue the testing of the modified scheme COS2, (\ref{SA30}), using the
shock tube problem with Lax initial data 
\begin{equation}
\mathbf{u}_{L}=\left\{ 
\begin{array}{c}
0.445 \\ 
0.311 \\ 
8.928
\end{array}
\right\} ,\quad \mathbf{u}_{R}=\left\{ 
\begin{array}{c}
0.5 \\ 
0 \\ 
1.4275
\end{array}
\right\} .  \label{VE170}
\end{equation}
The results of simulations with the two-stage modification under $C_{r}=0.9$%
, $\aleph _{\min }=0.3$ (left column) and $\aleph _{\min }=0.001$ (right
column) are depicted in Figure \ref{C09LA010}. 

One can readily see (Figure \ref{C09LA010}) that the scheme COS2 gives a not
worse resolution than that obtained by the schemes studied in \cite{Liu and
Tadmor 1998} and \cite[p. 472]{Balaguer and Conde 2005} without, in fact,
spurious oscillations.

Thus, the algorithm based on Theorem \ref{Necessary conditions for
GOS-monotonicity} shows a possibility to avoid spurious numerical
oscillations in a computed solution totally, and hence the second order
scheme, COS2, is found to be accurate and robust. However this algorithm can
lead to a smearing effect that partly reduces accuracy. There are several
reasons for this side effect of the algorithm. Particularly, the algorithm
is not optimal, i.e. it gives no more than rough approximations from below
to the values of $\aleph $. Moreover, in this algorithm, the value of $%
\aleph $ at a grid node $i$ is considered as a scalar, i.e. it is assumed
that the monotonicity parameter, $\aleph $, is common to all coordinates of
the vector $\mathbf{v}_{i}$. Thus, the algorithm lacks flexibility, and
hence the accuracy of the scheme COS2, (\ref{SA30}), can be reduced. Notice,
the conditions of Theorem \ref{Necessary conditions for GOS-monotonicity}
are not necessary and sufficient conditions for monotonicity (i.e., absence
of spurious oscillations) of the difference scheme COS2, and hence the total
elimination of spurious oscillations on the basis of Theorem \ref{Necessary
conditions for GOS-monotonicity} can lead to an undesirable smearing effect.
All this must be given proper weight in designing of the difference schemes.

\section{Appendix\label{Appendix1}}

\begin{proposition}
\label{Approximate derivative}Let us find the order of accuracy, $r$, in (%
\ref{C110}) if $d_{i}$ will be approximated by $\widetilde{d}_{i}$ with the
order of accuracy $s$, i.e. let 
\begin{equation}
d_{i}=\widetilde{d}_{i}+O\left( \left( \Delta x\right) ^{s}\right) .
\label{G90}
\end{equation}
Let $U\left( x\right) $ be sufficiently smooth, then we can write 
\begin{equation}
U_{i+1}=U_{i+05}+U_{i+05}^{\prime }\frac{\Delta x}{2}+\frac{1}{2}%
U_{i+05}^{\prime \prime }\left( \frac{\Delta x}{2}\right) ^{2}+O\left(
\left( \Delta x\right) ^{3}\right) ,  \label{HA300}
\end{equation}
\begin{equation}
U_{i}=U_{i+05}-U_{i+05}^{\prime }\frac{\Delta x}{2}+\frac{1}{2}%
U_{i+05}^{\prime \prime }\left( \frac{\Delta x}{2}\right) ^{2}+O\left(
\left( \Delta x\right) ^{3}\right) .  \label{HA310}
\end{equation}
Combining the equalities (\ref{HA300}) and \ref{HA310} we obtain 
\begin{equation}
U_{i+1}+U_{i}=2U_{i+05}+\left. \frac{\partial ^{2}U}{\partial x^{2}}\right|
_{i+05}\left( \frac{\Delta x}{2}\right) ^{2}+O\left( \left( \Delta x\right)
^{3}\right) .  \label{HA320}
\end{equation}
In a similar manner we write: 
\begin{equation}
d_{i+1}=U_{i+05}^{\prime }+U_{i+05}^{\prime \prime }\frac{\Delta x}{2}+\frac{%
1}{2}U_{i+05}^{\prime \prime \prime }\left( \frac{\Delta x}{2}\right)
^{2}+O\left( \left( \Delta x\right) ^{3}\right) ,  \label{HA330}
\end{equation}
\begin{equation}
d_{i}=U_{i+05}^{\prime }-U_{i+05}^{\prime \prime }\frac{\Delta x}{2}+\frac{1%
}{2}U_{i+05}^{\prime \prime \prime }\left( \frac{\Delta x}{2}\right)
^{2}+O\left( \left( \Delta x\right) ^{3}\right) .  \label{HA340}
\end{equation}
Subtracting the equations (\ref{HA330}) and (\ref{HA340}), we obtain 
\begin{equation}
\left. \frac{\partial ^{2}U}{\partial x^{2}}\right| _{i+05}=\frac{%
d_{i+1}-d_{i}}{\Delta x}+O\left( \left( \Delta x\right) ^{2}\right) .
\label{HA350}
\end{equation}
In view of (\ref{HA350}) and (\ref{G90}) we obtain from (\ref{HA320}) the
following interpolation formula 
\begin{equation}
U_{i+05}=\frac{1}{2}\left( U_{i+1}+U_{i}\right) -\frac{\Delta x}{8}\left( 
\widetilde{d}_{i+1}-\widetilde{d}_{i}\right) +O\left( \left( \Delta x\right)
^{4}+\left( \Delta x\right) ^{s+1}\right) .  \label{HA360}
\end{equation}
In view of (\ref{HA360}) we obtain that $r=\min \left( 4,s+1\right) .$
\end{proposition}

\begin{proposition}
\label{First to Second order scheme}As applied to central schemes (see
Section \ref{COS1}), we will construct a second order scheme based on
operator-splitting techniques. We will, in fact, use the summarized (summed)
approximation method \cite[Section 9.3]{Samarskii 2001} to estimate order of
approximation. Consider the following equation 
\begin{equation}
\mathcal{P}\mathbf{u}\equiv \mathcal{P}_{1}\mathbf{u}+\mathcal{P}_{2}\mathbf{%
u}\equiv \frac{\partial \mathbf{u}}{\partial t}-L\mathbf{u}=0\mathbf{,\quad }%
\mathcal{P}_{k}\mathbf{u}\equiv \frac{1}{2}\frac{\partial \mathbf{u}}{%
\partial t}-L_{k}\mathbf{u},\ k=1,2,  \label{FS10}
\end{equation}
where $L_{k}$ is an operator, e.g. a differential operator, a real analytic
function, etc', acting on $\mathbf{u}\left( x,t\right) $. We approximate (%
\ref{FS10}) on the cell $\left[ x_{i},x_{i+1}\right] \times \left[
t_{n},t_{n+0.5}\right] $ by the following difference equation with the
accuracy $O(\left( \Delta x\right) ^{2}+\left( \Delta t\right) ^{2})$%
\begin{equation}
\Pi \mathbf{v}\equiv \frac{\mathbf{v}_{i+0.5}^{n+0.5}-\mathbf{v}_{i+0.5}^{n}%
}{0.5\Delta t}-\Lambda _{1}\mathbf{v}^{n+0.25}-\Lambda _{2}\mathbf{v}%
^{n+0.25}=0,  \label{FS15}
\end{equation}
where it is assumed that the operator $L_{k}\mathbf{u}$ is approximated by
the operator $\Lambda _{k}\mathbf{u}$ with the accuracy $O(\left( \Delta
x\right) ^{2})$, i.e. 
\begin{equation}
\Lambda _{k}\mathbf{u}^{n+0.25}=\left( L_{k}\mathbf{u}\right)
_{i+0.5}^{n+0.25}+O\left( \left( \Delta x\right) ^{2}\right) .  \label{FS17}
\end{equation}
In view of the operator splitting idea, to the problem (\ref{FS15}) there
corresponds the following chain of difference schemes 
\begin{equation}
\Pi _{1}\mathbf{w}\equiv \frac{1}{2}\frac{\mathbf{w}_{i+0.5}^{n+0.25}-%
\mathbf{w}_{i+0.5}^{n}}{0.25\Delta t}-\Lambda _{1}\mathbf{w}_{1}^{n+0.25}=0,
\label{FS30}
\end{equation}
\begin{equation}
\Pi _{2}\mathbf{w}\equiv \frac{1}{2}\frac{\mathbf{w}_{i+0.5}^{n+0.5}-\mathbf{%
w}_{i+0.5}^{n+0.25}}{0.25\Delta t}-\Lambda _{2}\mathbf{w}_{2}^{n+0.25}=0.
\label{FS40}
\end{equation}
One can see from the above that the operator $\mathcal{P}_{k}\mathbf{u}$ is
approximated by $\Pi _{k}\mathbf{u}$ with the accuracy $O(\Delta t+\left(
\Delta x\right) ^{2})$%
\begin{equation}
\Pi _{1}\mathbf{u}_{i+0.5}^{n+0.25}=\left( \mathcal{P}_{1}\mathbf{u}\right)
_{i+0.5}^{n+0.25}-\frac{\Delta t}{16}\left( \frac{\partial ^{2}\mathbf{u}}{%
\partial t^{2}}\right) _{i+0.5}^{n+0.25}+O\left( \left( \Delta t\right)
^{2}+\left( \Delta x\right) ^{2}\right) ,  \label{FS60}
\end{equation}
\begin{equation}
\Pi _{2}\mathbf{u}_{i+0.5}^{n+0.25}=\left( \mathcal{P}_{2}\mathbf{u}\right)
_{i+0.5}^{n+0.25}+\frac{\Delta t}{16}\left( \frac{\partial ^{2}\mathbf{u}}{%
\partial t^{2}}\right) _{i+0.5}^{n+0.25}+O\left( \left( \Delta t\right)
^{2}+\left( \Delta x\right) ^{2}\right) .  \label{FS70}
\end{equation}
In view of (\ref{FS60})-(\ref{FS70}), the local truncation error \cite[p.
142]{LeVeque 2002}, $\psi $, on a sufficiently smooth solution $\mathbf{u}%
(x,t)$ to (\ref{FS10}) is found to be 
\begin{equation}
\psi =\Pi \mathbf{u}=\Pi _{1}\mathbf{u+}\Pi _{2}\mathbf{u}=  \label{FS75}
\end{equation}
\begin{equation}
\left( \mathcal{P}_{1}\mathbf{u+}\mathcal{P}_{2}\mathbf{u}\right)
_{i+0.5}^{n+0.25}+O\left( \left( \Delta t\right) ^{2}+\left( \Delta x\right)
^{2}\right) =O\left( \left( \Delta t\right) ^{2}+\left( \Delta x\right)
^{2}\right) .  \label{FS80}
\end{equation}
Thus, implicit Scheme (\ref{FS30}) together with explicit Scheme (\ref{FS40}%
) approximate (\ref{FS10}) with the second order.
\end{proposition}

\end{document}